\documentclass[fleqn,usenatbib]{mnras}

\usepackage{newtxtext,newtxmath}

\usepackage[T1]{fontenc}
\usepackage{ae,aecompl}
\usepackage{lastpage}
\usepackage[nolist]{acronym}

\usepackage{graphicx}	
\usepackage{amsmath}	
\usepackage{amssymb}	
\usepackage{xspace}   
\usepackage{hyperref}

\newcommand{\cxo}{{\it Chandra}}

\def\xmm{\textit{XMM--Newton}}
\newcommand{\ergcm}[1]{erg~cm$^{-2}$ s$^{-1}$}

\newcommand{\HI}{{H\,{\sc i}}\xspace}
\newcommand{\HII}{{H\,{\sc ii}}\xspace}

\newcommand{\D}{$^\circ$}

\begin{acronym}
\acro{AGN}{Active Galactic Nuclei}
\acro{ALMA}{Atacama Large Millimeter Array}
\acro{ATCA}{Australia Telescope Compact Array}
\acro{ATNF}{Australia Telescope National Facility}
\acro{ATOA}{Australia Telescope Online Archive}
\acro{AT20G}{Australia Telescope 20 GHz Survey}
\acro{ASKAP}{Australian Square Kilometre Array Pathfinder}
\acro{CASS}{CSIRO Astronomy and Space Science}
\acro{CABB}{Compact Array Broadband Backend}
\acro{CSIRO}{Australian Commonwealth Scientific and Industrial Research Organisation}
\acro{CSS}{Compact Steep Spectrum}
\acro{CTA}{Cherenkov Telescope Array}
\acro{EMU}{Evolutionary Map of the Universe}
\acro{ESP}{Early Science Project}
\acro{FWHM}{Full Width at Half-Maximum power}
\acro{GPS}{Gigahertz Peak Spectrum}
\acro{HFP}{High Frequency Peaker}
\acro{ISM}{Interstellar Medium}
\acro{LMC}{Large Magellanic Cloud}
\acro{MCs}{Small and Large Magellanic Clouds}
\acro{MCELS}{Magellanic Cloud Emission Line Survey}
\acro{MIRIAD}[{\tt MIRIAD}]{Multichannel Image Reconstruction, Image Analysis and Display}
\acro{MOST}{Molonglo Observatory Synthesis Telescope}
\acro{MRC}{Molonglo Reference Catalogue of Radio Sources}
\acro{MWA}{Murchison Widefield Array}
\acro{NRAO}{National Radio Astronomy Observatory}
\acro{NVSS}{NRAO VLA Sky Survey}
\acro{OPAL}{Online Proposal Applications \& Links}
\acro{pc}{parsec\acroextra{: 1 pc $\simeq 3.09 \times 10^{16}$ m}}
\acro{PMN}{Parkes-MIT-NRAO}
\acro{PN}{planetary nebula}
\acro{PWN}{Pulsar Wind Nebulae}
\acro{PNe}{planetary nebulae}
\acro{POSSUM}{Polarisation Sky Survey. of the Universe's Magnetism}
\acro{RM}{Rotation Measure}
\acro{QSO}{Quasi-Stellar Objects}
\acro{RFI}{Radio-Frequency Interference}
\acro{RMS}{Root Mean Squared}
\acro{SCEM}{School of Computing, Engineering and Mathematics}
\acro{SED}{Spectral Energy Distribution}
\acro{SFG}{Star Forming Galaxies}
\acro{SI}[$\alpha$]{spectral index\acroextra{, $S \propto \nu^\alpha$}}
\acro{SKA}{Square Kilometre Array}
\acro{SMBH}{Super Massive Black Hole}
\acro{SMC}{Small Magellanic Cloud}
\acro{SN}{supernova}
\acro{SNRs}{supernova remnants}
\acro{SNR}{supernova semnant}
\acro{SUMSS}{Sydney University Molonglo Sky Survey}
\acro{WISE}{Wide-Field Infrared Survey Explorer}
\acro{VLA}{Very Large Array} 
\acro{VLBI}{Very Long Baseline Interferometry} 
\acro{MYSO}{Massive Young Stellar Object}
\acro{HFPs}{High Frequency Peakers}
\acro{ESP}{Early Science Project}
\acro{YSOs}{young stellar objects}
\acro{IRAC}{Infrared Array Camera}
\acro{BL-Lac}{BL~Lacertae}
\acro{FSRQs}{Flat Spectrum Radio Quasars}
\acro{IR}{Infrared}
\acro{WAT}{Wide-Angle Tail}
\end{acronym}

\title[The ASKAP EMU ESP: Radio Continuum Survey of the SMC]{The ASKAP-EMU Early Science Project: \\
Radio Continuum Survey of the Small Magellanic Cloud}

\author[T. D. Joseph et al.]{T. D. Joseph,$^{1}$\thanks{E-mail: tana.joseph@manchester.ac.uk (TDJ)}
M. D. Filipovi\' c,$^{2}$
E. J. Crawford,$^{2}$
I. {Boji{\v c}i{\'c}},$^{2}$
E. L. Alexander,$^{1}$
\newauthor
G. F. Wong,$^{2,3}$
H. Andernach,$^{4}$  
H. Leverenz,$^{2}$ 
R. P. Norris,$^{2,3}$
R. Z. E. Alsaberi,$^{2}$ 
\newauthor
C. Anderson,$^{3}$ 
L. A. Barnes,$^{2}$ 
L. M. Bozzetto,$^{2}$ 
F. Bufano,$^{6}$ 
J. D. Bunton,$^{5}$ 
F. Cavallaro,$^{6}$ 
\newauthor
J. D. Collier,$^{2,7}$ 
H. D\'{e}nes,$^{8}$ 
Y. Fukui,$^{9,10}$ 
T. Galvin,$^{2,3}$ 
F. Haberl,$^{11}$ 
A. Ingallinera,$^{6}$ 
\newauthor
A. D. Kapinska,$^{12}$ 
B. S. Koribalski,$^{3}$ 
R. Kothes,$^{13}$ 
D. Li,$^{14,15}$ 
P. Maggi,$^{16}$ 
C. Maitra,$^{11}$ 
\newauthor
P. Manojlovi\'c,$^{2,3}$ 
J. Marvil,$^{12}$ 
N. I. Maxted,$^{2,17}$ 
A. N. O'Brien,$^{2,3}$ 
J. M. Oliveira,$^{18}$ 
\newauthor
C. M. Pennock,$^{18}$ 
S. Riggi,$^{6}$ 
G. Rowell,$^{19}$ 
L. Rudnick,$^{20}$ 
H. Sano,$^{9,10}$ 
M. Sasaki,$^{21}$ 
\newauthor 
N. Seymour,$^{22}$ 
R. Soria,$^{15,23}$ 
M. Stupar,$^{2}$ 
N. F. H. Tothill,$^{2}$ 
C. Trigilio,$^{6}$ 
K. Tsuge,$^{10}$ 
\newauthor
G. Umana,$^{6}$ 
D. Uro\v sevi\' c,$^{24,25}$ 
J. Th. van Loon,$^{18}$ 
E. Vardoulaki,$^{26}$ 
V. Velovi\'c,$^{2}$ 
\newauthor
M. Yew,$^{2}$ 
D. Leahy,$^{27}$ 
Y.-H. Chu,$^{28}$ 
M. J. Micha\l{}owski,$^{29}$ 
P. J. Kavanagh,$^{30}$ 
\newauthor
K. R. Grieve$^{2}$
 \\ \\
$^{1}$Department of Physics and Astronomy, University of Manchester, Oxford Road, Manchester, M13 9PL, UK\\
$^{2}$Western Sydney University, Locked Bag 1797, Penrith South DC, NSW 2751, Australia\\
$^{3}$CSIRO Astronomy and Space Science, PO Box 76, Epping, NSW 1710, Australia\\
$^{4}$Depto. de Astronom\'ia, DCNE, Universidad de Guanajuato, Apdo. Postal 144, Guanajuato, CP 36000, Gto., Mexico\\
$^{5}$School of Physics, The University of New South Wales, Sydney, 2052, Australia\\
$^{6}$INAF Osservatorio Astrofisico di Catania, via Santa Sofia 78, I-95123 Catania \\
$^{7}$The Inter-University Institute for Data Intensive Astronomy (IDIA), Department of Astronomy, University of Cape Town,\\ Rondebosch, 7701, South Africa\\
$^{8}$ASTRON - Netherlands Institute for Radio Astronomy, 7991 PD, Dwingeloo, The Netherlands\\
$^{9}$Institute for Advanced Research, Nagoya University, Furo-cho, Chikusa-ku, Nagoya 464-8601, Japan\\
$^{10}$Department of Physics, Nagoya University, Furo-cho, Chikusa-ku, Nagoya 464-8601, Japan\\
$^{11}$Max-Planck-Institut f\"{u}r extraterrestrische Physik, Giessenbachstra\ss e, D-85748 Garching, Germany\\
$^{12}$National Radio Astronomy Observatory, 1003 Lopezville Rd., Socorro, NM 87801, USA\\
$^{13}$Dominion Radio Astrophysical Observatory, Herzberg Programs in Astronomy and Astrophysics, National Research Council Canada,\\ PO Box 248, Penticton, BC V2A 6J9, Canada \\
$^{14}$CAS Key Laboratory of FAST, National Astronomical Observatories, Chinese Academy of Sciences, Beijing, 100101,  China\\
$^{15}$School of Astronomy and Space Sciences, University of Chinese Academy of Sciences, Beijing 100049, China\\
$^{16}$Observatoire Astronomique de Strasbourg, Universit\'e de Strasbourg, CNRS, 11 rue de l'Universit\'e, F-67000 Strasbourg, France\\
$^{17}$School of Science, The University of New South Wales, Australian Defence Force Academy, Canberra, 2600, Australia\\
$^{18}$Lennard-Jones Laboratories, Keele University, ST5 5BG, UK\\
$^{19}$School of Physical Sciences, The University of Adelaide, Adelaide 5005, Australia\\
$^{20}$School of Physics and Astronomy, University of Minnesota, Minneapolis, MN 55455, USA\\
$^{21}$Remeis Observatory and ECAP, Universit\"{a}t Erlangen-N\"{u}rnberg, Sternwartstr. 7, D-96049 Bamberg, Germany\\
$^{22}$International Centre for Radio Astronomy Research, Curtin University, Bentley, WA 6102, Australia\\
$^{23}$Sydney Institute for Astronomy, School of Physics A28, The University of Sydney, Sydney, NSW 2006, Australia \\
$^{24}$Department of Astronomy, Faculty of Mathematics, University of Belgrade, Studentski trg 16, 11000 Belgrade, Serbia\\
$^{25}$Isaac Newton Institute of Chile, Yugoslavia Branch\\
$^{26}$Argelander-Institut f\"{u}r Astronomie, Auf dem H\"{u}gel 71, D-53121 Bonn, Germany \\
$^{27}$Department of Physics and Astronomy, University of Calgary, University of Calgary, Calgary, Alberta, T2N 1N4, Canada\\
$^{28}$Institute of Astronomy and Astrophysics, Academia Sinica (ASIAA), Taipei 10617, Taiwan\\
$^{29}$Astronomical Observatory Institute, Faculty of Physics, Adam Mickiewicz University, ul. S\l{}oneczna 36, PL-60-286 Pozna\'n, Poland\\
$^{30}$School of Cosmic Physics, Dublin Institute for Advanced Studies, 31 Fitzwillam Place, Dublin 2, Ireland\\
}
\date{Accepted XXX. Received YYY; in original form ZZZ}
\pubyear{2019}
\hypersetup{draft}
\begin{document}
\label{firstpage}
\pagerange{\pageref{firstpage}--\pageref{lastpage}}
\maketitle

\clearpage
\newpage
\begin{abstract}
We present two new radio continuum images from the \ac{ASKAP} survey in the direction of the \ac{SMC}. These images are part of the \ac{EMU} \ac{ESP} survey of the Small and Large Magellanic Clouds. The two new source lists produced from these images contain radio continuum sources observed at 960\,MHz (4489 sources) and 1320\,MHz (5954 sources) with a bandwidth of 192\,MHz and beam sizes of 30.0\arcsec$\times$30.0\arcsec\ and 16.3\arcsec$\times$15.1\arcsec, respectively. The median \ac{RMS} noise values are 186\,$\mu$Jy\,beam$^{-1}$ (960\,MHz) and 165\,$\mu$Jy\,beam$^{-1}$ (1320\,MHz). To create point source catalogues, we use these two source lists, together with the previously published \ac{MOST} and the \ac{ATCA} point source catalogues to estimate spectral indices for the whole population of radio point sources found in the survey region. Combining our \ac{ASKAP} catalogues with these radio continuum surveys, we found 7736 point-like sources in common over an area of 30\,deg$^2$. In addition, we report the detection of two new, low surface brightness supernova remnant candidates in the SMC. The high sensitivity of the new \ac{ASKAP} \ac{ESP} survey also enabled us to detect the bright end of the \ac{SMC} planetary nebula sample, with 22 out of 102 optically known planetary nebulae showing point-like radio continuum emission. Lastly, we present several morphologically interesting background radio galaxies.  
\end{abstract}




\section{Introduction}
\label{intro}

This is an exciting time for the study of nearby galaxies. These nearby external galaxies offer an ideal laboratory, since they are close enough to be resolved, yet located at relatively well known distances \citep[see e.g.][]{2019Natur.567..200P}. New generations of Magellanic Cloud (MC) surveys across the entire electromagnetic spectrum reflect a major opportunity to study different objects and processes in the elemental enrichment of the \ac{ISM}. The study of these interactions in different domains, including radio, optical and X-ray, allow a better understanding of objects such as \ac{SNRs}, \ac{PNe}, (Super)Bubbles and their environments, \ac{YSOs}, symbiotic (accreting compact object) binaries and Wolf-Rayet (WR) wind-wind-collision binaries.

Various new high resolution ($\sim$1\arcsec) and high sensitivity surveys of the \ac{MCs}, such as \xmm\ and \cxo\ \citep[X-rays; see e.g.][]{2012A&A...545A.128H}, \emph{Herschel} \citep{2011AJ....142..102G} and \emph{Spitzer} \citep[IR;][]{2006AJ....132.2268M}, UM/CTIO Magellanic Cloud Emission Line Survey \citep[MCELS, optical;][]{2005AAS...20713203W} and ATCA/MOST (radio), provide a solid base for detailed multi-wavelength studies of radio objects within and behind the \ac{MCs}.

Our main area of interest is the radio objects natal to the \ac{MCs}, particularly \ac{SNRs} and \ac{PNe}. To date, some 85 \ac{SNRs} in the \ac{MCs} have been identified, with a further 20 candidates awaiting confirmation \citep{2016A&A...585A.162M,2017ApJS..230....2B}. Similarly, over 50 \ac{PNe} \citep{2009MNRAS.399..769F,2010SerAJ.181...63B,2016Ap&SS.361..108L,2017MNRAS.468.1794L} and hundreds of \HII\ regions and \ac{YSOs} have been identified \citep[see for example][]{2013MNRAS.428.3001O}. Over 8500 radio sources have also been detected in the region of the Clouds -- mainly AGN, radio galaxies and quasars \citep[][Grieve et al. in prep.]{2012SerAJ.185...53W,2016PhDT.......266C}. Additionally, some comprehensive studies of the magnetic fields of the \ac{MCs} have been undertaken with the present generation of radio continuum surveys \citep[ATCA;][]{2005Sci...307.1610G,2008ApJ...688.1029M,2012ApJ...759...25M}.

In this paper, we focus on the Small Magellanic Cloud (\ac{SMC}), a dwarf irregular galaxy. Its proximity \citep[$\sim $60\,kpc;][]{2005MNRAS.357..304H} enables us to conduct detailed radio frequency studies of its gas and stellar content, without the complication of the foreground emission and absorption we encounter when working within our own Galaxy. For these reasons, the \ac{SMC} has been the subject of many radio studies over several decades. 

Starting in the mid 1970s, the \ac{SMC} has been the subject of both single dish and interferometric radio continuum surveys. These monitoring campaigns have produced over a dozen catalogues of sources towards the \ac{SMC} \citep{1976MNRAS.174..393C,1976AuJPh..29..329M,1986A&A...159...22H,1990PKS...C......0W,1997A&AS..121..321F,1998PASA...15..280T,1998A&AS..130..421F,1997A&AS..121..321F,2002MNRAS.335.1085F,2004MNRAS.355...44P,2005MNRAS.364..217F,2006MNRAS.367.1379R,2007MNRAS.376.1793P,2011SerAJ.182...43W,2011SerAJ.183...95C,2011SerAJ.183..103W,2012SerAJ.184...93W,2012SerAJ.185...53W,2018MNRAS.480.2743F} \citep[see also Table 1 in][for details]{2011SerAJ.183..103W}.

For the reasons mentioned above, the SMC was also selected as a prime target for the Early Science Project (\ac{ESP}) of the newly built Australian Square Kilometre Array Pathfinder \citep[\ac{ASKAP};][]{2008ExA....22..151J}. \ac{ASKAP} is a radio interferometer that allows us to survey the \ac{SMC} with regularly sampled observations. \ac{ASKAP} also provides sensitivity down to the $\mu$Jy range as well as a large field of view of 30\,deg$^2$ \citep{2013PASA...30....6M}. The goal of this project is to produce  high sensitivity and high resolution continuum images of the \ac{MCs} as well as to catalogue discrete radio continuum sources.

The \ac{ASKAP} \ac{EMU} \ac{ESP} survey will be a good complement to and, in some cases, a significant improvement on previous similar studies of the southern skies. For instance, the Australia Telescope Large Area Survey \citep[ATLAS;][]{2006AJ....132.2409N,2008AJ....135.1276M} was a 1400\,MHz radio survey covering a total of roughly 6\,deg$^2$ on the sky, down to an \ac{RMS} noise level of $<$30\,$\mu$Jy, requiring 380 hours of observation time. This survey uncovered over 3000 distinct radio sources out to a redshift of 2. ASKAP's higher resolution and increased sensitivity will be able to achieve such results on a much shorter time scale \citep[see Fig.~1 in][]{2015MNRAS.453.4020F}.

Another obvious advantage of \ac{ASKAP} is the size of the field of view. For example, the \ac{SUMSS} would need $\sim$16 fields and $\sim$192~hours to cover the \ac{ASKAP} \ac{EMU} \ac{SMC} survey area to the required sensitivity \citep[see][]{2003MNRAS.342.1117M}; in contrast the \ac{ASKAP} observations were composed of eight fields of about 12 hours each (a total of 96 hours). 

In this paper we present two new catalogues from the \ac{ASKAP} \ac{ESP} surveys for different types of radio continuum sources towards the \ac{SMC}. These catalogues were obtained from images taken at 960\,MHz ($\lambda = 32$\,cm) and 1320\,MHz ($\lambda = 23$\,cm). For the point source catalogue, we combine the \ac{ASKAP} data with the previously published \ac{MOST} catalogue \citep{1998PASA...15..280T,2011SerAJ.183..103W} and the \ac{ATCA} $\lambda$= 20, 13, 6 and 3\,cm catalogues \citep[][and references therein]{ 2011SerAJ.183..103W,2012SerAJ.184...93W}. 

The paper is laid out as follows: Section~\ref{data} describes the data used to create the source lists. In section~\ref{sdet} we describe the source detection methods used, section~\ref{cat} describes the new \ac{ASKAP} source catalogues and in section~\ref{scompare} we compare our work to previous catalogues of point sources towards the \ac{SMC}. Sections~\ref{SNR_samp} and \ref{PNR_samp} describe the latest \ac{ASKAP} \ac{SMC} populations of \ac{SNRs} and \ac{PNe}, respectively. In Section~\ref{Other_sources}, we briefly discuss other sources of interest, including those behind the SMC.

\section{Data, Observing and Processing}
\label{data}

\begin{figure*}
	\includegraphics[scale=0.9,angle=-90, trim=73 125 60 0, clip]{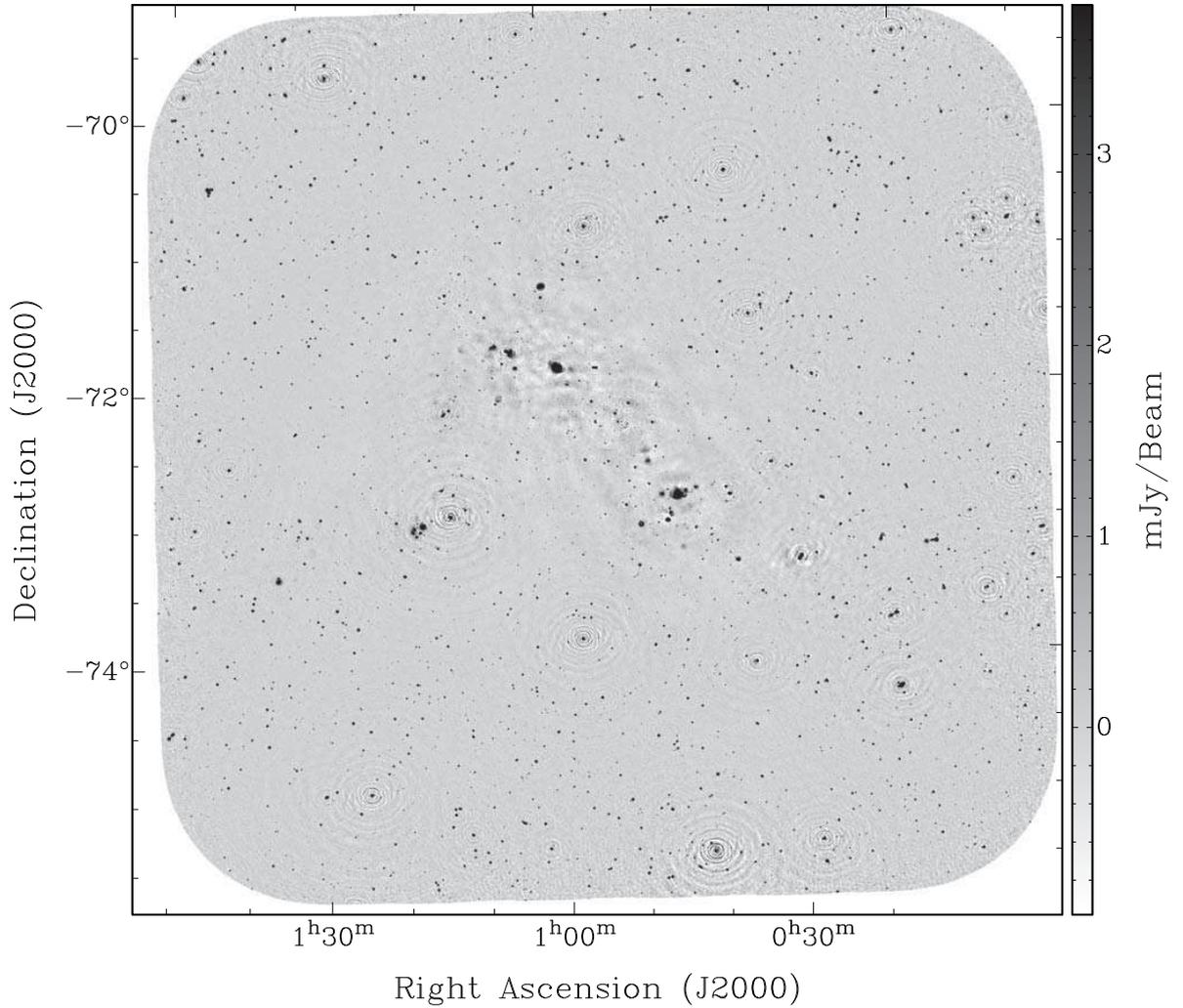}
	\caption{\ac{ASKAP} \ac{ESP} image of the \ac{SMC} at 960\,MHz. The beam size is 30.0\arcsec\ $\times$ 30.0\arcsec\ and the side scale bar represents the image grey scale intensity range.}
	\label{img_960}
\end{figure*}

\begin{figure*}
	\includegraphics[scale=0.8,angle=-90, trim=23 115 50 0, clip]{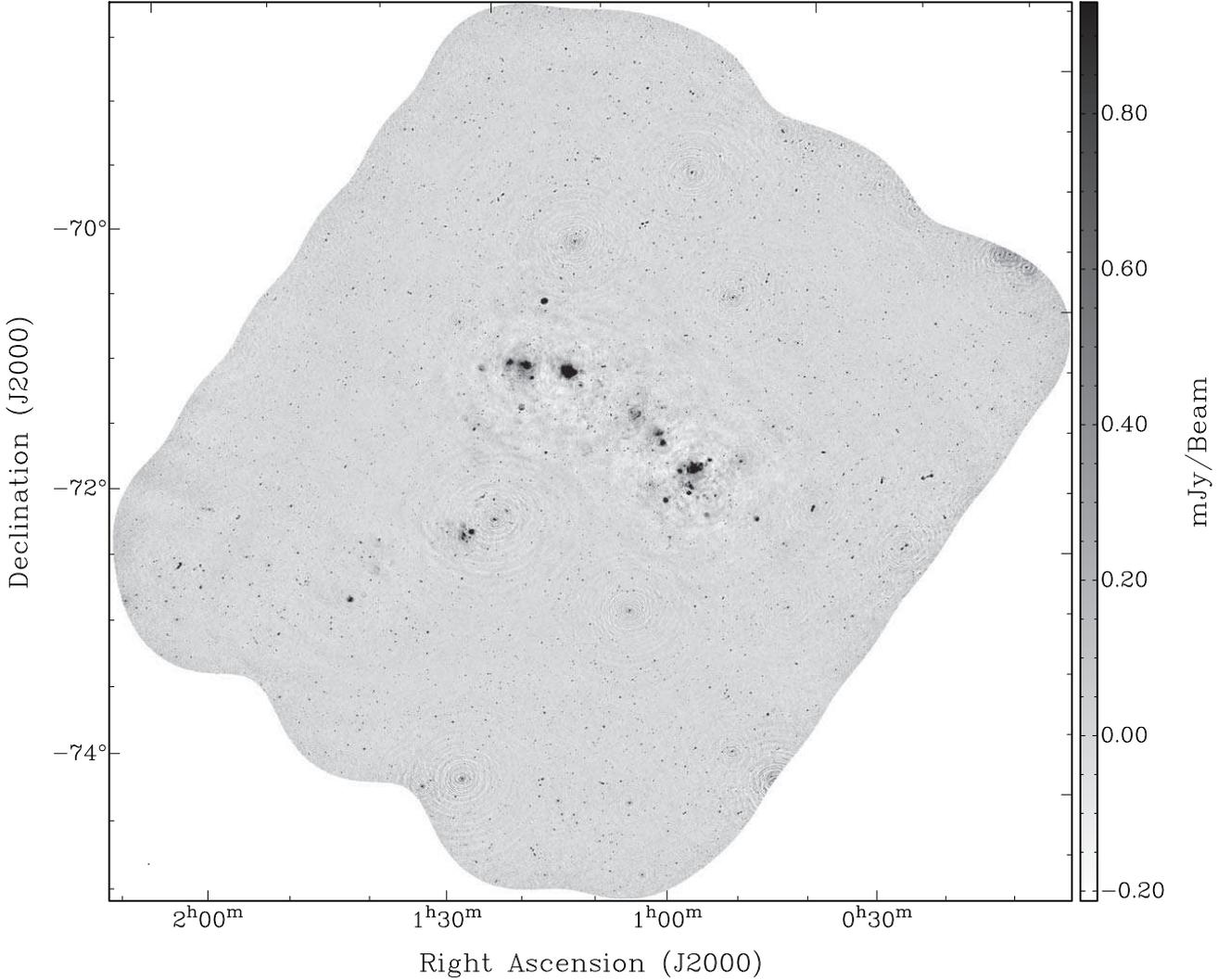}
	\caption{\ac{ASKAP} \ac{ESP} image of the \ac{SMC} at 1320\,MHz. The beam size is 16.3\arcsec\ $\times$ 15.1\arcsec\ and the side scale bar represents the image grey scale intensity range.}
	\label{img_1320}
\end{figure*}

The \ac{SMC} was observed as part of the \ac{ASKAP} commissioning and early science verification \citep{2009IEEEP..97.1507D, 2014PASA...31...41H, 2016PASA...33...42M}. Here, we present observations at 960\,MHz taken on 2017~September~3 (Figure\,\ref{img_960}; using 12 antennas: 2, 3, 4, 6, 12, 14, 16, 17, 19, 27, 28, and 30), and  at 1320\,MHz on 2017~November~3~-~5 (Figure~\ref{img_1320}, using 16 antennas: 1, 2, 3, 4, 5, 6, 10, 12, 14, 16, 17, 19, 24, 27, 28, 30). The \HI\ spectral and dynamical analyses of the 1320\,MHz data have been presented in \citet{2018NatAs...2..901M} and \citet{2019MNRAS.483..392D} respectively. 

We note that the current observations were made with only 33\,per\,cent and 44\,per\,cent (for 960\,MHz and 1320\,MHz respectively) of the full \ac{ASKAP} antenna configuration and 66\,per\,cent of the final bandwidth that will be available in the final array. We believe that with the full array, we will be able to achieve a factor of two increase in sensitivity compared to what is currently possible.

A bandwidth of 192\,MHz was used and the maximum baseline for these observations was 2.3\,km. The observations cover a total field of view of 30\,deg$^2$, with exposure times of 10 to 11 hours per pointing. To optimise sensitivity and survey speed, the 36 beams on each antenna were configured in a hexagonal grid on the sky \citep{mcconnell17}. The source 1934-638 was observed and used for the flux density calibration of all images.

The data calibration, processing, and imaging were carried out using the ASKAPsoft pipeline \citep{Cornwell11}. For both sets of images we processed the data with the multiscale clean algorithm, noting from our previous work \citep{2011SerAJ.182...43W} that the largest detectable features were $\sim$192\arcsec. Therefore, we selected spacial scales of 192\arcsec, 96\arcsec\ and 48\arcsec\ as a geometric progression. We also noted features on the scale of 16\arcsec, and so this spatial scale was also selected. The 1320\,MHz image was cleaned and then mosaiced. For the 960\,MHz image, we  set the pixel size to 6\arcsec, and set the restoring beam to $30\arcsec\times 30\arcsec$ in order to maximise our resolution and sensitivity and to more easily compare these new results with other \ac{SMC} surveys referenced in this work.

\begin{figure*}
	\includegraphics[scale=0.78,angle=-90, trim=0 55 00 0, clip]{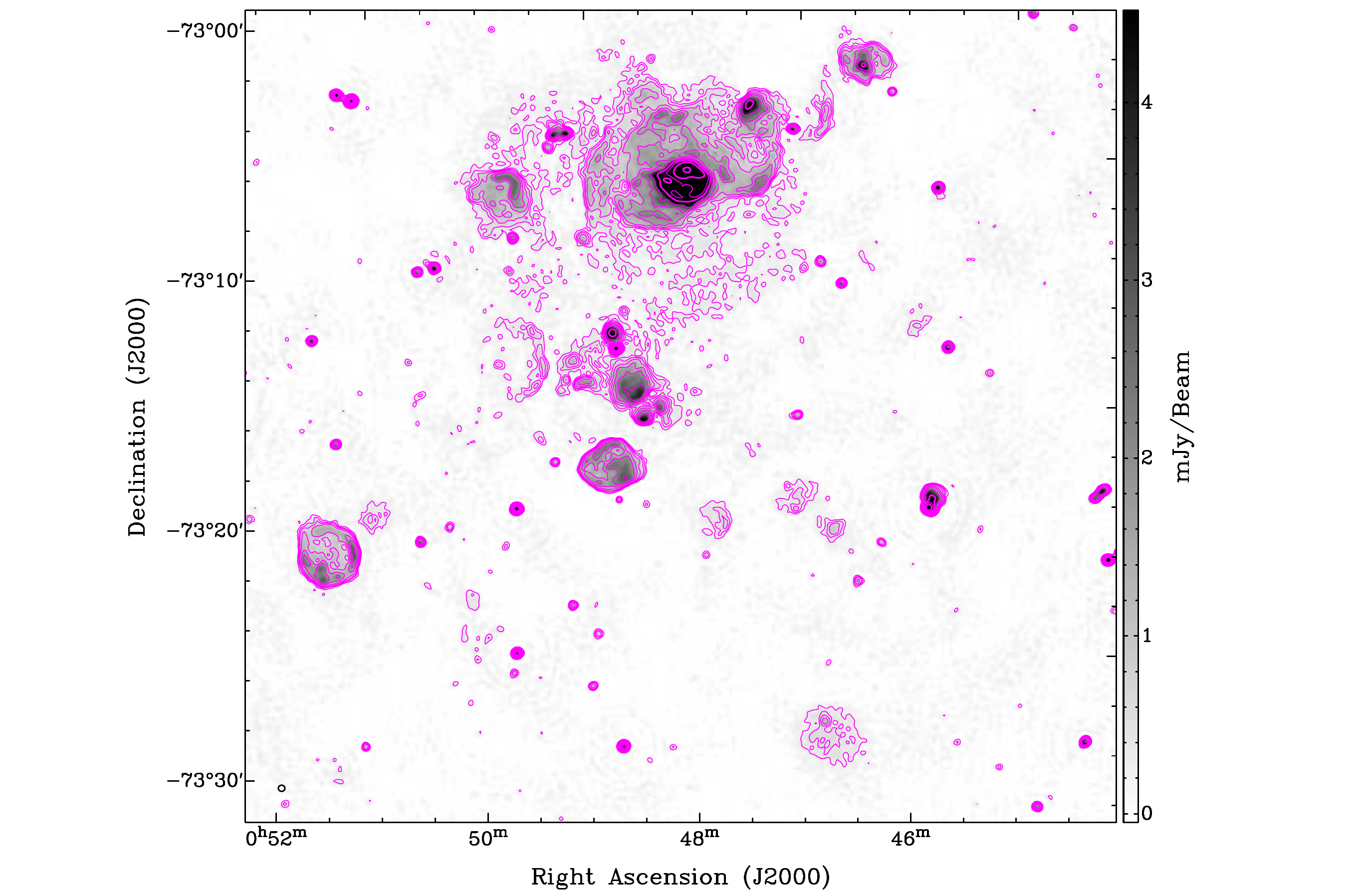}
	\caption{\ac{ASKAP} \ac{ESP} image of the \ac{SMC} N\,19 region at 1320\,MHz (grey scale and contours). Magenta contours are: 0.3, 0.5, 0.7, 1, 1.5, 2, 3, 5, 7, 10 and 15~mJy~beam$^{-1}$. The beam size of 16.3\arcsec\ $\times$ 15.1\arcsec\ is shown as a small black ellipse in the lower left corner.}
	\label{img_SMC_N19}
\end{figure*}

\begin{figure*}
	\includegraphics[scale=0.8,angle=-90, trim=90 75 90 0, clip]{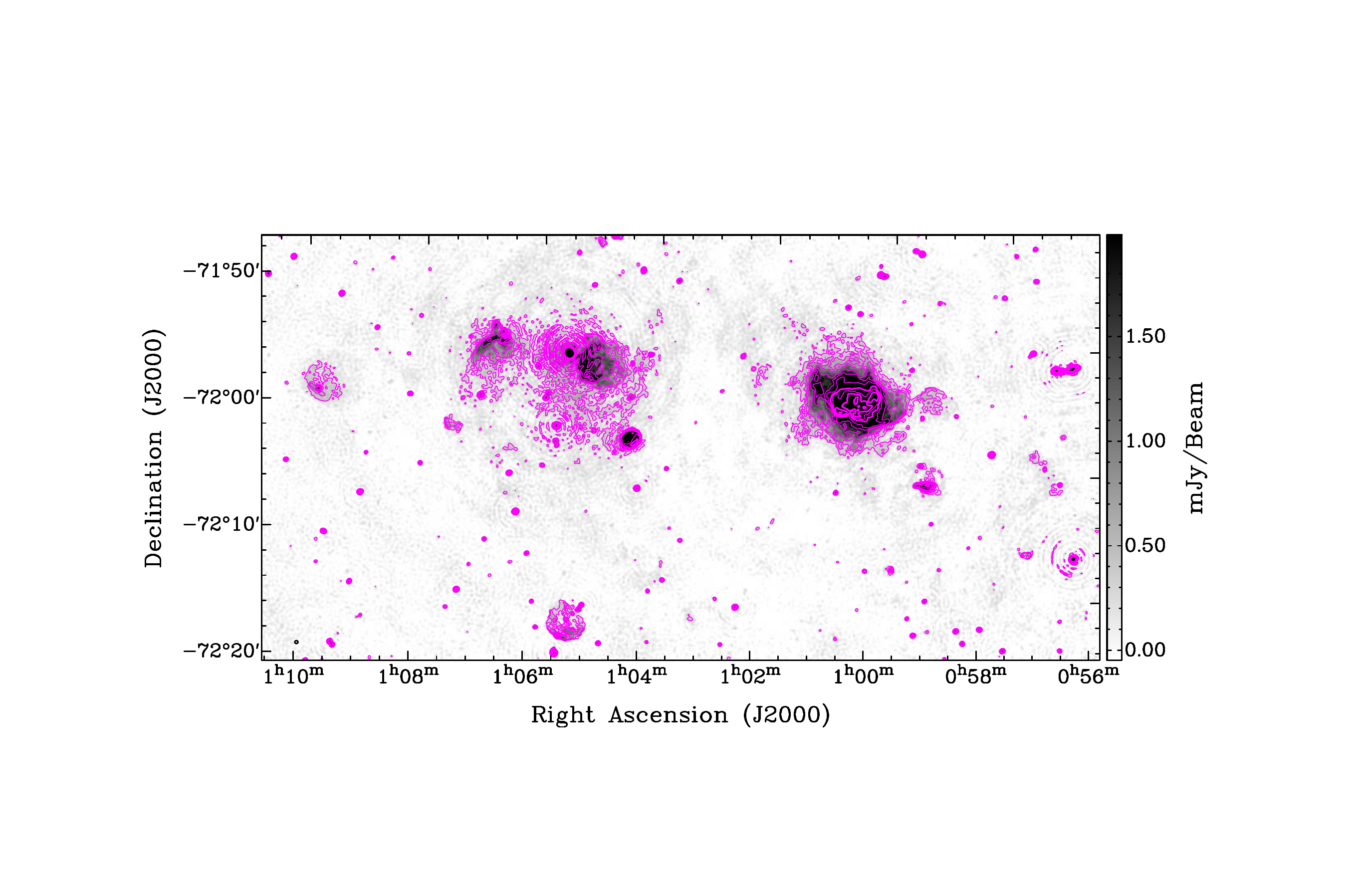}
	\caption{\ac{ASKAP} \ac{ESP} image of the \ac{SMC} N\,66 and SNR\,1E0102-72 region at 1320\,MHz (grey scale and contours). Magenta contours are: 0.3, 0.5, 0.7, 1, 1.5, 2, 3, 5, 7, 10 and 15~mJy~beam$^{-1}$. The beam size of 16.3\arcsec\ $\times$ 15.1\arcsec\ is shown as a small black ellipse in the lower left corner.}
	\label{img_SMC_N66}
\end{figure*}

\begin{figure}
	\includegraphics[width=0.8\columnwidth,trim=90 140 80 80,clip,angle=-90]{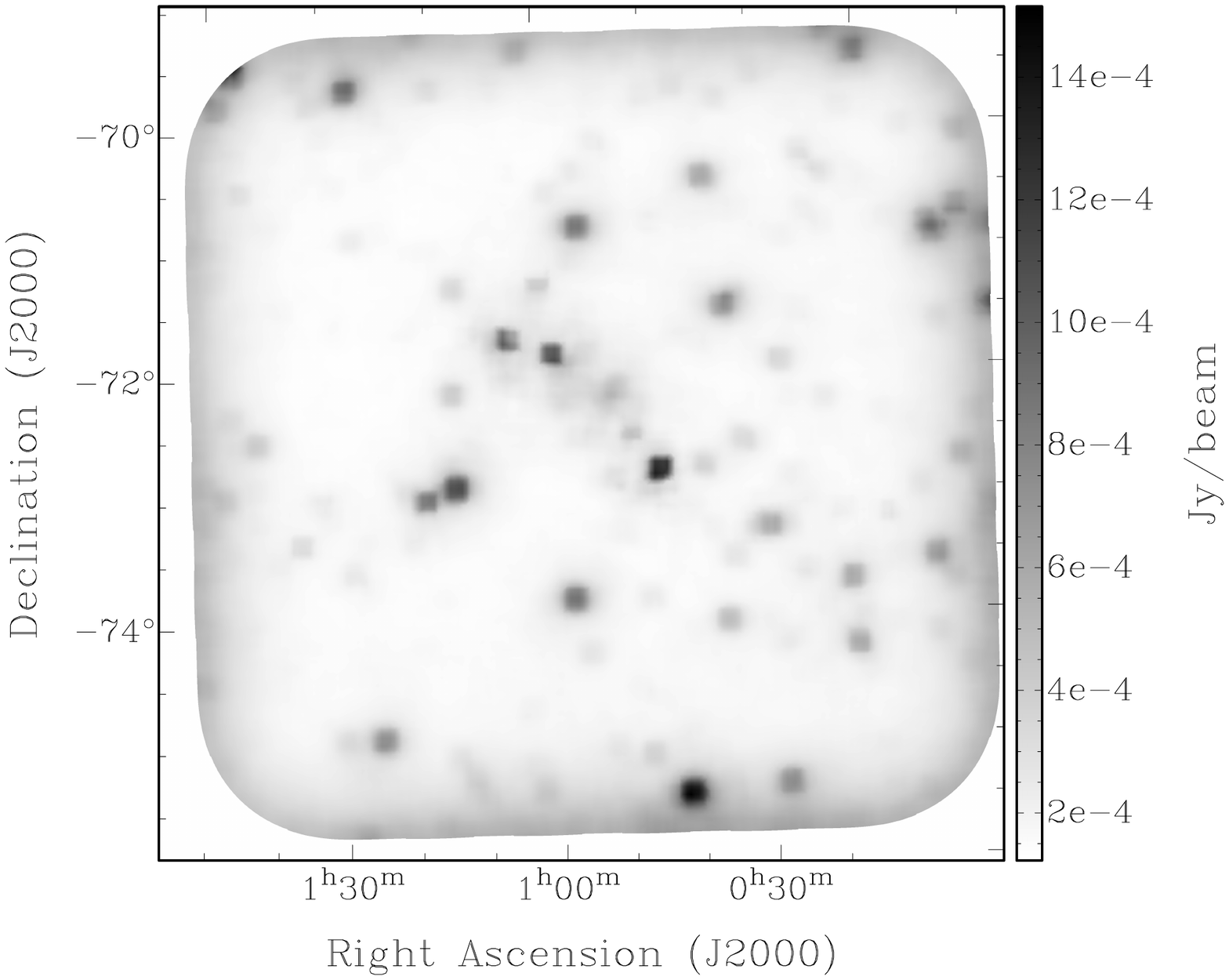}
	\caption{\ac{RMS} map of the 960\,MHz \ac{ASKAP} observations, produced by \texttt{BANE} with the default parameters. The image is of the same pixel scale as in Figure~\ref{img_960}. Higher RMS levels are found at the edge of the field (where only one beam is present) and around the brighter sources.}
	\label{fig:960MHz_rms}
\end{figure}

\begin{figure}
	\includegraphics[width=0.78\columnwidth,trim=10 0 10 0,clip,angle=-90]{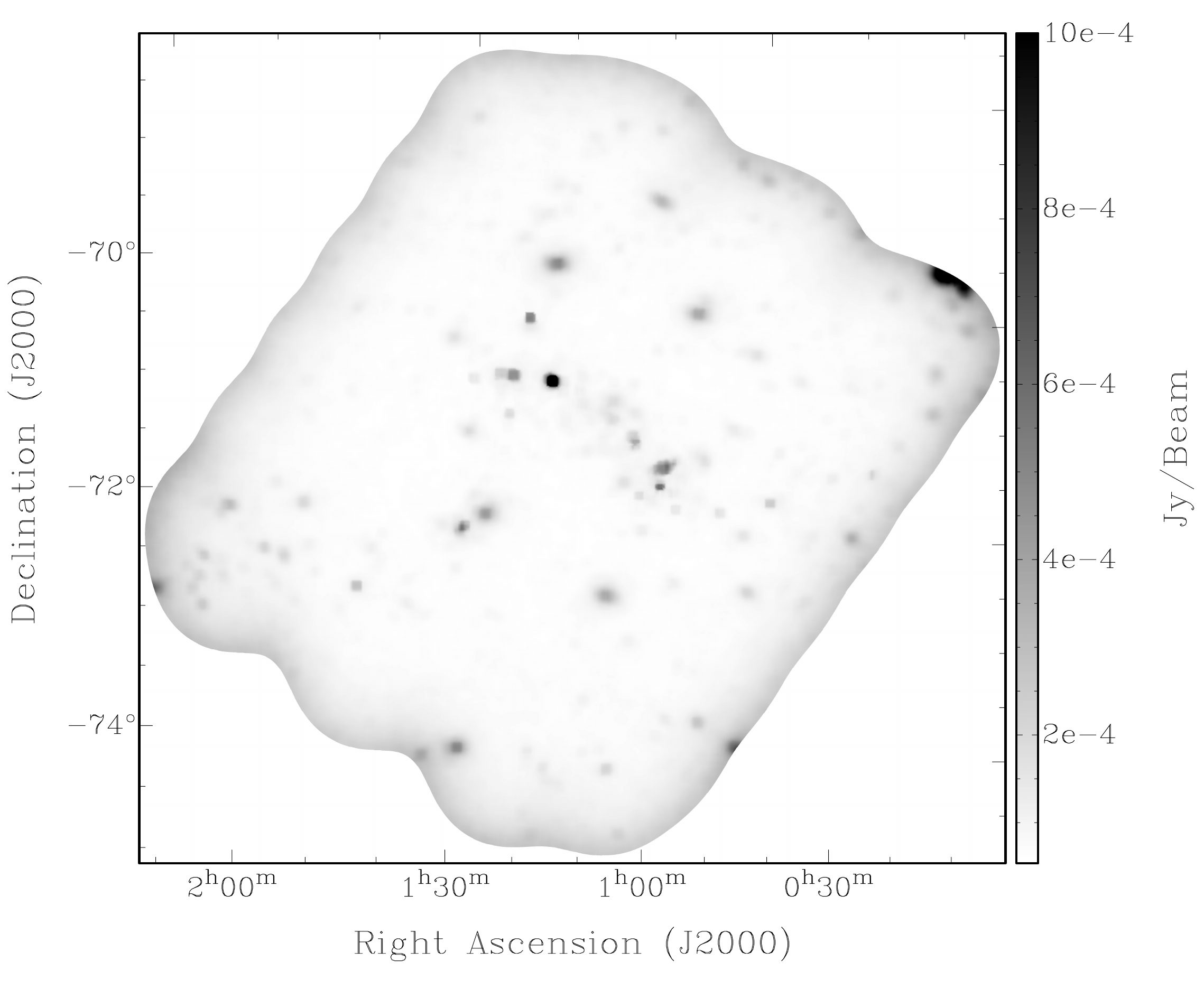}
	\caption{\ac{RMS} map of the 1320\,MHz \ac{ASKAP} observations, produced by \texttt{BANE} with the default parameters. The image is of the same pixel scale as in Figure~\ref{img_1320}. Higher RMS levels are found at the edge of the field (where only one beam is present) and around the brighter sources.}
	\label{fig:1320MHz_rms}
\end{figure}

The properties of the 960\,MHz and 1320\,MHz images are summarised in Table\,\ref{Img_data_table}. These two new \ac{ASKAP} images are shown in Figures~\ref{img_960} and \ref{img_1320}, with zoomed in views showing the resolved structure of the emission in Figures.~\ref{img_SMC_N19} and \ref{img_SMC_N66}. Figures~\ref{fig:960MHz_rms} and \ref{fig:1320MHz_rms} show the \ac{RMS} maps generated by the source finding software, \textsc{aegean} \citep{2012ascl.soft12009H,2018PASA...35...11H} for the 960\,MHz and 1320\,MHz images respectively. 

We note that our \ac{ESP} 960\,MHz image was made at very early stages of the \ac{ASKAP} testing and a range of issues, such as positional accuracy and calibration, were discovered. We have made every effort to identify and correct these problems. The 1320\,MHz image as made at a later date when these issues were already known and could therefore be avoided, mitigated or corrected as needed.

\begin{table*}
\centering
	\caption{Properties of the 960\,MHz and 1320\,MHz radio continuum images as well other MOST/ATCA surveys used in this study.}
	\label{Img_data_table}
	\begin{tabular}{cccccccl} 
		\hline
		$\nu$     & $\lambda$ & Telescope & Median \ac{RMS}            & Best \ac{RMS}              & Beam Size          & Total number & Reference \\
       (MHz)      & (cm)      &           & ($\mu$Jy~beam$^{-1}$) & ($\mu$Jy~beam$^{-1}$) & (arcsec)           & of point sources   & \\
		\hline
       1320       & 23        & ASKAP     & 165                   & 55                    & 16.3 $\times$ 15.1 & 5954 & This work \\
       960        & 32        & ASKAP     & 186                   & 110                   & 30.0 $\times$ 30.0 & 4489 & This work\\
		\hline
       843        & 36        & MOST      & 700                   & 500                   & 40.0 $\times$ 40.0 & 1689 & \cite{2011SerAJ.183..103W} \\
       1400       & 20        & ATCA      & 700                   & 600                   & 17.8 $\times$ 12.2 & 1560 & \cite{2011SerAJ.183..103W} \\
       2370       & 13        & ATCA      & 400                   & 300                   & 45.0 $\times$ 45.0 & 742  & \cite{2011SerAJ.183..103W} \\
       4800       & 6         & ATCA      & 700                   & 500                   & 30.0 $\times$ 30.0 & 601  & \cite{2012SerAJ.184...93W} \\
       8640       & 3         & ATCA      & 800                   & 700                   & 20.0 $\times$ 20.0 & 457  &  \cite{2012SerAJ.184...93W} \\
		\hline
	\end{tabular}
\end{table*}

\section{ASKAP ESP SMC Source catalogues}

\subsection{Source detection}
\label{sdet}

The \textsc{aegean} source finding software was used to create an overall catalogue of sources from the \ac{ASKAP} images. Due to the combination of the multiple beams and artefacts from bright sources, images from \ac{ASKAP} have variable noise across the field. This variable noise must be parameterised before source-finding to ensure that accurate source thresholds are determined. To do this, noise (\ac{RMS}) and background level maps were made using the \texttt{BANE} routine in \textsc{aegean}, with its default parameters. \texttt{BANE} uses a grid algorithm with a sliding box-car and sigma-clipping approach, with the resulting maps being at the same pixel scale as the input images \citep[for further detail, see][]{2018PASA...35...11H}. The maps were then used with the default parameters in \textsc{aegean} to create the initial source lists at 5$\sigma$ level. Visual inspection of the sources was carried out to verify detections from the initial source lists. 

\subsection{Source Catalogues}
\label{cat}

In total, we found 4489 and 5954 point sources in our new \ac{ASKAP} 960\,MHz and 1320\,MHz images, respectively (see Tables~\ref{tab:960_point} and \ref{tab:1320_point}). There are 3536 unique sources that have both \ac{ASKAP} 960\,MHz and 1320\,MHz flux densities. This catalogue excludes known \ac{SMC} \ac{SNRs}, \ac{PNe} and \HII\ regions which are listed separately (see Sections~\ref{SNR_samp} and \ref{PNR_samp}).

We combine our two new \ac{ASKAP} catalogues of point sources with previously published source lists from \ac{MOST} (at 843\,MHz) and \ac{ATCA} (1400, 2370, 4800 and 8640\,MHz). To do this, we used a 10\arcsec\ search radius to find common sources and found a total of 7736 discrete sources which we list in Table~\ref{tab:allsources}. Out of these 7736 sources, there are 659 sources that do not have any \ac{ASKAP} flux densities and 112 that do not have \ac{MOST}/\ac{SUMSS} flux densities. 

Where possible, we also list the estimated \ac{SI}\footnote{Defined as $S_{\nu}$~$\propto$~$\nu^\alpha$, where: $S_{\nu}$ is flux density, $\nu$ is frequency, and $\alpha$ is spectral index.} of the source including error (Table~\ref{tab:allsources}; Col.~12). We also note that there are 49 ($\sim$0.5 per cent of the total population) sources in Table~\ref{tab:allsources} with questionable \ac{SI} estimates of $\alpha<-4$ and $\alpha>+2.5$. Where the \ac{SI} values are extreme we flag those sources to emphasis caution. The reasons behind such unrealistic \ac{SI} for these few sources ($<$0.3 per cent out of our 7736 sources) are twofold. One is that the flux density measurements are made between only two nearby frequency bands (such as for example 1400/1320\,MHz or 960/843\,MHz) where a small change (or error) in size or flux density leads to large changes and unrealistic estimates in \ac{SI}. The second issue is that almost all of such sources lie near near the edges of the field where $uv$ coverage and sensitivity are significantly poorer.

\begin{table}
\centering
\caption{Point source catalogue derived from our \ac{ASKAP} 960~MHz image. The catalogue at 960\,MHz consists of 4489 point sources. The full table is available in the online version of the article.}
\begin{tabular}{llccl}
\hline
Source  & Name & RA (J2000)  & Dec (J2000)           & S$_{960 \rm{MHz}}$  \\
No.     & ASKAP     & hh mm ss    & \D\ \arcmin\ \arcsec\ & (mJy)\\  
\hline
1 &  J000437--744211 & 00:04:36.52 & --74:42:11.0 & 4.0$\pm$0.5 \\
2 &  J000506--751559 & 00:05:05.77 & --75:15:58.6 & 4.0$\pm$0.5\\
3 &  J000508--745454 & 00:05:07.97 & --74:54:53.6 & 16.0$\pm$0.5\\
4 &  J000545--741232 & 00:05:44.94 & --74:12:31.6 & 16.9$\pm$0.6\\
5 &  J000550--744806 & 00:05:49.77 & --74:48:05.9 & 27.7$\pm$0.5\\
6 &  J000550--742134 & 00:05:50.46 & --74:21:34.4 & 3.6$\pm$0.5\\
7 &  J000603--743754 & 00:06:03.48 & --74:37:54.1 & 10.9$\pm$0.5\\
8 &  J000608--740148 & 00:06:08.34 & --74:01:47.6 & 8.2$\pm$0.6\\
9 &  J000608--740240 & 00:06:08.50 & --74:02:40.2 & 9.4$\pm$0.6\\
10 & J000609--740538 & 00:06:09.42 & --74:05:38.3 & 2.9$\pm$0.6\\
\hline
\end{tabular}
 \label{tab:960_point}
\end{table}

\begin{table}
\centering
\caption{Point source catalogue derived from our \ac{ASKAP} 1320~MHz image. The catalogue at 1320\,MHz consists of 5954 point sources. The full table is available in the online version of the article.}
\begin{tabular}{llccl}
\hline
Source  & Name       & RA (J2000)  & Dec (J2000)           & S$_{1320 \rm{MHz}}$  \\
No.     & ASKAP      & hh mm ss    & \D\ \arcmin\ \arcsec\ & (mJy)\\  
\hline
1 &  J000537--715839 & 00:05:36.81 & --71:58:39.2 & 5.3$\pm$0.9\\
2 &  J000547--722502 & 00:05:46.55 & --72:25:01.9 & 3.8$\pm$0.5\\
3 &  J000646--720801 & 00:06:45.91 & --72:08:01.2 & 2.1$\pm$0.3\\
4 &  J000648--722252 & 00:06:48.25 & --72:22:51.8 & 10.1$\pm$0.3\\
5 &  J000653--715740 & 00:06:52.59 & --71:57:40.2 & 36.3$\pm$0.4\\
6 &  J000654--722034 & 00:06:54.07 & --72:20:34.2 & 2.1$\pm$0.4\\
7 &  J000713--714611 & 00:07:12.94 & --71:46:10.6 & 4.1$\pm$0.6\\
8 &  J000726--720631 & 00:07:26.15 & --72:06:30.8 & 3.4$\pm$0.2\\
9 &  J000732--720732 & 00:07:31.97 & --72:07:32.2 & 1.4$\pm$0.2\\
10&  J000739--721026 & 00:07:38.64 & --72:10:26.1 & 2.1$\pm$0.4\\
\hline
\end{tabular}
 \label{tab:1320_point}
\end{table}

\begin{table*}
\centering
\scriptsize
\caption{An excerpt from the combined catalogue of point sources. These sources should be referred to as EMU-ESP-SMC-ffff NNNN. The full table is available in the online version of the article}.
\begin{tabular}{rcccccccccccccc}
\hline
(1) & (2) & (3) & (4) & (5) & (6) & (7) & (8) & (9) & (10) & (11) & (12) & (13) & (14) & (15) \\ 
No &  RA (J2000)   & DEC (J2000)   & S$_{843\,\rm{MHz}}$ & S$_{960\,\rm{MHz}}$  & S$_{1320\,\rm{MHz}}$  & S$_{1400\,\rm{MHz}}$ & S$_{2370\,\rm{MHz}}$ & S$_{4800\,\rm{MHz}}$ & S$_{8640\,\rm{MHz}}$ & No. & $\alpha \pm \Delta \alpha$ &  S$_{1\,\rm{GHz}}$ & Cat. No. & Cat. No. \\
   & hh:mm:ss.s        & dd:mm:ss.s & (mJy) &  (mJy) & (mJy) & (mJy) & (mJy) & (mJy) & (mJy) & points  &  & (mJy) & 960 & 1320 \\
\hline
3165 &	00:51:40.16 & --72:38:16.5 &   6.19 & 	5.5 & 	4.1  & 	...  & 	...  & 	... & 	... & 	3 & 	--0.94 $\pm$ 0.01 & 	5.3 & 	2021 & 	2276\\
3166 &	00:51:41.38 & --73:13:36.9 &  13.96 &  10.0 & 	...  & 	10.6 & 	19.5 & 10.0 & 11.10 & 	6 & 	--0.10 $\pm$ 0.10 &    12.4 & 	2022 & 	...\\
3167 &	00:51:41.65 & --70:28:46.3 & 	... & 	1.1 & 	0.82 & 	...  & 	...  & 	... & 	... & 	2 & 	--0.92            & 	1.1 & 	2025 & 	2278\\
3168 &	00:51:41.85 & --69:45:10.2 & 	... & 	3.5 & 	2.7  & 	...  & 	...  & 	... & 	... & 	2 & 	--0.81            & 	3.4 & 	2024 & 	2279\\
3169 &	00:51:42.02 & --72:55:56.3 &  71.88 &  61.3 & 	...  & 38.42 & 	42.6 & 21.3 &  7.26 & 	6 & 	--0.90 $\pm$ 0.10 &    61.7 & 	2023 & 	...\\
3170 &	00:51:42.12 & --73:45:04.7 & 	... & 	3.4 & 	2.72 & 	...  & 	...  & 	... & 	... & 	2 & 	--0.7             & 	3.3 & 	2026 & 	2280\\
3171 &	00:51:45.11 & --75:22:31.3 & 	... & 	... & 	0.53 & 	...  & 	...  & 	... & 	... & 	1 & 	...               & 	... & 	...  & 	2281\\
3172 &	00:51:45.98 & --69:28:14.6 & 	... & 	... & 	4.5  & 	...  & 	...  & 	... & 	... & 	1 & 	...               & 	... & 	...  & 	2282\\
3173 &	00:51:46.61 & --75:32:16.4 & 	... & 	4.2 & 	2.7  & 	...  & 	...  & 	... & 	... & 	2 & 	--1.4             & 	4.0 & 	2027 & 	2283\\
3174 &	00:51:47.31 & --71:03:01.6 & 	... & 	... & 	0.52 & 	...  & 	...  & 	... & 	... & 	1 & 	...               & 	... & 	...  & 	2284\\
3175 &	00:51:47.84 & --73:19:33.4 & 	... & 	1.1 & 	0.9  & 	...  & 	...  & 	... & 	... & 	2 & 	--0.63            & 	1.1 & 	2028 & 	2285\\
3176 &	00:51:47.89 & --73:04:54.0 &  20.24 & 	19.7& 	12.9 & 12.68 & 	6.2  & 	2.1 & 	... & 	6 & 	--1.32 $\pm$ 0.06 &    17.9 & 	2029 & 	2286\\
3177 &	00:51:48.39 & --72:50:48.3 &   9.63 & 	8.1 & 	...  & 	8.29 & 	10.3 & 	... & 	... & 	4 & 	0.1 $\pm$ 0.2     & 	8.9 & 	2030 & 	...\\
3178 &	00:51:49.52 & --73:38:39.4 & 	... & 	... & 	0.6  & 	...  & 	...  & 	... & 	... & 	1 & 	...               & 	... &    ... & 	2287\\
3179 &	00:51:50.04 & --74:54:40.4 & 	... & 	0.9 & 	1.3  & 	...  & 	...  & 	... & 	... & 	2 & 	1.2               & 	0.9 & 	2031 & 	2288\\
3180 &	00:51:51.24 & --72:55:37.8 & 	... & 	... & 	...  & 	...  & 	...  & 	... &  4.87 & 	1 & 	...               & 	... & 	...  & 	...\\
3181 &	00:51:51.24 & --74:11:15.2 & 	... & 	... & 	0.96 & 	...  & 	...  & 	... & 	... & 	1 & 	...               & 	... & 	...  & 	2289\\
3182 &	00:51:51.46 & --72:05:53.6 & 	... & 	... & 	0.51 & 	...  & 	...  & 	... & 	... & 	1 & 	...               & 	... & 	...  & 	2290\\
3183 &	00:51:53.37 & --73:31:10.9 & 	... & 	... & 	0.9  & 	...  & 	...  & 	... & 	... & 	1 & 	...               & 	... & 	...  & 	2291\\
3184 &	00:51:53.67 & --73:45:21.6 & 5.90   & 	5.4 & 	3.96 & 	4.21 & 	...  & 	... & 	... & 	4 & 	--0.80 $\pm$ 0.10 & 	5.2 & 	2032 & 	2292\\
\hline
\end{tabular}
 \label{tab:allsources}
\end{table*}

Non-point sources, such as blended and extended sources, were flagged and excised to leave only a catalogue of point sources. Although not used in the further analysis, we provide estimates of positions and flux densities for detected non-point sources. We present the results from both catalogues in Tables~\ref{tab:960} and \ref{tab:1320} where a total of 282 and 641 non-point sources are found at 960\,MHz and 1320\,MHz surveys, respectively. Because of the different resolution across the various \ac{SMC} surveys, some of these listed non-point sources could be resolved in one survey but could appear as a point source in another and as such they would not be listed in Tables~\ref{tab:960} or \ref{tab:1320}.

\subsection{Comparison with previous catalogues}
\label{scompare}

\begin{figure}
	\includegraphics[width=\columnwidth,trim=0 0 0 0,clip]{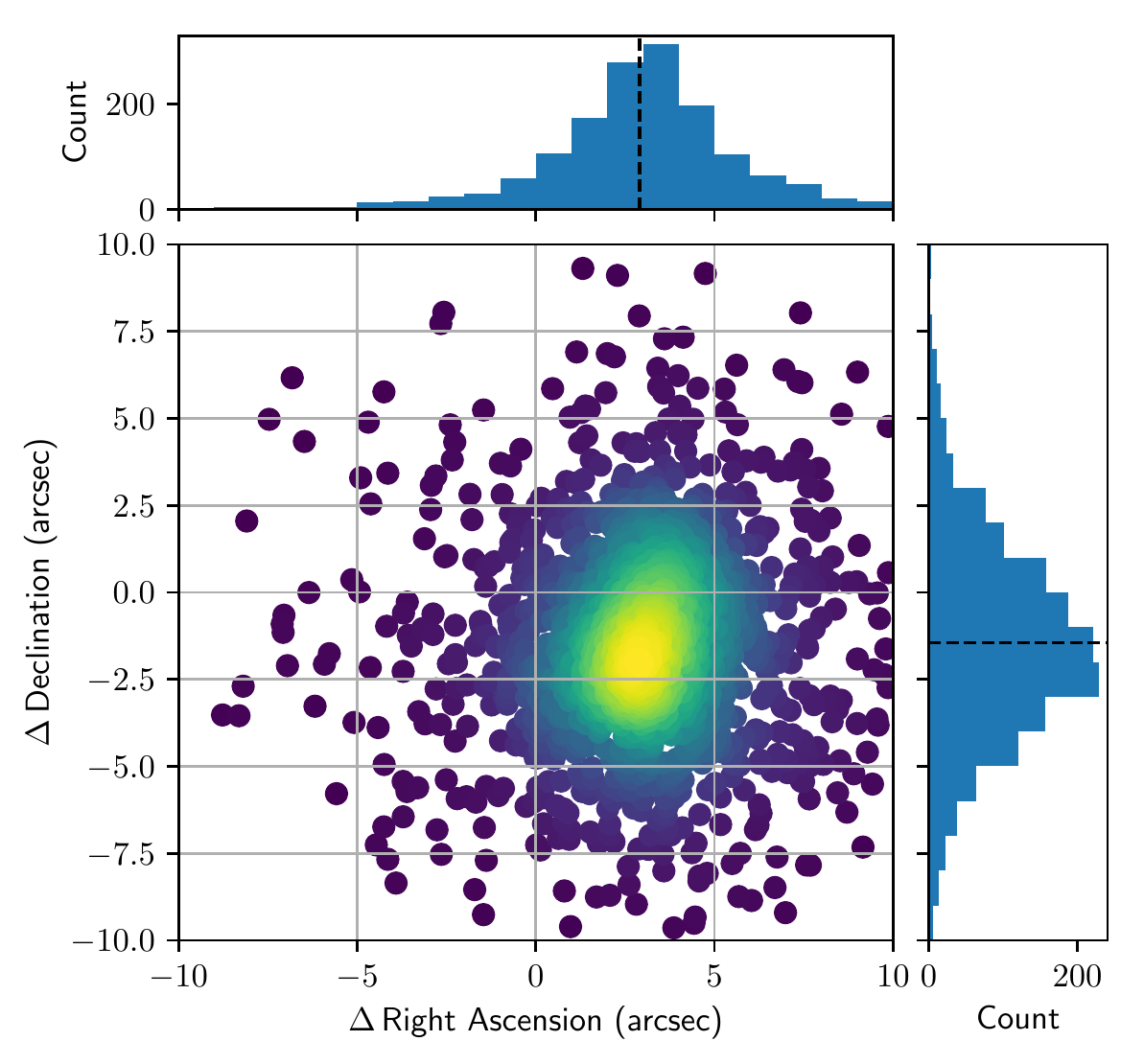}
	\caption{Positional difference (\ac{MOST} -- \ac{ASKAP}) of 1509 sources found in both the 843\,MHz (MOST) and the 960\,MHz catalogues. The mean offsets are $\Delta$RA=+2.90\arcsec\ (SD=2.65) and $\Delta$DEC=--1.45\arcsec\ (SD=2.92). }
    \label{fig:843_vs_960}
\end{figure}

\begin{figure}
	\includegraphics[width=\columnwidth,trim=0 0 0 0,clip]{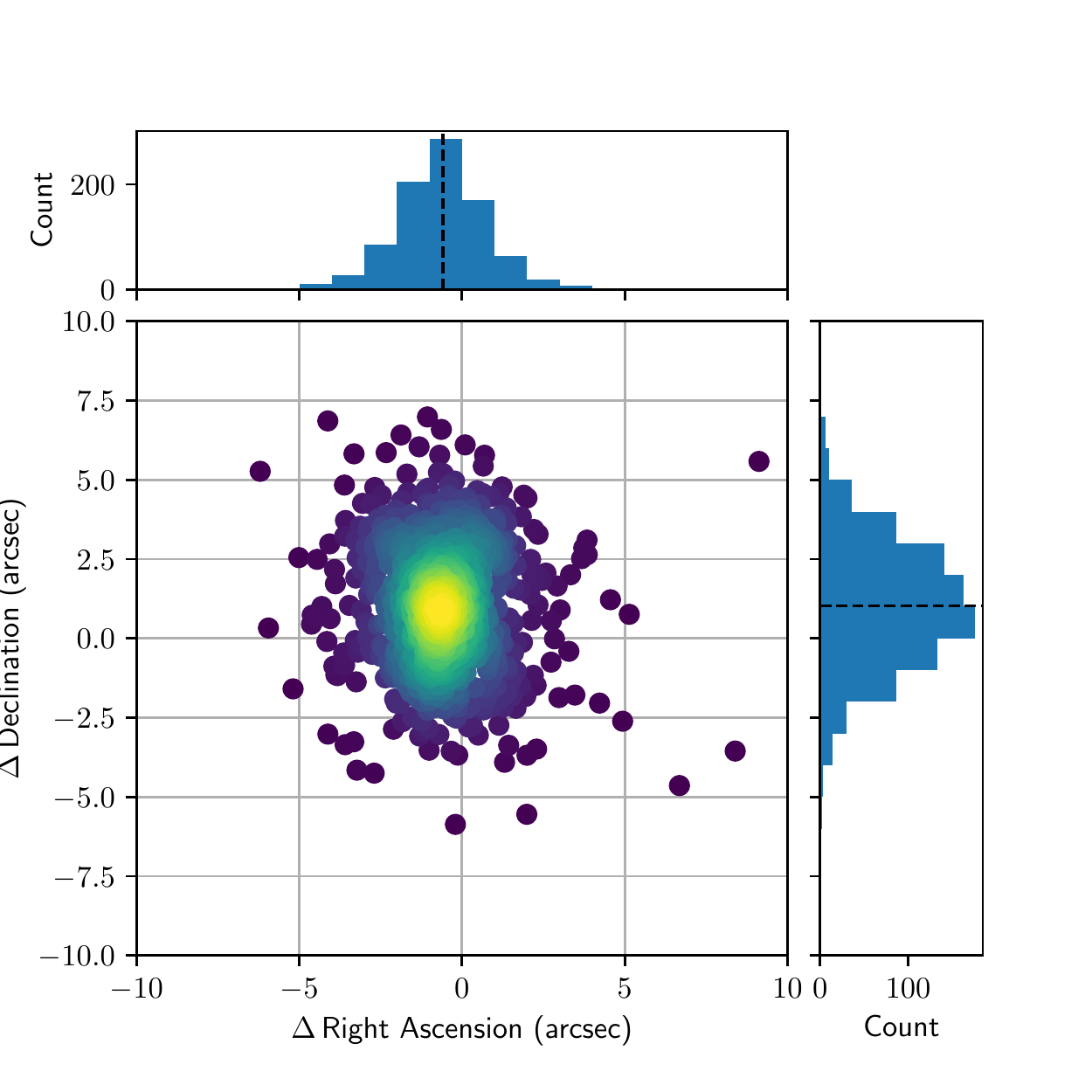}
	\caption{Positional difference (\ac{ATCA} -- \ac{ASKAP}) of 889 sources found in both the 1320\,MHz and the 1400\,MHz (ATCA) catalogues. The mean offsets are} $\Delta$RA=--0.58\arcsec\ (SD=1.50) and $\Delta$DEC=+1.03\arcsec\ (SD=1.95).
    \label{fig:1320_vs_1400}
\end{figure}

We compare position differences ($\Delta$RA and $\Delta$DEC) between our new \ac{ASKAP} images and previous catalogues at 843\,MHz (see Figure~\ref{fig:843_vs_960}) and 1400\,MHz (see Figure~\ref{fig:1320_vs_1400}). We did not find any significant shift in position in our 1320\,MHz vs. 1400\,MHz position comparison. For the 889 sources in common, we found that the $\Delta$RA=$-$0.58\arcsec\ (SD=1.50\arcsec) and $\Delta$DEC=+1.03\arcsec\ (SD=1.95\arcsec). Somewhat worse results are reported for the 843\,MHz vs. 960\,MHz comparison of 1509 sources with $\Delta$RA=+2.90\arcsec\ (SD=2.65\arcsec) and $\Delta$DEC=$-$1.45\arcsec\ (SD=2.92\arcsec). These position differences are only a small fraction of the beamsize at the given frequency.

Positional shifts of $\sim$3\arcsec\ in our 960\,MHz image are not insignificant, especially if we want to look for multiband counterparts. The reason for the discrepancy lies in the fact that this image comes from the \ac{ASKAP} testing and early operation period where a number of issues were found and acknowledged. Specifically, throughout the paper we use the coordinates from other \ac{SMC} surveys for the various sources wherever possible. An excerpt of the combined point source catalogue is shown in Table\,\ref{tab:allsources}.

\begin{figure}
	\includegraphics[width=\columnwidth,trim=0 0 0 0,clip]{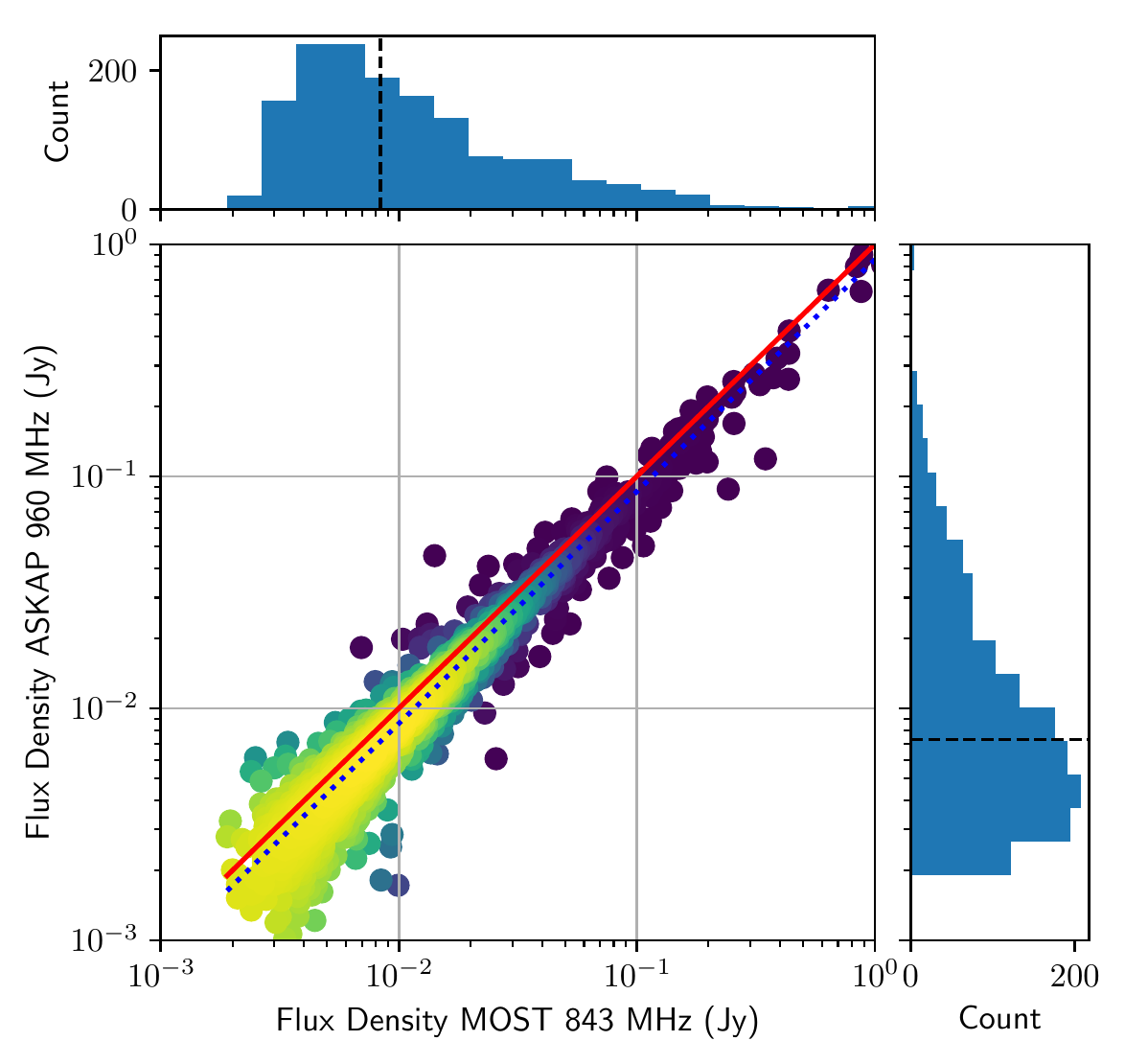}
	\caption{Integrated flux density comparison of sources found in both the 960\,MHz and the 843\,MHz catalogues. The best fit slope (linear) is 0.89$\pm$0.01 (dotted blue) while the red line represent 1-to-1 ratio (see Section~\ref{scompare}). The points are colour coded to indicate local density, yellow for high density through to purple for low density. The source integrated flux density distributions are shown in the side and top panels, with the black dashed line at the median integrated flux density.}
	\label{fig:FLUX_843_vs_960}
\end{figure}

\begin{figure}
	\includegraphics[width=\columnwidth,trim=0 0 0 0,clip]{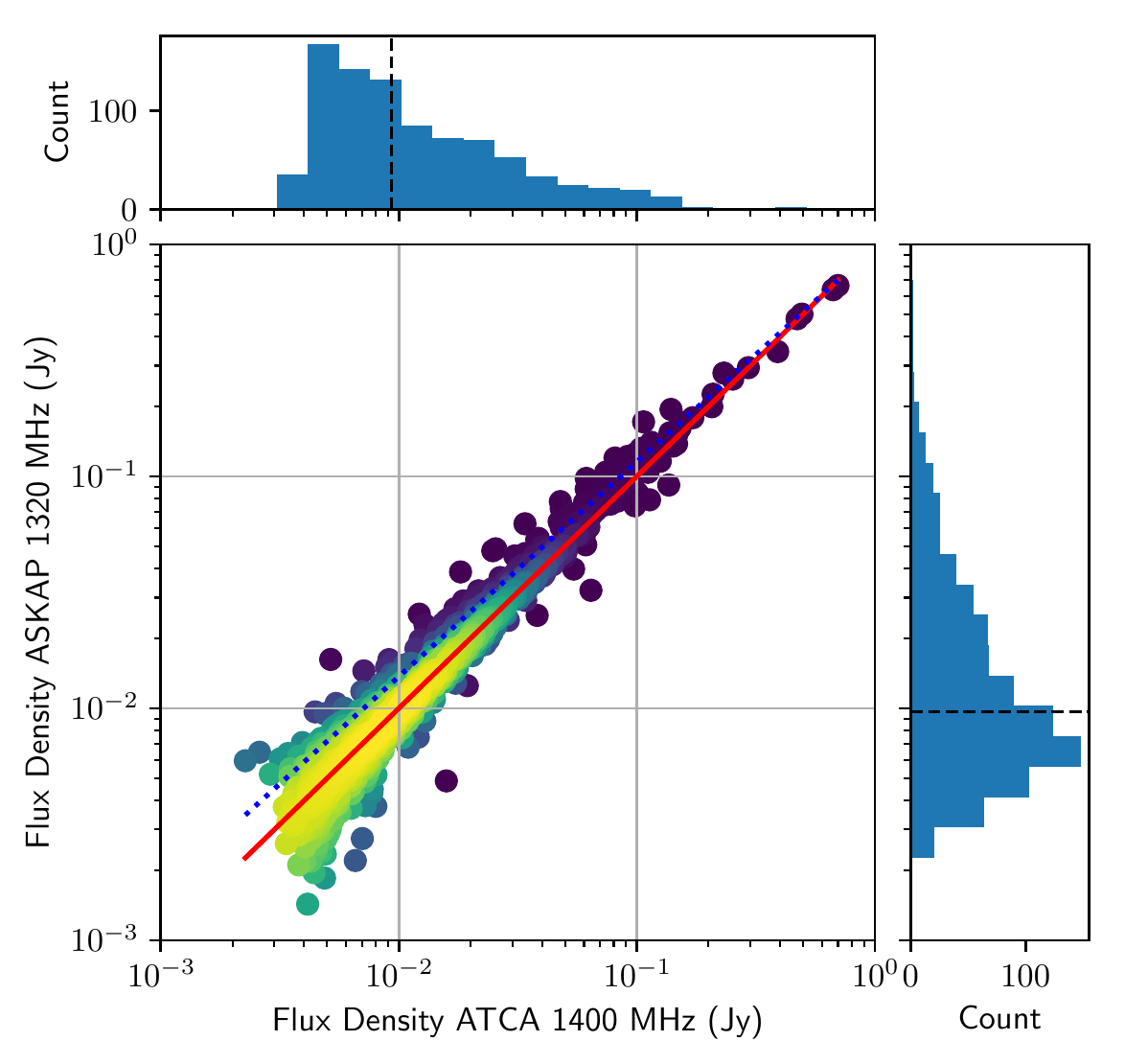}
	\caption{Integrated flux density comparison of sources found in both the 1320\,MHz and the 1400\,MHz catalogues. The best fit slope (linear) is 0.99$\pm$0.01 (dotted blue) while red line represent 1-to-1 ratio (see Section~\ref{scompare}). The points are colour coded to indicate local density, yellow for high density through to purple for low density. The source integrated flux density distributions are shown in the side and top panels, with the black dashed line at the median integrated flux density.}
	\label{fig:FLUX_1320_vs_1400}
\end{figure}

In order to assess the reliability of our integrated flux values, we compared the values on compact (extended \HII\ regions are excluded) sources to catalogue values from other nearby frequencies. We performed two sets of comparisons: our ASKAP 960\,MHz values with the values from MOST at 843\,MHz and our ASKAP 1320\,MHz values with the ATCA 1400\,MHz values. The agreement is excellent, as can be seen in Figures~\ref{fig:FLUX_843_vs_960} and \ref{fig:FLUX_1320_vs_1400}. 

As a quick check on flux density scales, we fit S$_{\rm ASKAP} = k\times \rm S_{other} + z$, allowing for some small zero level offsets (z). For the S$_{\rm 960\,MHz}$/S$_{\rm 843\,MHz}$ and S$_{\rm 1320\,MHz}$/S$_{\rm 1400\,MHz}$ comparison, we find a slope ($k$) of 0.89 and 0.99 respectively, which corresponds to an average \ac{SI} of --0.9 and --0.2 respectively. Given that the average \ac{SI} for the majority of sources in our field of view is around --0.8, we would expect that the integrated flux density at 843\,MHz would be $\sim$10~per cent higher than at 960\,MHz. Similarly, the difference between 1320\,MHz and 1400\,MHz would cause the average flux density in our \ac{ASKAP} 1320\,MHz image to be higher by about 4.5~per cent. The S$_{\rm 960\,MHz}$/S$_{\rm 843\,MHz}$ value is somewhat steeper than the average \ac{SI} calculated for each source individually across larger frequency ranges. The S$_{\rm 1320\,MHz}$/S$_{\rm 1400\,MHz}$ spectrum suggests a possible flux density scale inconsistency at the 5~per cent level, within the uncertainty expectations. However, the high quality of these data indicate that with the full \ac{ASKAP} array and final calibration, it may be possible to tie the flux density scales at different frequencies to much higher accuracy than currently possible.

In order to estimate the number of matches between these two new \ac{ASKAP} catalogues and other combined catalogues which could arise purely by chance, we produced artificial source catalogues with positions shifted from the real position. Positions from the final catalogue were shifted by $\pm10$~arcmin in RA and DEC (4 different positions) and used as input for \textsc{aegean}'s prioritised fitting method \citep{2018PASA...35...11H}. Only cross-matches within half the synthesised beam \ac{FWHM} (for each survey) were considered matches. We found the average number of chance coincidences to be 53 for the 960\,MHz image and 60 for the 1320\,MHz image (out of total 7736 sources from the point source catalogue Table~\ref{tab:allsources} or $\sim$0.7~per cent). This result implies that the large fraction of correlations between two \ac{ASKAP} catalogues are highly likely to be real.

\begin{table}
\centering
\caption{Non-point source catalogue derived from our \ac{ASKAP} 960\,MHz image. The catalogue at 960\,MHz consists of 282 non-point sources. } The flags are coded as: 2 partially blended source, 3 fully blended or extended source, 4 source is very likely a part of a larger structure. The full table is available in the online version of the article.
\begin{tabular}{rcccl}
\hline
Source  & RA (J2000)  & Dec (J2000)  & S$_{960 \rm{MHz}}$ & Flag \\
number    & hh mm ss    & \D\ \arcmin\ \arcsec\ & (mJy) &  \\  
\hline
  1 & 00:09:39.65 & --73:08:16.6 & 61.3 $\pm$ 6.2  & 3,4\\
  2 & 00:09:57.31 & --73:08:48.8 & 74.4 $\pm$ 7.5  & 3,4\\
  3 & 00:10:12.51 & --73:21:23.9 & 110  $\pm$ 11   & 3\\
  4 & 00:11:25.26 & --74:22:36.1 & 2.52 $\pm$ 0.38 & 3\\
  5 & 00:12:15.73 & --75:36:56.8 & 5.97 $\pm$ 0.73 & 3\\
  6 & 00:14:22.33 & --75:18:40.2 & 3.04 $\pm$ 0.38 & 3\\
  7 & 00:14:24.67 & --72:17:22.5 & 2.13 $\pm$ 0.48 & 3\\
  8 & 00:14:29.91 & --72:17:21.5 & 2.13 $\pm$ 0.48 & 3\\
  9 & 00:14:36.23 & --70:53:34.9 & 119  $\pm$ 12   & 3\\
 10 & 00:14:47.72 & --70:53:25.5 & 154  $\pm$ 15   & 3\\
\hline
\end{tabular}
 \label{tab:960}
\end{table}

\begin{table}
\centering
\caption{Non-point source catalogue derived from our \ac{ASKAP} 1320\,MHz image. The catalogue at 1320\,MHz consists of 641 non-point sources. The flags are coded as in Table~\ref{tab:960}. The full table is available in the online version of the article.}
\begin{tabular}{rcccl}
\hline
Source & RA (J2000) & Dec (J2000) & S$_{1320 \rm{MHz}}$ & Flag \\
number   & hh mm ss   & \D\ \arcmin\ \arcsec\ & (mJy) &  \\  
\hline
  1 & 00:07:36.88 & --72:12:00.6 & 12.4 $\pm$ 1.3  & 3 \\
  2 & 00:09:47.69 & --72:44:48.6 & 20.8 $\pm$ 2.1  & 3 \\
  3 & 00:10:24.52 & --72:00:37.6 & 5.13 $\pm$ 0.56 & 2 \\
  4 & 00:10:28.35 & --72:00:23.0 & 6.23 $\pm$ 0.66 & 2 \\
  5 & 00:11:58.28 & --72:00:48.5 & 22.7 $\pm$ 2.3  & 3 \\
  6 & 00:12:47.55 & --73:12:57.6 & 12.6 $\pm$ 1.3  & 3 \\
  7 & 00:14:25.83 & --72:17:20.8 & 3.06 $\pm$ 0.35 & 3 \\
  8 & 00:14:31.23 & --72:09:54.5 & 4.57 $\pm$ 0.48 & 3 \\
  9 & 00:15:09.85 & --72:48:06.5 & 3.72 $\pm$ 0.40 & 2 \\
 10 & 00:15:24.82 & --72:17:43.3 & 2.74 $\pm$ 0.32 & 2 \\
\hline
\end{tabular}
 \label{tab:1320}
\end{table}

\begin{figure}
	\includegraphics[width=\columnwidth,trim=0 0 0 0,clip]{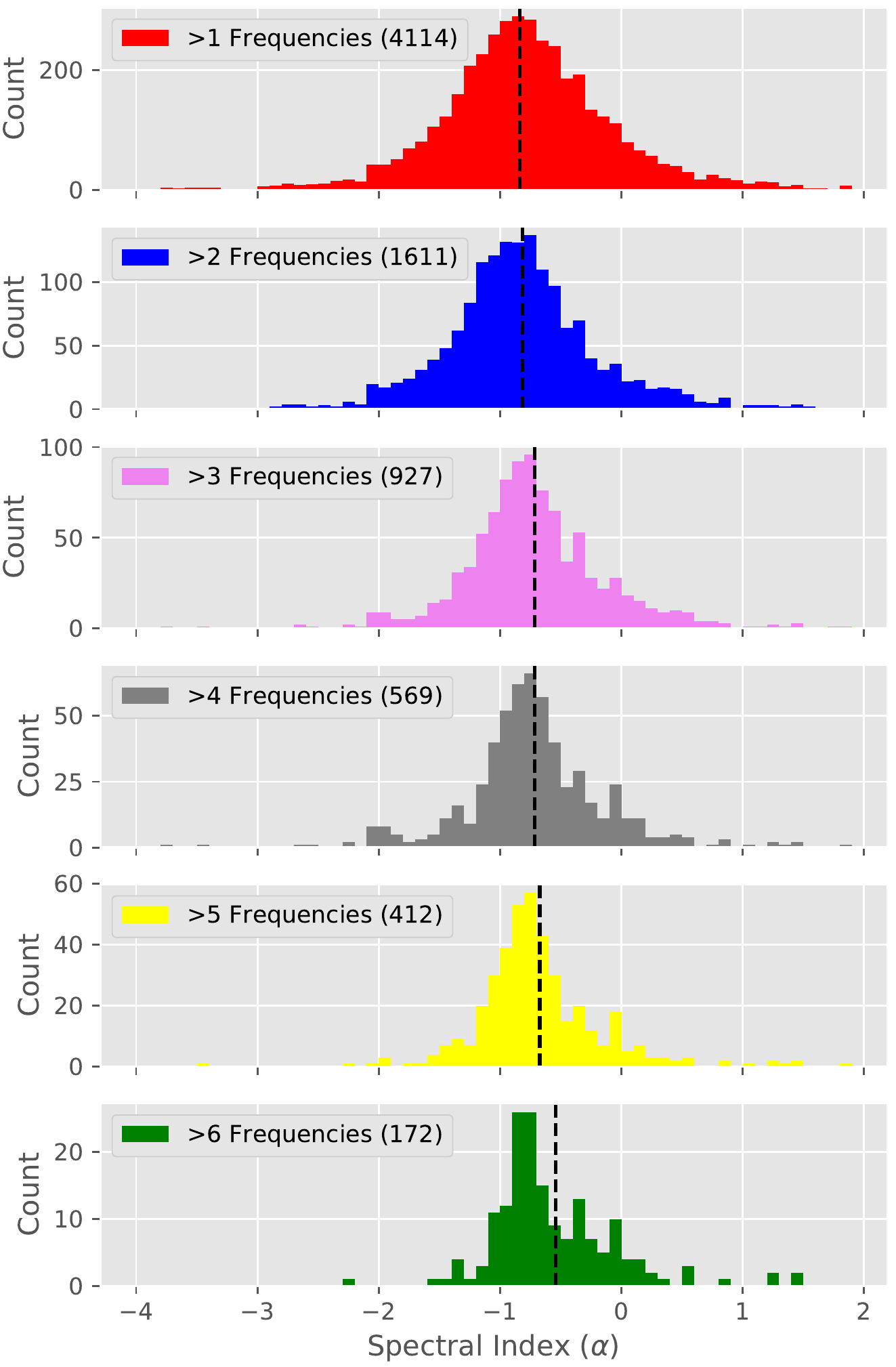}
	\caption{Spectral index distribution of all sources in the field of \ac{SMC} binned at 0.1. The vertical dashed line represents the mean \ac{SI} of each panel, as discussed in Section~\ref{scompare}. The uppermost panel includes  all the sources of the other panels beneath. }
	\label{fig:spectral_index}
\end{figure}

Finally, we estimate the radio spectral index for all sources in common and show their distribution in Figure~\ref{fig:spectral_index}. There are 4114 sources found at only two frequencies (marked in red; Figure~\ref{fig:spectral_index}) and for those we estimate a mean $\alpha$ of --0.84. For 1611 sources that are found in three different catalogues (marked in blue; Figure~\ref{fig:spectral_index}) we found a mean $\alpha$ of --0.81 (SD=1.35). We also estimate the average \ac{SI} for sources that are detected in four (927 sources; purple; Figure~\ref{fig:spectral_index}; $\alpha$=--0.71, SD=0.75), five (569 sources; grey; Figure~\ref{fig:spectral_index}; $\alpha$=--0.71, SD=0.59), six (412 sources; orange; Figure~\ref{fig:spectral_index}; $\alpha$=--0.67, SD=0.72) and seven (172 sources; green; Figure~\ref{fig:spectral_index}; $\alpha=-0.54$, SD=0.51) different frequencies. Given that our sample sizes of \ac{SNRs}, \ac{PNe} and \HII\ regions are around 100-150 (see Sections~\ref{SNR_samp} and \ref{PNR_samp}), this distribution is as expected and indicates that the vast majority of our sources from Table~\ref{tab:allsources} are most likely to be background objects \citep[see e.g.][]{1998A&AS..130..421F,2018MNRAS.477..578C,2018MNRAS.474..779G}. We note that some sources with flux density measurements at more than two frequencies might exhibit spectral curvature and therefore the fitted value of alpha would not represent a good estimate.

\section{\ac{ASKAP} \ac{SMC} supernova remnant Sample}
\label{SNR_samp}

Because of their proximity and location well away from the Galactic Plane, we are able to study the sources belonging to the \ac{MCs}, such as the \ac{SNR} population. Together, these galaxies offer the opportunity to produce a complete sample of \ac{SNRs} suitable for population studies focused on size, evolution, radio spectral index and beyond, as shown by \citet{2016A&A...585A.162M} and \citet{2017ApJS..230....2B}. To that end, one of our prime goals with the next generation of \ac{ASKAP} surveys is to detect new and predominantly low surface brightness \ac{SNRs}. Indeed, with its unique coverage and depth, this new \ac{ASKAP} \ac{ESP} survey allowed us to search for new \ac{SNRs} and at the same time, measure the physical properties of the already established \ac{SNRs}, examples of which are shown in Figures~\ref{img_SMC_N19} and \ref{img_SMC_N66}.

Previous studies of \ac{SNRs} in the \ac{SMC} \citep{2005MNRAS.364..217F,2007MNRAS.376.1793P,2011A&A...530A.132O,2012A&A...545A.128H,2014AJ....148...99C,2015ApJ...803..106R,2019MNRAS.486.2507A,2019MNRAS.485L...6G,2019arXiv190404836S} have established 19 objects as bona-fide \ac{SNRs} with two more considered as good candidates. These two \ac{SNR} candidates are not detected in our radio images and we will discuss them in our subsequent papers. 

Here, we present our radio continuum study results which suggest two new sources to be \ac{SNR} candidates (MCSNR~J0057--7211 and MCSNR~J0106--7242), bringing our sample of \ac{SMC} \ac{SNRs} and \ac{SNR} candidates to 23. At the same time, we measure integrated flux densities for 18 of the 19 known \ac{SMC} \ac{SNRs} (see Table~\ref{tbl:snr}) and present our integrated flux density estimates for the two new \ac{SMC} \ac{SNR} candidates found in our new \ac{ASKAP} \ac{SMC} surveys (Table~\ref{tbl:snrcan}). An in-depth study of the \ac{SMC} \ac{SNR} population will be presented in Maggi et al. (submitted, https://arxiv.org/abs/1908.11234).

These two new SNR candidates were initially selected purely based on their typical morphological appearance (circular shape). As our \ac{SMC} \ac{SNR} sample is morphologically diverse, various approaches (and initial parameters) were employed in order to measure the best \ac{SNR} flux densities. Namely, we used the \textsc{miriad} \citep{1995ASPC...77..433S} task \texttt{imfit} to extract integrated flux density, extensions (diameter/axes) and position angle for each radio detected \ac{SNR}. For cross checking and consistency, we also used \textsc{aegean} and found no significant difference in integrated flux density estimates. 

We used two methods: For \ac{SNRs} which are known point sources (such as SNR~1E\,0102.2--7219, which is not resolved in radio) we use simple Gaussian fitting which produced the best result. The second approach was applied to all resolved \ac{SNRs}. For those, we measured their local background noise (1$\sigma$) and carefully select the exact area of the \ac{SNR}. We then estimated the sum of all brightnesses above 5$\sigma$ of each individual pixel within that area and converted it to \ac{SNR} integrated flux density following \citet[][eq.~24]{1966ARA&A...4...77F}. We also made corrections for an extended background where applicable i.e. for sources where nearby extended object such as \HII\ region(s) is evident. However, for the most of our \ac{SMC} \ac{SNRs} this extended background contribution is minimal. 

\begin{figure}
    \includegraphics[width=\columnwidth]{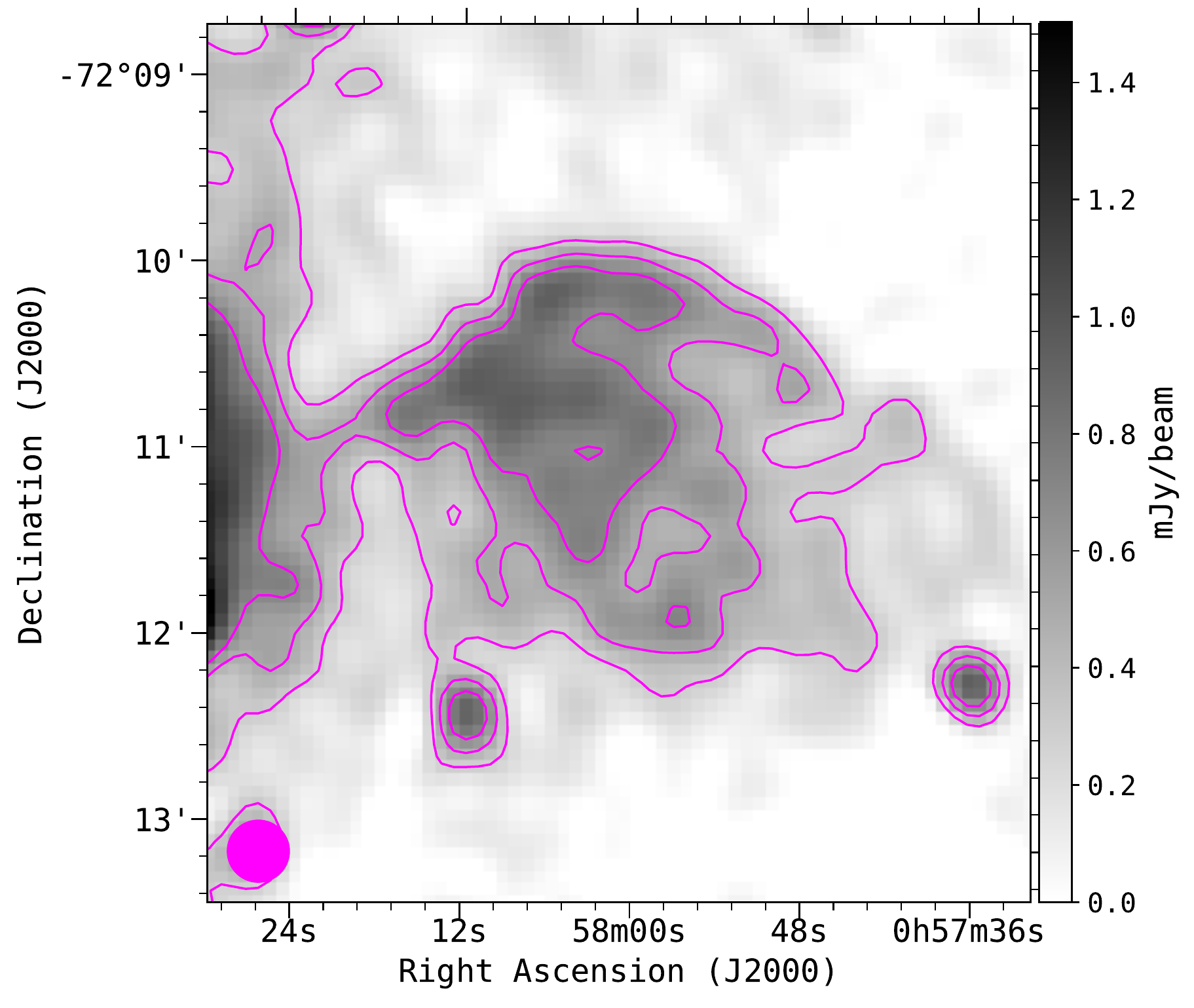}
	\caption{\ac{ASKAP} \ac{ESP} image of the new \ac{SMC} SNR candidate MCSNR\,J0057-7211 at 1320\,MHz (grey scale and contours) smoothed to a resolution of 20\arcsec $\times$ 20\arcsec. Magenta contours are: 0.3, 0.5, and 0.7~mJy~beam$^{-1}$. The smoothed beam is shown as a filled magenta circle in the lower left corner. Two point-like sources in the lower right and left corner are unrelated background sources. The local RMS noise is 0.1~mJy~beam$^{-1}$. }
	\label{img_MCSNRJ0057}
\end{figure}

\begin{figure}
	\includegraphics[width=\columnwidth]{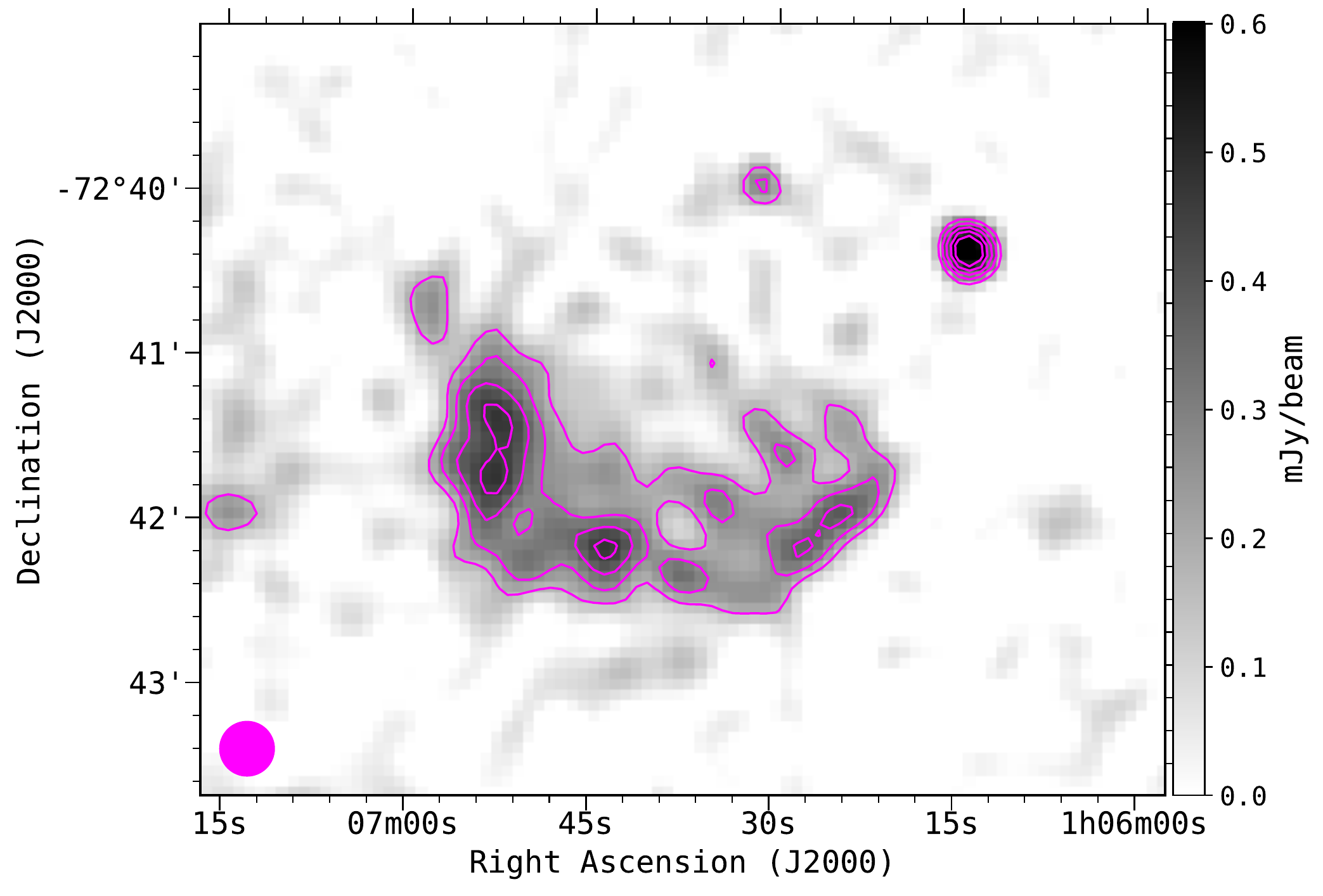}
	\caption{\ac{ASKAP} \ac{ESP} image of the new low surface brightness \ac{SMC} SNR candidate MCSNR\,J0106-7242 at 1320\,MHz (grey scale and contours) smoothed to a resolution of 20\arcsec $\times$ 20\arcsec. Magenta contours are: 0.18, 0.27, 0.36, 0.45 and 0.54~mJy~beam$^{-1}$. The smoothed beam is shown as a filled magenta circle in the lower left corner. The point-like source in the upper right corner is an unrelated background source. The local RMS noise is 0.06~mJy~beam$^{-1}$. }
	\label{img_MCSNRJ0106}
\end{figure}

The two new \ac{SNR} candidates are shown in Figs.~\ref{img_MCSNRJ0057} and \ref{img_MCSNRJ0106} and their integrated flux density measurements in Table~\ref{tbl:snrcan}. These two new \ac{SNR} candidates display approximately semi-circular structures consistent with a typical spherical morphology. As expected, they are both of low radio surface brightness, which is the main reason for their previous non-detection. We estimate the spectral index for both objects (Table~\ref{tbl:snrcan}) and they are consistent with typical SNR spectra, as found in, for example, the larger \ac{LMC} population \citep[see Fig.~13 in][]{2017ApJS..230....2B}. Therefore, in addition to their typical morphology, their radio spectral index points to a non-thermal origin which further supports that these objects be classified as \ac{SNR} candidates. Neither of these two \ac{SNR} candidates are detected at optical or \ac{IR} wave bands, which is not unusual given that a number of previously known bona-fide \ac{SNRs} have only been seen at one wavelength \citep{2008A&A...485...63F}. 

New \ac{ASKAP} SNR candidate MCSNR~J0057$-$7211 \citep[also see ][]{1991MNRAS.249..722Y} is located inside the ellipse around XMMU~J0057.7--7213 \citep[on the northern side, see Fig.~6 in][]{2012A&A...545A.128H}. The nearby point source XMMU~J005802.4--721205 is listed as an \ac{AGN} candidate \citep{2013A&A...558A...3S}. Also, there is a moderately bright, point-like X-ray source at 00:58:02.604, $-$72:12:06.7 with a non-thermal spectrum and \mbox{$L_{\rm{X}} \sim 10^{34}$~erg~s$^{-1}$} \citep{2012A&A...545A.128H}.

On inspection of present generation \xmm\ mosaic images, we find diffuse emission at the position of the second \ac{ASKAP} \ac{SMC} \ac{SNR} candidate -- MCSNR~J0106$-$7242. A more comprehensive study of the whole \ac{SMC} \ac{SNR} population will be presented in an upcoming study by Maggi~et~al.~(in prep.). 

We also use the equipartition formulae\footnote{\url{http://poincare.matf.bg.ac.rs/~arbo/eqp/}} \citep{Arbutina_2012,2013ApJ...777...31A,Uro_evi__2018} to estimate the magnetic field strength for these two \ac{SNR} candidates. While this derivation is purely analytical, we emphasise that it is formulated especially for the estimation of the magnetic field strength in \ac{SNRs}. The average equipartition field over the whole shell of MCSNR~J0057--7211 is $\sim$15\,$\mu$G while estimates for MCSNR~J0106--7242 are around $\sim$8\,$\mu$G, with an estimated minimum energy\footnote{We use the following values: $\theta$=1.37\arcmin\ and 1.29\arcmin; $\kappa=0$; S$_{\rm 1\,GHz}$=0.0307\,Jy and 0.02363\,Jy; and f=0.25.} of E$_{\rm min}$=6$\times10^{49}$\,erg and E$_{\rm min}$=1.5$\times10^{49}$\,erg, respectively. These values are typical of older \ac{SNRs} at the end of the Sedov phase where the magnetic field is three to four times more compressed than that of middle-age \ac{SNRs}. 

The position of these two \ac{SNR} candidates on the surface brightness to diameter ($\Sigma$--\textit{D}) diagram ($\Sigma$= 6.38$\times10^{-22}$\,W\,m$^{-2}$\,Hz$^{-1}$\,sr$^{-1}$ and 5.38$\times10^{-22}$\,W\,m$^{-2}$\,Hz$^{-1}$\,sr$^{-1}$, D=47\,pc and 44.9\,pc, respectively) by \cite{2018ApJ...852...84P}, suggests that these remnants are in the late Sedov phase, with an explosion energy of 1--2$\times10^{51}$\,erg, which evolves in an environment with a density of 0.02--0.2\,cm$^{-3}$.

\begin{table*}
\centering
\caption{19 \ac{SNRs} in the \ac{SMC}. Only MCSNRJ0103--7201 is not detected in our \ac{ASKAP} \ac{ESP} images. The integrated flux density errors are $<$10~per cent. Column~2 (Other Name) abbreviations are: DEM\,S: \citet{1976MmRAS..81...89D}, [HFP2000]: \citet{2000A&AS..142...41H}, IKT: \citet{1983IAUS..101..535I}, SXP: \citet{2012A&A...537L...1H}. } 
 \begin{tabular}{llccccccccccccccc}
 \hline
MCSNR				&	Other 		&	RA			&	DEC				& $S_{960\,\mathrm{MHz}}$	&	$S_{1320\,\mathrm{MHz}}$ 	\\
Name				&	Name		&	(J2000)		&	(J2000)			&	(Jy)					&	(Jy)	\\
\hline
J0041--7336			&	DEM\,S5    	&	00 41 01.7	&	$-$73 36 30.4	&	0.138					&	0.130	\\
J0046--7308			&[HFP2000]\,414	&	00 46 40.6	&	$-$73 08 14.9	&	0.111					&	0.110	\\
J0047--7308			&	IKT\,2     	&	00 47 16.6	&	$-$73 08 36.5	&	0.441					&	0.381	\\
J0047--7309			&	        	&	00 47 36.5	&	$-$73 09 20.0	&	0.201					&	0.185	\\
\smallskip
J0048--7319			&	IKT\,4     	&	00 48 19.6	&	$-$73 19 39.6	&	0.121					&	0.092	\\
J0049--7314			&	IKT\,5     	&	00 49 07.7	&	$-$73 14 45.0	&	0.068					&	0.060	\\
J0051--7321			&	IKT\,6     	&	00 51 06.7 	&	$-$73 21 26.4	&	0.085					&	0.096	\\
J0052--7236			&	DEM\,S68 	&	00 52 59.9	&	$-$72 36 47.0	&	0.091					&	0.081	\\
J0058--7217			&	IKT\,16    	&	00 58 22.4 	&	$-$72 17 52.0	&	0.079					&	0.070	\\
\smallskip
J0059--7210			&	IKT\,18    	&	00 59 27.7 	&	$-$72 10 09.8	&	0.559					&	0.502	\\
J0100--7133			&	DEM\,S108	&	01 00 23.9 	&	$-$71 33 41.1	&	0.161					&	0.146	\\
J0103--7209			&	IKT\,21    	&	01 03 17.0 	&	$-$72 09 42.5	&	0.100					&	0.085	\\
J0103--7247			&[HFP2000]\,334 &	01 03 29.1	&	$-$72 47 32.6	&	0.0288					&	0.025	\\
J0103--7201			&				&	01 03 36.6	&	$-$72 01 35.1	&	--						&	--	\\
\smallskip
J0104--7201			&1E\,0102.2-7219&	01 04 01.2 	&	$-$72 01 52.3	&	0.402					&	0.272	\\
J0105--7223			&	IKT\,23    	&	01 05 04.2 	&	$-$72 23 10.5	&	0.102					&	0.095	\\
J0105--7210			&	DEM\,S128  	&	01 05 30.5	&	$-$72 10 40.4	&	--						&	0.050	\\
J0106--7205			&	IKT\,25    	&	01 06 17.5 	&	$-$72 05 34.5	&	0.0095					&	0.0090	\\
\smallskip
J0127--7333			&	SXP\,1062	&	01 27 44.1	&	$-$73 33 01.6	&	0.0072					&	0.0068	\\
\hline
 \end{tabular}
 \smallskip
 \flushleft
\label{tbl:snr}
\end{table*}

\begin{table*}
\centering
\caption{Details of two new \ac{ASKAP} \ac{SNRs} candidates in the \ac{SMC}. The integrated flux density errors are $<$10~per cent. N\,S abbreviation stands as at \citet{1956ApJS....2..315H}}. \\
 \begin{tabular}{llccccccccccccccc}
 \hline
MCSNR				&	Other 		&	RA			&	DEC				& $S_{960\,\mathrm{MHz}}$	&	$S_{1320\,\mathrm{MHz}}$ 	& $\alpha\pm \Delta \alpha$ \\
Name				&	Name		&	(J2000)		&	(J2000)			&	(Jy)					&	(Jy)	\\
\hline
J0057--7211      	&	N\,S66D    	&	00 57 49.9 	&	$-$72 11 47.1	&	0.030					&	0.0244	& --0.75$\pm$0.04\\
J0106--7242      	&				&	01 06 32.1	&	$-$72 42 17.0	&	0.024					&	0.020	& --0.55$\pm$0.02\\
\hline
 \end{tabular}
 \smallskip
 \flushleft
\label{tbl:snrcan}
\end{table*}

\section{\ac{ASKAP} \ac{SMC} Planetary Nebula Sample}
\label{PNR_samp}

The location and proximity of the \ac{SMC} also provides an opportunity to create a complete sample of radio continuum detected planetary nebulae (\ac{PNe}) in that nearby galaxy. \ac{PNe} are important for studies of the chemical, atomic, molecular and solid-state galactic ISM enrichment \citep{2005JKAS...38..271K,2015HiA....16..623K}. The next generation \ac{ASKAP} surveys aim to provide detection of lower surface brightness \ac{PN} to help complete the \ac{SMC} PN sample.

\begin{figure*}			
	\includegraphics[width=2.1\columnwidth]{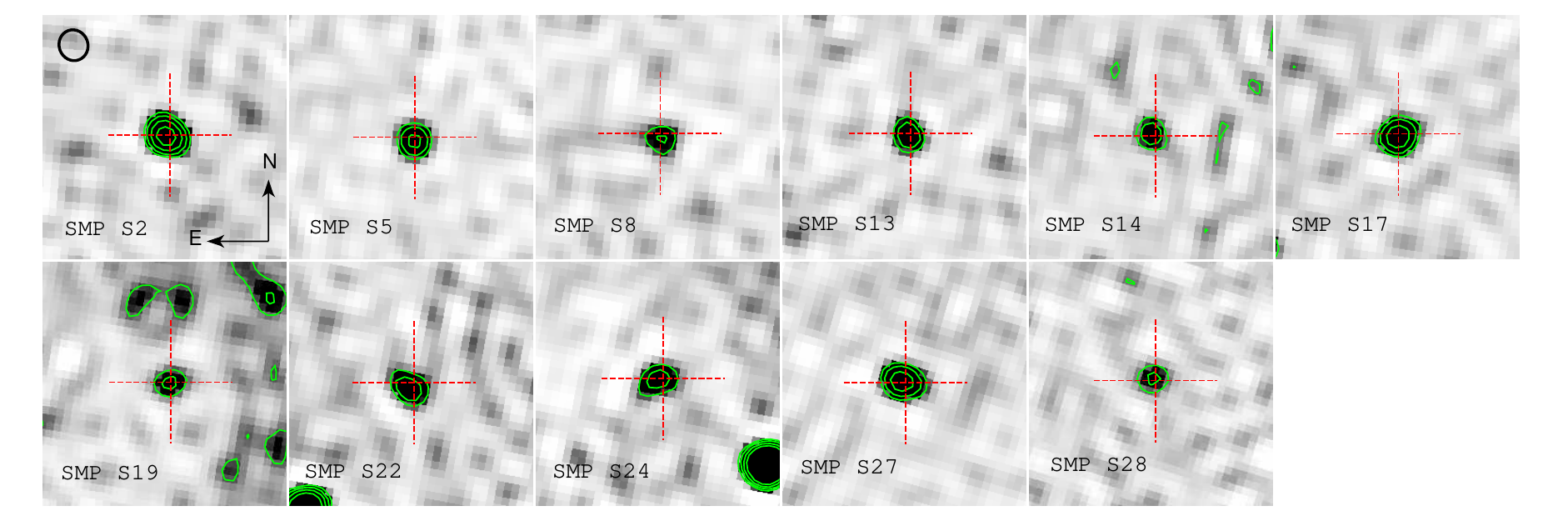}
	\caption{Finding charts of 11 \ac{SMC} \ac{PNe} with positive detection at 1320\,MHz. Each field is 2~arcmin in size and the grey-scale uses the same {\it sinh} stretching. The approximate shape of the synthesized beam and orientation of each chart is displayed in the upper left corner. Red cross represents the catalogued position of a PN and green contours are radio continuum intensity at 3, 5, 8 and 12$\times$ \ac{RMS} noise measured in the vicinity of the object. Here, we present only objects with 5$\sigma$ detections for which we measured accurate flux densities. N is up and E is left in each panel.}
	\label{pn_fcharts}
\end{figure*}

Previous searches for radio \ac{PNe} in the \ac{SMC} \citep{2008SerAJ.176...65P,2009MNRAS.399..769F,2010SerAJ.181...63B,2016Ap&SS.361..108L} yielded 16 bona-fide \ac{PN} detections. Our \ac{ASKAP} ESP survey has revealed 6 new PN radio detections (see Table~\ref{tbl:MeasurePNe}) reported here for the first time (Figure~\ref{pn_fcharts}), bringing the total number of known \ac{SMC} \ac{PNe} detected in radio to 22. Our new data contribute 18 new accurate radio continuum flux density measurements from \ac{ASKAP} on this sample (excluding dubious detections and upper flux limits), of which 7 are at 960\,MHz and 11 at 1320\,MHz.

All finding charts created here have been visually inspected for a possible detection. Of 102 true, likely and possible \ac{SMC} \ac{PNe} in our base catalogue we have matched 17 radio counterparts with peak emission over three times the local noise in the 1320\,MHz map and 8 in the 960\,MHz map. The flux densities were measured using the Gaussian fitting method \texttt{imfit} from \textsc{casa}\footnote{We also used \textsc{aegean}, \textsc{miriad} and \textsc{Selavy} software packages to check for consistency and we found no noticeable discrepancy between various source finders.} \citep{mcmullin2007casa}. Since none of the \ac{SMC} \ac{PNe} are expected to be resolved based on their known optical size, the Gaussian fitting was constrained to the beam size, effectively measuring the peak of the emission. Calculations of uncertainties for this method are based on \cite{1997PASP..109..166C} and have been adopted directly from \texttt{imfit}'s output. We visually inspected all possible detections with a peak brightness over $F\geq3\sigma$ using a comparison between the original and the residual maps. 

The results are presented in Table~\ref{tbl:MeasurePNe} and Figure~\ref{pn_fcharts}. Out of 17 detections at 1320\,MHz, we measured accurate flux densities for 11 \ac{PNe} with peak brightness over 5$\sigma$. Likewise, in the 960\,MHz band we accurately measured 7 out of 8 detected \ac{PNe}. We flagged \ac{PNe} with the peak brightness below 5$\sigma$ in Table~\ref{tbl:MeasurePNe} with a value in parentheses. The flux density estimates for these \ac{PNe} can be considered only as upper limits.

We modelled 5~GHz flux densities for all detected \ac{PNe} in order to construct a radio continuum spectrum distribution of radio-detected \ac{SMC} \ac{PNe}. The flux modelling was performed as follows: {\it a}) if more than 2 data points were available we apply free-free emission spectral energy modelling (\ac{SED}; see further text), {\it b}) if only one or two data points were measured, we estimated the 5~GHz integrated flux density from the measurements at the frequency or frequencies available by applying a simple power law approximation i.e. $S_{5GHz}=S_{\nu}\cdot(5/\nu[\text{GHz}])^{-0.1}$. 

For \ac{SED} modelling we used a spherical shell model  with a constant electron density in the shell ($n_{e}$), outer radius ($R_{\text{out}}$) and inner radius ($R_{\text{in}}$). The model can now be applied to measured data points with:

\begin{equation}
S_{\nu} = \frac{4\pi k T_e \nu^2}{c^2 D^2}R_{\text{out}}^2\int_{0}^{\infty} x (1 - e^{-\tau_{\nu}\cdot g_1(x)})dx
\label{eq:flux_nu}
\end{equation}
\noindent where $\tau_{\nu}$ is the optical thickness through the centre of the nebula at frequency $\nu$ which, for an assumption of $n_{e}=const$ and a pure hydrogen isothermal plasma, can be approximated with $\tau_{\nu}\approx8.235 \cdot 10^{-2}  T_e^{-1.35}  \nu^{-2.1}  n_e^2\cdot 2(R_{\text{out}}-R_{\text{in}})$. Finally, the function $g(x)$ describes the geometry of the nebula \citep[see][for more details]{1975AandA....39..217O}. For this model $g(x)$ has a form:

\begin{equation}
\begin{split}
g_1(x) 	& = \sqrt{1-x^2} - \sqrt{\mu^2-x^2}	& \text{for}~x < \mu, \\
		& = \sqrt{1-x^2}					& \text{for}~\mu\leq x < 1~\text{and} \\
		& = 0						&\text{for}~x \geq 1, \\
\end{split}
\end{equation}

\noindent where $\mu=R_{\text{in}}/R_{\text{out}}$ i.e. inner to outer radii ratio. We fixed the electron temperature to its canonical value ($T_e=10^4$~K) and $\mu=0.4$ as this is found to be the expected average value for majority of Galactic \ac{PNe} \citep{2007AandA...473..467S,2001AandA...378..958M}. With an assumed distance to the \ac{SMC} of 60~kpc we fit the two free parameters, $R_{\text{out}}$ and the emission measure ($EM$), through the centre of the nebula. Finally, the model shown here has been used to estimate the integrated flux density at 5~GHz.

\begin{figure*}			
	\includegraphics[width=2.1\columnwidth]{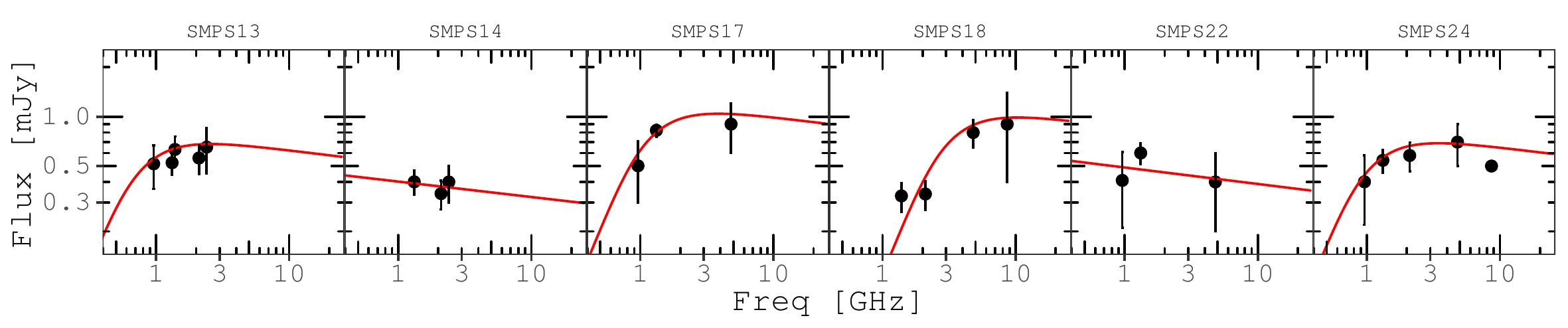}
	\caption{Best fit model \ac{SED}s to the observed flux densities for 6 \ac{SMC} \ac{PNe} with three or more available and good data points.}
	\label{pn_seds}
\end{figure*}

In Fig.~\ref{pn_seds} we show graphical results of the \ac{SED} fitting. From the six \ac{PNe} with an adequate number of data points to apply our spherical shell model, only four converged to acceptable values of $R_{\text{out}}$ and $EM$. For two \ac{PNe} (SMP~S14 and SMP~S22) the model failed to converge and the data were fitted with the simple power law $S_\nu\sim\nu^{\alpha}$. The spectral indices ($\alpha$) obtained are --0.09 and --0.1 for SMP~S14 and SMP~S22, respectively. 

\begin{figure}			
	\includegraphics[width=\columnwidth]{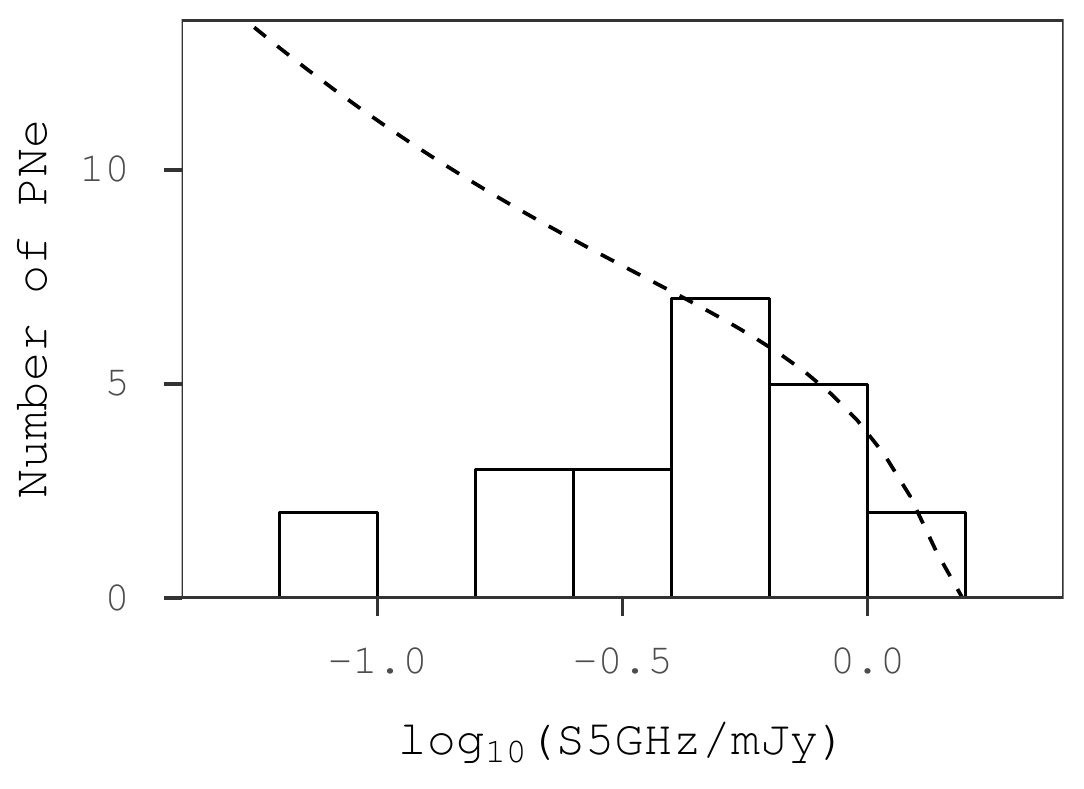}
	\caption{Planetary nebulae luminosity function (PNLF) for the \ac{SMC}. The dashed line represents the theoretical PNLF estimated assuming the sample is complete down to 0.6\,mJy. }
	\label{pnlf}
\end{figure}

We present the modelled 5~GHz total flux densities in Table~\ref{tbl:MeasurePNe} (Column~11). The distribution of the modelled 5~GHz total flux densities for the detected sample is presented in Fig.~\ref{pnlf}. It can be seen that the number of \ac{PNe} drops down below 0.6~mJy which is approximately the detection limit for \ac{ASKAP} \ac{ESP} data. Objects detected below this limit are either upper limits or detections originating from high sensitivity \ac{ATCA} observations \citep{2011SerAJ.183..103W}. Therefore, we believe that our sample of radio detected \ac{SMC} \ac{PNe} is now complete down to $\sim$0.6~mJy. We have used this distribution to roughly estimate the number of \ac{SMC} \ac{PNe} which will be detectable in future \ac{ASKAP} observations of the \ac{SMC}. 

With the approximation that \ac{PNe} are fully ionised spherical shells of  constant mass, expanding with  constant velocity and ionised by a non-evolving central star \citep{1963ApJ...137..747H} the optically thin radio continuum flux would behave as $F\propto~R(t)^{-3}\propto~t^{-3}$. Although simplistic, this approximation has proven to be quite effective in describing changes in flux from Balmer lines during the expansion phase in a large number of observationally constructed PN luminosity functions \citep{2010MNRAS.405.1349R,2010PASA...27..149C}. 

Using a sample of radio catalogued Galactic Bulge \ac{PNe}, \cite{2010PhDT.......399B} showed that the theoretical shape of the PNLF \citep{1989ApJ...339...53C} effectively describes the distribution of radio flux densities of \ac{PNe} at a known distance. Using our assumption that the \ac{SMC} \ac{PN} radio sample is now complete down to 0.6~mJy, we have used the theoretical shape of the PNLF to estimate the distribution of 5~GHz integrated flux densities below the \ac{ASKAP} \ac{ESP} detection limit (more details in Boji\v ci\' c et al. 2019, in prep.). We fit the truncated exponential function \citep{1989ApJ...339...53C} to the obtained distribution of log$_{10}(S_{\rm{5 GHz}})$ fluxes (in mJy). The data is binned to 0.2 dex in log flux density and we have used only bins containing PNe with $S_{\rm{5 GHz}}>0.6$~mJy for the fit. The estimated rough model is over-plotted on the resulting histogram (Figure~\ref{pnlf}; dashed line). Finally, we anticipate that increasing the sensitivity by an order of magnitude would allow detection of another 20 \ac{SMC} \ac{PNe}, while reaching a 10~$\mu$Jy~beam$^{-1}$ \citep{2011PASA...28..215N} will allow us to increase the number of detections to $\approx120$ \ac{PNe} i.e. over 50~per\,cent of the expected \ac{SMC} \ac{PNe} population \citep{2002AJ....123..269J}.

\begin{table*}
\centering
\scriptsize
\caption{Radio continuum population of \ac{PNe} in the \ac{SMC}. The new \ac{ASKAP} radio continuum detection and integrated flux density measurements of the \ac{SMC} \ac{PNe} are indicated with $^{\dag}$.
Uncertain detections and upper and lower limits in flux estimates are indicated in flux columns with :, $<$ and $>$, respectively. The integrated flux density errors are $<$10~per cent unless otherwise stated.} 
 \begin{tabular}{lccccccccccccccc}
 \hline
 					&				&				&ATCA			&ATCA			&ATCA			&ATCA-CABB		&	ATCA	&ASKAP			&ASKAP	& model\\
Other 					&RA    			&DEC    		&$S_{3\,\mathrm{cm}}$  &$S_{6\,\mathrm{cm}}$  &$S_{13\,\mathrm{cm}}$ &$S_{2.1\,\mathrm{GHz}}$ &$S_{20\,\mathrm{cm}}$ &$S_{23\,\mathrm{cm}}$ 		&$S_{32\,\mathrm{cm}}$ &$S_{6\,\mathrm{cm}}$\\
Name 				& (J2000)		& 	(J2000)		&8640\,MHz 	&4800\,MHz 		&2400\,MHz 		&2100\,MHz 		&1388\,MHz 	&1320\,MHz 					&960\,MHz 		&5000\,MHz\\   
					& 				& 				&(mJy) 			&(mJy)	 		&(mJy)	 		&(mJy)	 		&(mJy)	 	&(mJy)	 				&(mJy)			&(mJy)\\ 
	(1)  			&(2)			&(3)			&(4)			&(5)			&(6)			&(7)			&(8)		&(9)	  				&(10)			&(11)\\ 
\hline
SMP\,S2$^{\dag}$			 	&00:32:39 		&$-$71:41:59.5  	&...			&... 			&...	 		&... 			&(2)			&1.25$\pm$0.08 			&1.1$\pm$0.2	&1.1\\
SMP\,S3$^{\dag}$ 	&00:34:22 		&$-$73:13:21.5  	&... 			&... 			&... 			&... 			&... 		& (0.3) 					& (0.4) 			&0.3\\
SMP\,S5$^{\dag}$ 	&00:41:22 		&$-$72:45:16.8  	&... 			&... 			&... 			&... 			&... 		&0.67$\pm$0.08 			&0.3$\pm$0.2 	&0.6\\
SMP\,S6 			&00:41:28 		&$-$73:47:06.4 	&1.1$\pm0.5$ 	&1.3$\pm0.1$ 	&... 			&...			&>0.2 		&... 					&... 			&1.2\\
\smallskip
SMP\,S8$^{\dag}$ 	&00:43:25 		&$-$72:38:18.8  	&...			&... 			&... 			&... 			&... 		&0.43$\pm$0.08 			&... 			&0.4\\
SMP\,S9	 			&00:45:21 		&$-$73:24:10.0  	&... 			&... 			&... 			&0.15			&... 		&...					&...			&0.1\\
SMP\,S10$^{\dag}$ 	&00:47:00 		&$-$72:49:16.6  	&... 			&... 			&... 			&... 			&(0.3) 		& (0.4) 					&... 			&0.2\\
SMP\,S13$^{\dag}$	&00:49:52 		&$-$73:44:21.7  	&... 			&...			&0.7$\pm0.2$	&0.56 			&... 		&0.52$\pm$0.08 			&0.52$\pm$0.15 	&0.6\\
SMP\,S14$^{\dag}$	&00:50:35 		&$-$73:42:57.9 	&... 			&... 			&0.4$\pm0.1$ 	&0.34 			&... 		&0.40$\pm$0.06 			&... 			&0.4\\
\smallskip	
SMP\,S16$^{\dag}$	&00:51:27 		&$-$72:26:11.7 	&... 			&0.6$\pm0.1$ 	&... 			&... 			&... 		&<0.4 					&... 			&0.6\\
J18 				&00:51:43 		&$-$73:00:54.5 	&... 			&... 			&0.24 			&... 			&... 		&... 					&... 			&0.2\\
SMP\,S17$^{\dag}$	&00:51:56 		&$-$71:24:44.2 	&... 			&0.9$\pm0.3$ 	&... 			&... 			&... 		&0.82$\pm$0.07 			&0.5$\pm$0.2 	&0.9\\
SMP\,S18$^{\dag}$	&00:51:58 		&$-$73:20:31.9	&0.9$\pm0.5$ 	&0.8$\pm0.15$ 	&... 			&0.34 			&0.3		& (0.3) 					&... 			&0.8\\
SMP\,S19$^{\dag}$	&00:53:11 		&$-$72:45:07.6 	&... 			&... 			&0.6$\pm0.2$ 	&... 			&... 		&0.36$\pm$0.08 			&... 			&0.6\\
\smallskip
MA891		 		&00:55:59 		&$-$72:14:00.3  	&... 			&... 			&... 			&0.92 			&... 		&... 					&... 			&0.8\\
LIN 302$^{\dag}$ 	&00:56:19 		&$-$72:06:58.5  	&... 			&... 			&... 			&0.11 			&... 		& (0.3) 					&... 			&0.1\\
SMP\,S21		 	&00:56:31 		&$-$72:27:02.0  	&... 			&... 			&... 			&0.21 			&... 		&... 					&... 			&0.2\\
SMP\,S22$^{\dag}$	&00:58:37 		&$-$71:35:48.8 	&... 			&0.4$\pm0.2$ 	&... 			&... 			&... 		&0.60$\pm$0.08 			&0.4$\pm$0.2 	&0.4\\
SMP\,S23$^{\dag}$ 	&00:58:42 		&$-$72:56:59.9  	&... 			&... 			&... 			&... 			&... 		& (0.4) 					&... 			&0.3\\
\smallskip
SMP\,S24$^{\dag}$	&00:59:16 		&$-$72:01:59.8 	&0.5 			&0.7$\pm0.2$ 	&... 			&0.58 			&... 		&0.54$\pm$0.09 			&0.4$\pm$0.2 	&0.7\\
SMP\,S27$^{\dag}$ 	&01:21:11 		&$-$73:14:34.8  	&... 			&... 			&... 			&... 			&... 		&0.88$\pm$0.08 			&0.68$\pm$0.10 	&0.8\\
SMP\,S28$^{\dag}$ 	&01:24:12 		&$-$74:02:32.3  	&... 			&... 			&... 			&... 			&... 		&0.32$\pm$0.07			&... 			&0.3\\
\hline
 \end{tabular}
\label{tbl:MeasurePNe}
\end{table*}

\section{Other interesting sources} 
 \label{Other_sources}
 
In Sections~\ref{SNR_samp} and \ref{PNR_samp} we investigated \ac{SNRs} and \ac{PN} populations within the \ac{SMC}. Large \ac{SMC} \HII\ region complexes N\,19 and N\,66 are shown in Figures~\ref{img_SMC_N19} and \ref{img_SMC_N66}. Together with other \ac{SMC} \HII\ regions and \ac{YSOs}, they will be further investigated in our subsequent papers.

\begin{figure*}
    \includegraphics[scale=0.88]{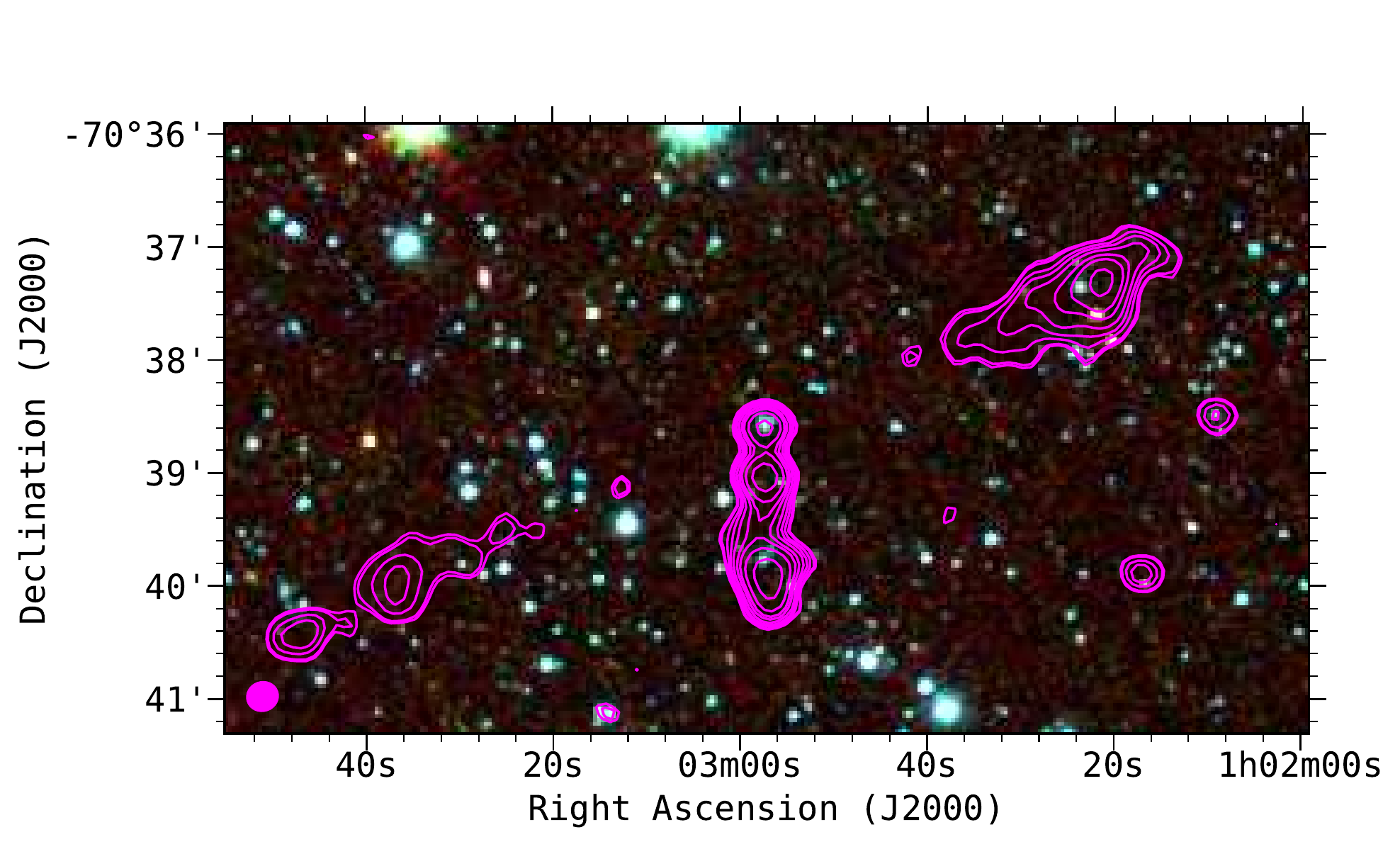}
	\caption{\ac{ASKAP} \ac{ESP} image (contours) of the possible double black hole \ac{AGN}. The background image is a three-colour \ac{IRAC} composite with 8.0, 4.5 and 3.6 \micron{} represented as red, green and blue, respectively. The magenta radio contours are from our 1320\,MHz survey drawn at 0.25, 0.3, 0.5, 0.7, 1, 1.5, 2, 3 and 5\,mJy\,beam$^{-1}$. The 1320\,MHz radio beam size of 16.3\arcsec\ $\times$ 15.1\arcsec\ is shown as a filled magenta ellipse in the lower left corner. 
    }
	\label{img_BBH_cand}
\end{figure*}

\begin{figure}
    \includegraphics[width=\columnwidth]{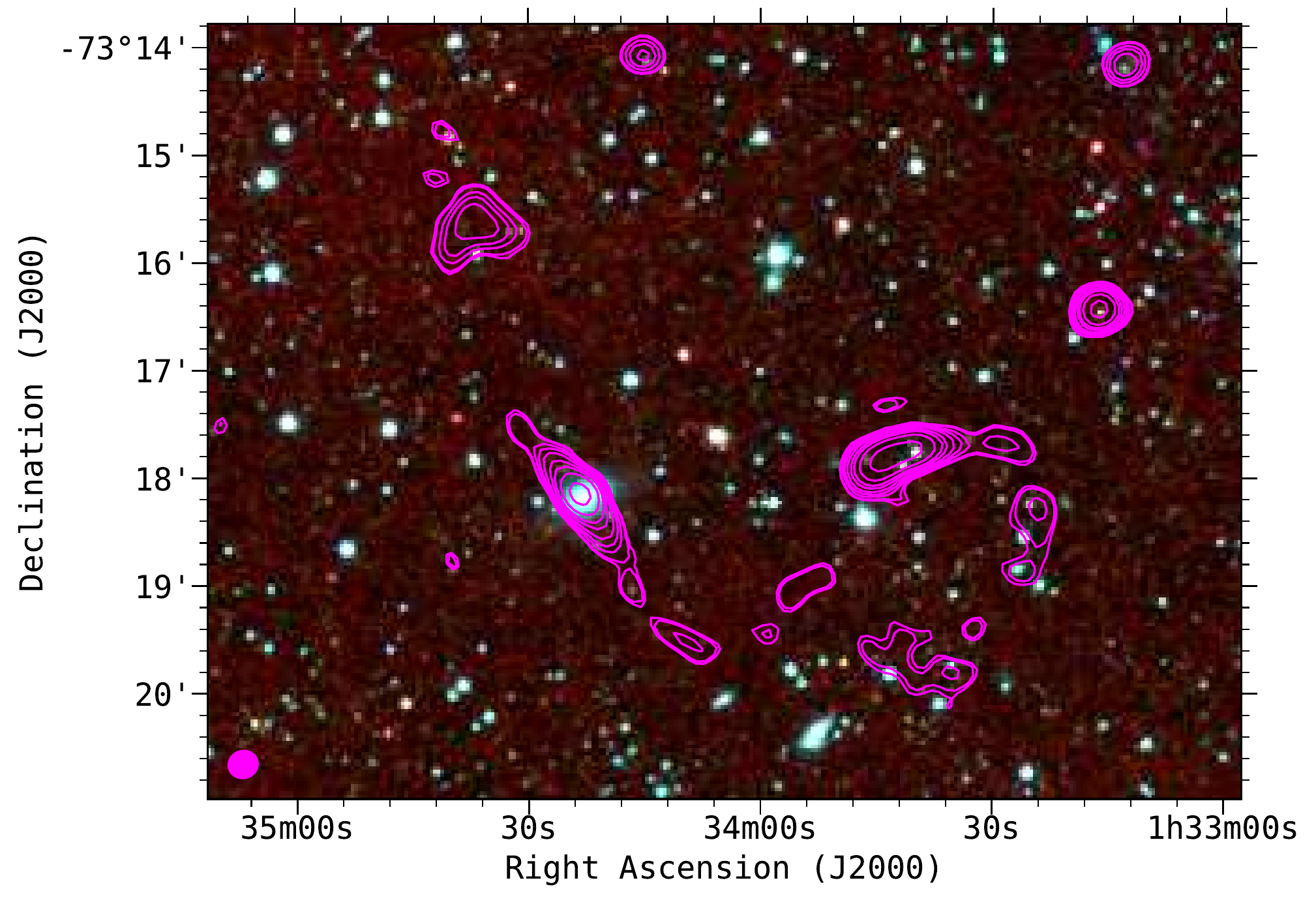}
	\caption{\ac{ASKAP} \ac{ESP} image (contours) shows a long, twisted structure that appears to be a highly distorted tailed radio galaxy associated with 2MASX~J01342297--7318113, a bright galaxy (V=15.9~mag) without spectroscopic redshift. Although the multiple bends might suggest that there is actually more than one radio source, there is no obvious second optical/IR host. The bright compact object near the centre of the radio source elongated E-W and $\sim$3~arcmin W of 2MASX~J01342297--7318113, is 2MASS J01334172--7317527, but GaiaDR2 \citep{2018A&A...616A...1G} shows it to be a star with significant parallax and proper motion. The background image is a three-colour IRAC composite with 8.0, 4.5 and 3.6 \micron{} represented as red, green and blue, respectively. The magenta radio contours are from our 1320\,MHz survey drawn at 0.25, 0.3, 0.5, 0.7, 1, 1.5, 2, 3 and 5\,mJy\,beam$^{-1}$. The 1320\,MHz beam size of 16.3\arcsec\ $\times$ 15.1\arcsec\ is shown as a filled magenta ellipse in the lower left corner.
	}
	\label{img_bckg2}
\end{figure}

\begin{figure}
    \includegraphics[width=\columnwidth]{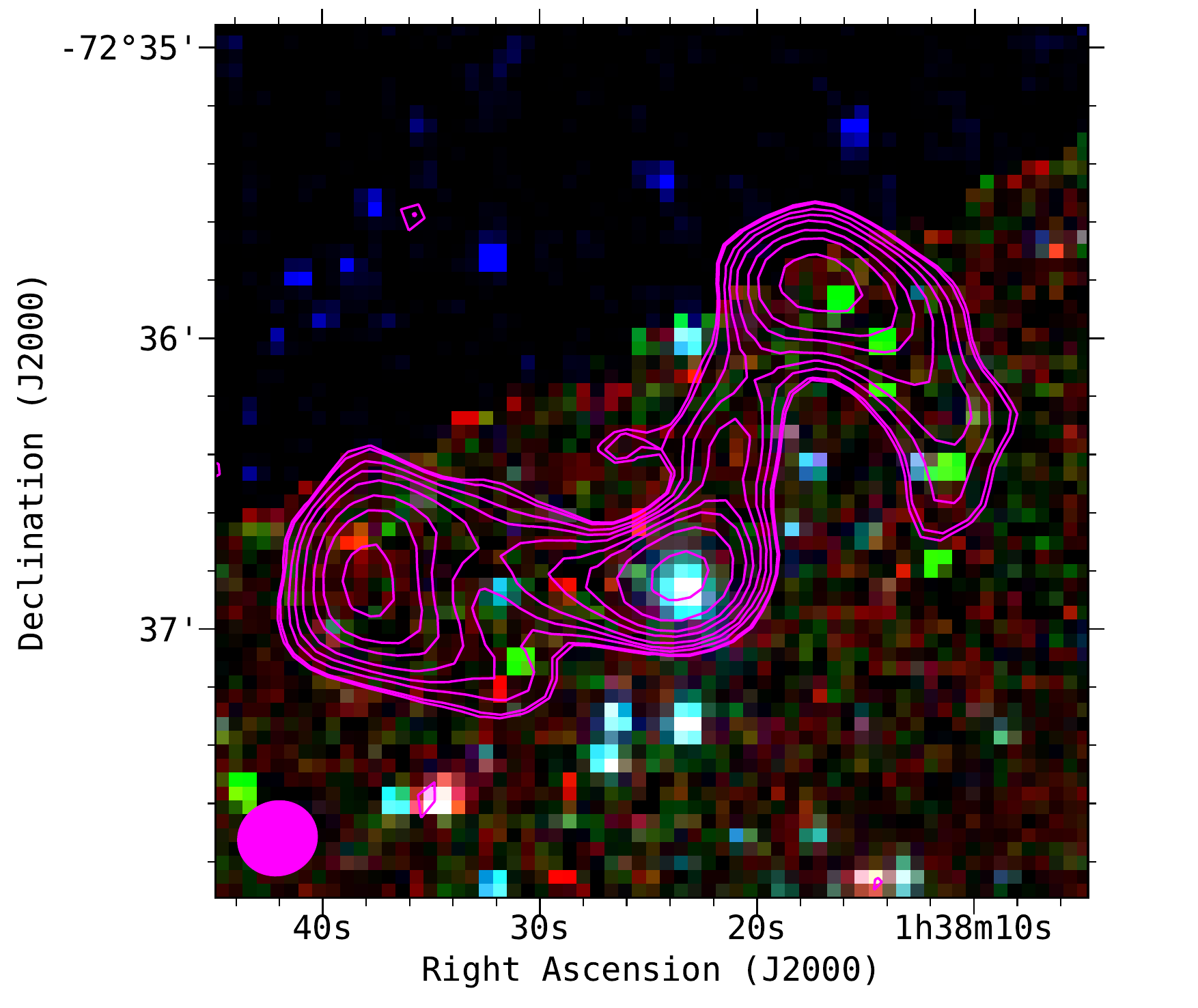}
	\caption{\ac{ASKAP} \ac{ESP} image of the ``duck'' AGN complex or possible bent-tail radio galaxy. The background image is a three-colour IRAC composite with 8.0, 4.5 and 3.6 \micron{} represented as red, green and blue, respectively. The magenta radio contours are from our 1320\,MHz survey drawn at 0.25, 0.3, 0.5, 0.7, 1, 1.5, 2, 3 and 5\,mJy\,beam$^{-1}$. The 1320\,MHz beam size of 16.3\arcsec\ $\times$ 15.1\arcsec\ is shown as a filled magenta ellipse in the lower left corner.}
	\label{img_bckg4}
\end{figure}

\begin{figure*}
    \includegraphics[scale=0.88]{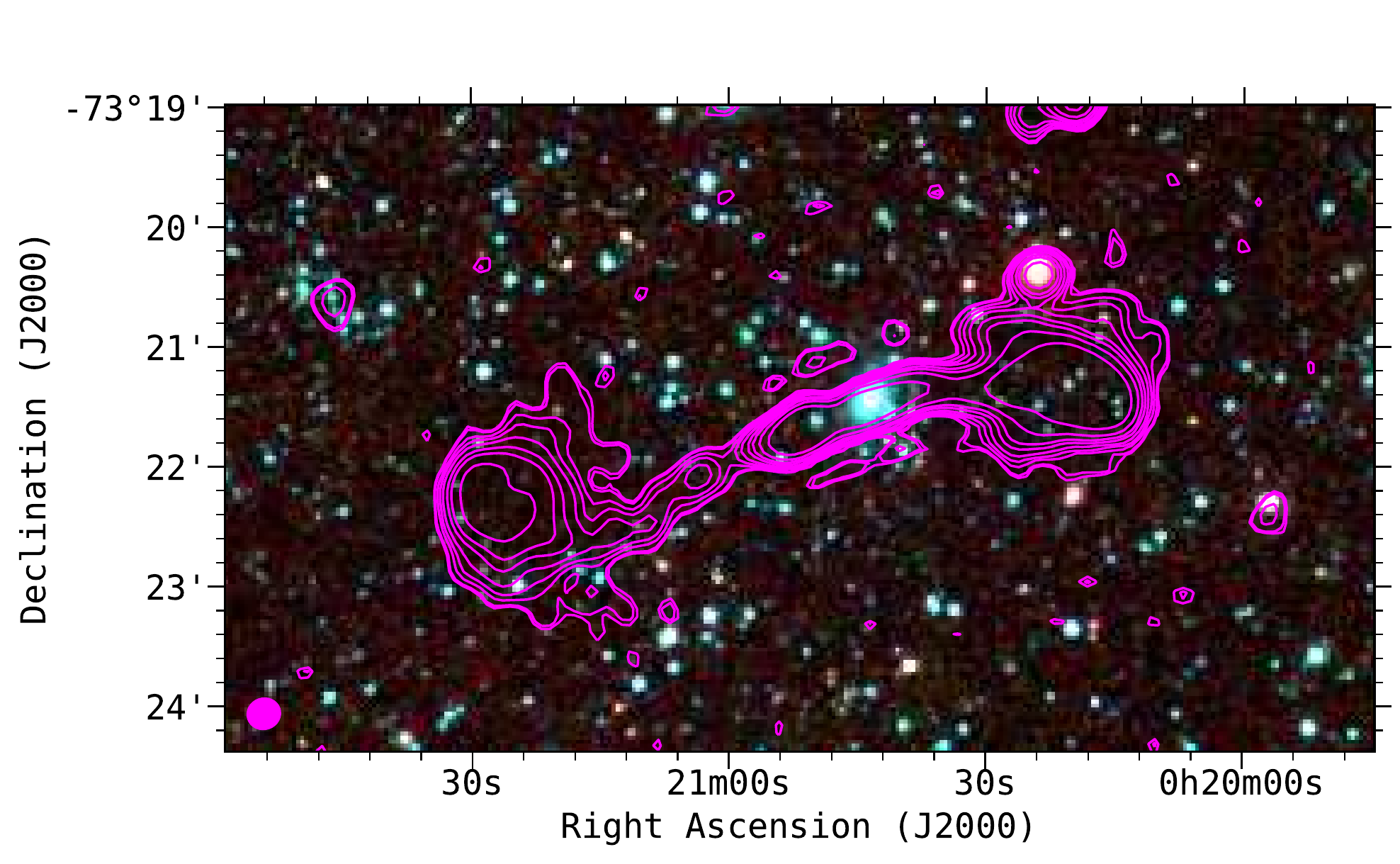}
    \vspace{-0.75cm}
	\caption{\ac{ASKAP} \ac{ESP} image of the FR-II \ac{AGN}. The background image is a three-colour IRAC composite with 8.0, 4.5 and 3.6 \micron{} represented as red, green and blue, respectively. The magenta radio contours are from our 1320\,MHz survey drawn at 0.25, 0.3, 0.5, 0.7, 1, 1.5, 2, 3 and 5\,mJy\,beam$^{-1}$. The 1320\,MHz beam size of 16.3\arcsec\ $\times$ 15.1\arcsec\ is shown as a filled magenta ellipse in the lower left corner.}
	\label{img_bckg3}
\end{figure*}

\begin{figure}
    \includegraphics[width=\columnwidth]{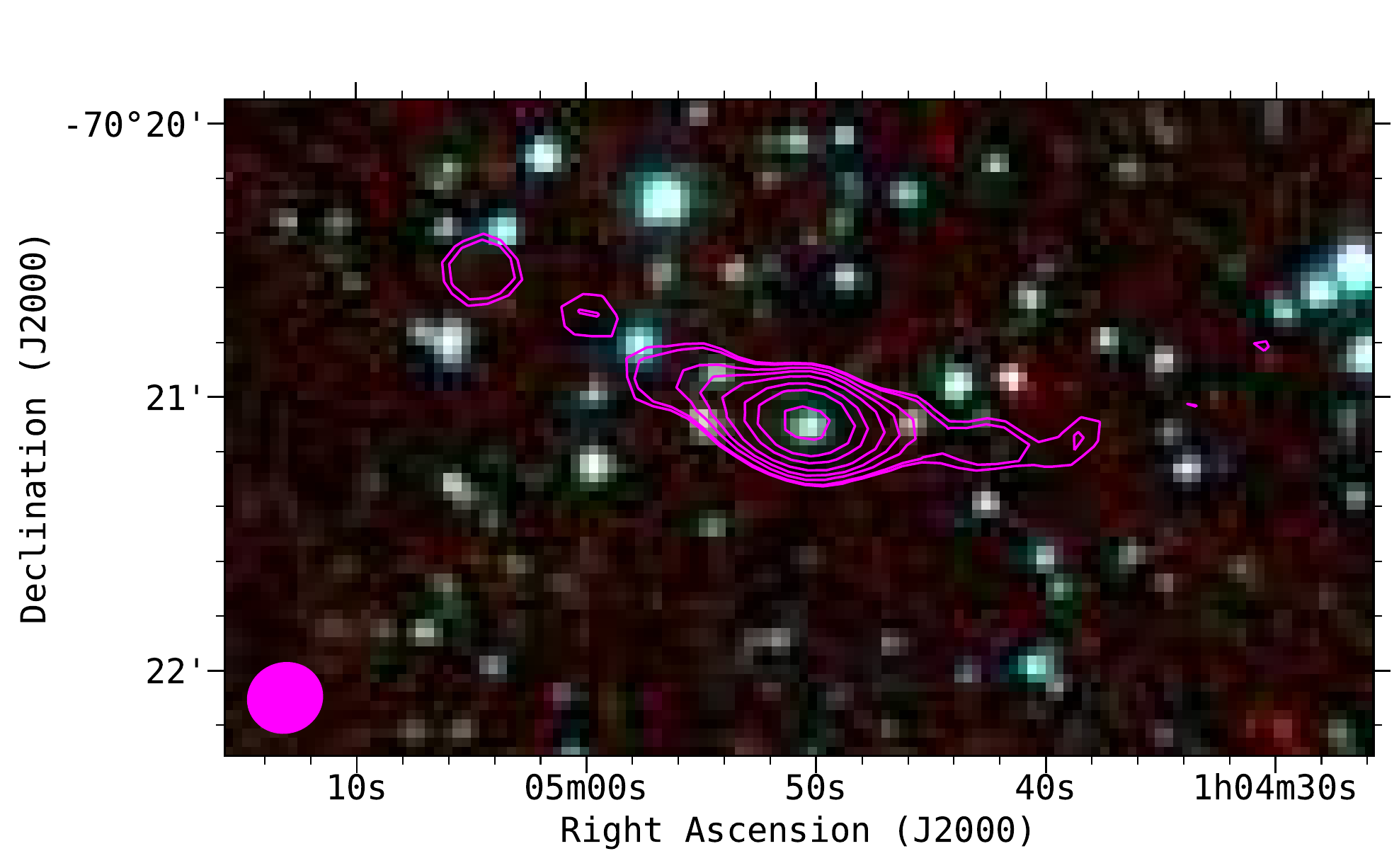}
	    \vspace{-0.75cm}
	    \caption{\ac{ASKAP} \ac{ESP} image of the FR-I AGN. The background image is a three-colour IRAC composite with 8.0, 4.5 and 3.6 \micron{} represented as red, green and blue, respectively. The magenta radio contours are from our 1320\,MHz survey drawn at 0.25, 0.3, 0.5, 0.7, 1, 1.5, 2, 3 and 5\,mJy\,beam$^{-1}$. The 1320\,MHz beam size of 16.3\arcsec\ $\times$ 15.1\arcsec\ is shown as a filled magenta ellipse in the lower left corner.}
	\label{img_bckg5}
\end{figure}

We would also like to highlight some sources of interest behind the \ac{SMC} that are worth following up. Due to their complex radio structure they are probes of galaxy interactions or interaction with the environment. These are presented in Figures~\ref{img_BBH_cand}, \ref{img_bckg2}, \ref{img_bckg4}, \ref{img_bckg3} and \ref{img_bckg5} and would fall into the category of extended radio \ac{AGN}. 

One of the most interesting sources behind the SMC revealed by our ASKAP observations is the radio \ac{AGN} shown in Figure~\ref{img_BBH_cand}. This object displays a set of radio lobes associated to the an infrared IRAC source (background of Figure~\ref{img_BBH_cand}). Also associated with the same source seems to be a radio jet with direction pointing towards the observer. Over the past year there has been a multi-wavelength effort to reveal of the true nature behind this peculiar radio structure, which might be linked to a binary supermassive black hole. Still we cannot rule out chance coincidence. Scheduled follow-up observations, with ATCA (PI: Vardoulaki) and SALT (PI: van Loon), will help shed light to the nature of this interesting radio source.

Other sources also show complex \ac{AGN} structures with various morphological types (Figures~\ref{img_bckg4} and \ref{img_bckg3}) \citep[see also][]{2018MNRAS.481.5247O} and sizes, including a bent source in a possible galaxy cluster (Figure~\ref{img_bckg2}). Such morphology of the extended radio emission, is expected from binary driven jets. A similar configuration is also seen toward other \ac{SMBH}s, such as OJ\,287 \citep{2019arXiv190110768K}. The object in Figure~\ref{img_bckg5} is a slightly bent FR-I type radio galaxy, possibly the central part of a \ac{WAT} in a cluster of galaxies.

\begin{figure}
    \includegraphics[width=\columnwidth]{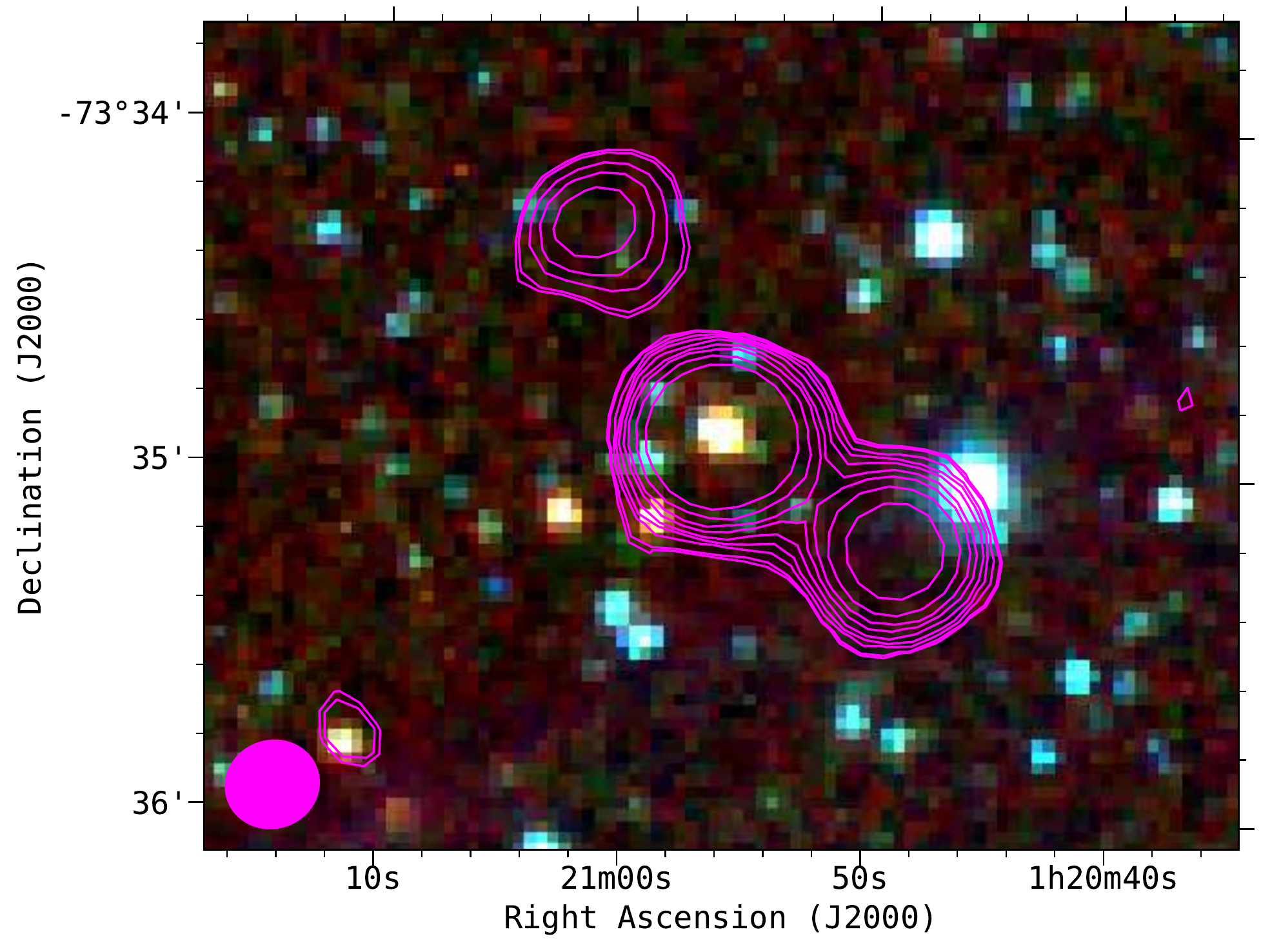}
	\caption{\ac{ASKAP} \ac{ESP} image of the possible AGN J0120$-$7334. The background image is a three-colour IRAC composite with 8.0, 4.5 and 3.6 \micron{} represented as red, green and blue, respectively. The yellow radio contours are from our 1320\,MHz survey drawn at 0.25, 0.3, 0.5, 1, 2, 3, 5, 10, 20 and 30\,mJy\,beam$^{-1}$. The 1320\,MHz beam size of 16.3\arcsec\ $\times$ 15.1\arcsec\ is shown as a filled magenta ellipse in the lower left corner.
	}
	\label{img_bckg23}
\end{figure}

\begin{figure*}
    \includegraphics[width=\textwidth]{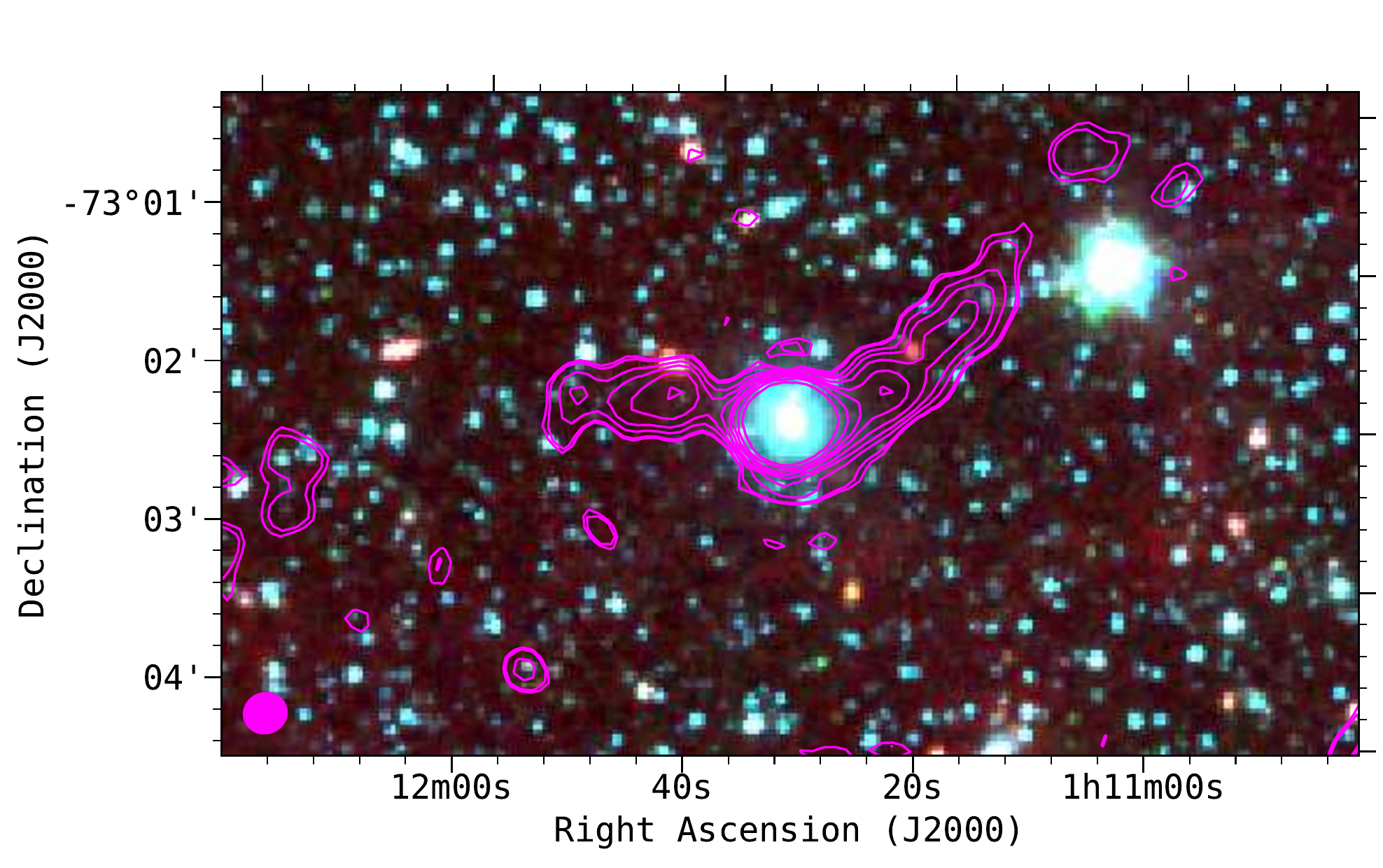}
	    \vspace{-0.75cm}
	    \caption{\ac{ASKAP} \ac{ESP} image of the AGN complex J0111$-$7302. The background image is a three-colour IRAC composite with 8.0, 4.5 and 3.6 \micron{} represented as red, green and blue, respectively. The yellow radio contours are from our 1320\,MHz survey drawn at 0.25, 0.3, 0.5, 1, 2, 3, 5, 10, 20, 30 and 50\,mJy\,beam$^{-1}$. The 1320\,MHz beam size of 16.3\arcsec\ $\times$ 15.1\arcsec\ is shown as a filled magenta ellipse in the lower left corner.
	}
	\label{img_bckg22}
\end{figure*}

We also examined seven \ac{FSRQs} and \ac{BL-Lac} candidates from \citet{2018ApJ...867..131Z} in our radio catalogues. We found that objects J0111$-$7302 (proposed \ac{BL-Lac}\footnote{We note that BL-Lac's with large extents are assumed to be compact which is in contrast to this object. We also note that, for example, \citet{2017A&A...603A.131H} show several known extended (even giant) \ac{BL-Lac}}.; Figure~\ref{img_bckg22}) and possibly J0120$-$7334 (proposed \ac{FSRQs}; Figure~\ref{img_bckg23}) exhibit typical FR-I morphology with complex but steep spectral indices which would argue for their \ac{AGN} nature. The other five sources listed in \citet{2018ApJ...867..131Z} are point-like radio sources in our catalogues: J0039$-$7356 (\ac{BL-Lac}; $\alpha=-1.1$), J0054$-$7248 (\ac{FSRQs}; detected only at 1320\,MHz), J0114$-$7320 and J0122$-$7152 (both proposed \ac{FSRQs} but we detect as a complex \ac{AGN} with jets) and J0123$-$7236 (\ac{BL-Lac}; $\alpha=-0.7$). In addition, we found four radio sources in our catalogue that correspond to the Visual and Infrared Survey Telescope for Astronomy \citep[VISTA;][]{2006Msngr.126...41E}, survey of the \ac{MCs} \citep[VMC;][]{2011A&A...527A.116C} and spectroscopically confirmed quasars \citep{2016A&A...588A..93I}. They are J0027$-$7223 (S$_{1320\,\rm{MHz}}$=0.265\,mJy), J0029$-$7146 ($\alpha=-1.0$), J0035$-$7201 ($\alpha=-0.5$) and J0119$-$7348 ($\alpha=-0.8$). While small, this sample exhibits steep spectral indices typical of the majority of background radio objects.

Finally, we note a radio detection of an ultra-bright submillimeter galaxy MM~J01071$-$7302  \citep{2013ApJ...774L..30T} and found a steep spectrum with $\alpha=-0.9$.

In total, we found 7736 point radio sources with fluxes over 5 times the local noise, the vast majority of which are likely to be in the background of the \ac{SMC}. Through absorption measurements, all these sources can provide excellent probes for the study of cold gas in both \ac{SMC} and the Galaxy \citep[e.g.][]{2018ApJS..235....1L,2015aska.confE.130M,2013PASA...30....3D}. A more detailed analysis of these background sources will be presented in Pennock et al. (in prep.).


\section{Conclusions}
In this paper we present the \ac{ASKAP} \ac{EMU} \ac{ESP} radio continuum survey of the \ac{SMC} taken at 960\,MHz and 1320\,MHz. Our findings can be summarised as follows:
\begin{itemize}
    \item  This new \ac{ASKAP} survey is a significant improvement (factor of $\sim$4 in the median RMS) compared to previous ATCA/MOST  surveys of the \ac{SMC}. 
    \item We identify 4489 and 5954 point sources at 960\,MHz and 1320\,MHz, respectively (Tables~\ref{tab:960_point} and \ref{tab:1320_point}), with the  majority of these sources detected above the 5$\sigma$ threshold in their respective bands. We also list non-point sources at both \ac{ASKAP} frequencies in Tables~\ref{tab:960} and \ref{tab:1320} (282 and 641, respectively).
    \item Combining our two new \ac{ASKAP} catalogues with other radio continuum surveys, we found 7736 point-like sources in common which we list in Table~\ref{tab:allsources}, together with spectral indices we determined from all available survey data.
    \item Two new low surface brightness \ac{SNR} candidates were discovered, bringing the total number of \ac{SNRs} and \ac{SNR} candidates in the \ac{SMC} to 23.
    \item Radio counterparts to 22 optically known \ac{PNe} were detected. This sample of \ac{PNe} is complete down to 0.6\,mJy.
\end{itemize}

\section*{Acknowledgements}

The Australian SKA Pathfinder (ASKAP) are part of the Australian Telescope which is funded by the Commonwealth of Australia for operation as National Facility managed by CSIRO. We used the \textsc{karma} and \textsc{miriad} software packages developed by the \ac{ATNF}. Operation of \ac{ASKAP} is funded by the Australian Government with support from the National Collaborative Research Infrastructure Strategy. \ac{ASKAP} uses the resources of the Pawsey Supercomputing Centre. Establishment of \ac{ASKAP}, the Murchison Radio--astronomy Observatory and the Pawsey Supercomputing Centre are initiatives of the Australian Government, with support from the Government of Western Australia and the Science and Industry Endowment Fund. We acknowledge the Wajarri Yamatji people as the traditional owners of the Observatory site.
T.D.J.~acknowledges support for this research from a Royal Society Newton International Fellowship, NF171032.
M.J.M.~acknowledges the support of the National Science Centre, Poland, through the SONATA BIS grant 2018/30/E/ST9/00208.
The National Radio Astronomy Observatory is a facility of the National Science Foundation operated under cooperative agreement by Associated Universities, Inc.
Partial support for L.R. comes from U.S. National Science Foundation grant AST1714205 to the University of Minnesota.
Project/paper is partially supported by NSFC No. 11690024, CAS International Partnership No. 114A11KYSB20160008.
This work is part of the project 176005 ``Emission nebulae: structure and evolution'' supported by the Ministry of Education, Science, and Technological Development of the Republic of Serbia.
H.A. benefited from project CIIC 218/2019 of University of Guanajuato.
The authors would like to thank the anonymous referee for a constructive report and useful comments. 





\bibliographystyle{mnras}
\bibliography{refs1.bib} 

\begin{thebibliography}{}
\makeatletter
\relax
\def\mn@urlcharsother{\let\do\@makeother \do\$\do\&\do\#\do\^\do\_\do\%\do\~}
\def\mn@doi{\begingroup\mn@urlcharsother \@ifnextchar [ {\mn@doi@}
  {\mn@doi@[]}}
\def\mn@doi@[#1]#2{\def\@tempa{#1}\ifx\@tempa\@empty \href
  {http://dx.doi.org/#2} {doi:#2}\else \href {http://dx.doi.org/#2} {#1}\fi
  \endgroup}
\def\mn@eprint#1#2{\mn@eprint@#1:#2::\@nil}
\def\mn@eprint@arXiv#1{\href {http://arxiv.org/abs/#1} {{\tt arXiv:#1}}}
\def\mn@eprint@dblp#1{\href {http://dblp.uni-trier.de/rec/bibtex/#1.xml}
  {dblp:#1}}
\def\mn@eprint@#1:#2:#3:#4\@nil{\def\@tempa {#1}\def\@tempb {#2}\def\@tempc
  {#3}\ifx \@tempc \@empty \let \@tempc \@tempb \let \@tempb \@tempa \fi \ifx
  \@tempb \@empty \def\@tempb {arXiv}\fi \@ifundefined
  {mn@eprint@\@tempb}{\@tempb:\@tempc}{\expandafter \expandafter \csname
  mn@eprint@\@tempb\endcsname \expandafter{\@tempc}}}

\bibitem[\protect\citeauthoryear{{Alsaberi} et~al.,}{{Alsaberi}
  et~al.}{2019}]{2019MNRAS.486.2507A}
{Alsaberi} R. Z.~E.,  et~al., 2019, \mn@doi [\mnras] {10.1093/mnras/stz971},
  \href {https://ui.adsabs.harvard.edu/abs/2019MNRAS.486.2507A} {486, 2507}

\bibitem[\protect\citeauthoryear{Arbutina, Uro{\v{s}}evi{\'{c}},
  Andjeli{\'{c}}, Pavlovi{\'{c}}  \& Vukoti{\'{c}}}{Arbutina
  et~al.}{2012}]{Arbutina_2012}
Arbutina B.,  Uro{\v{s}}evi{\'{c}} D.,  Andjeli{\'{c}} M.~M.,  Pavlovi{\'{c}}
  M.~Z.,   Vukoti{\'{c}} B.,  2012, \mn@doi [The Astrophysical Journal]
  {10.1088/0004-637x/746/1/79}, 746, 79

\bibitem[\protect\citeauthoryear{{Arbutina}, {Uro{\v s}evi{\'c}}, {Vu{\v
  c}eti{\'c}}, {Pavlovi{\'c}}  \& {Vukoti{\'c}}}{{Arbutina}
  et~al.}{2013}]{2013ApJ...777...31A}
{Arbutina} B.,  {Uro{\v s}evi{\'c}} D.,  {Vu{\v c}eti{\'c}} M.~M.,
  {Pavlovi{\'c}} M.~Z.,   {Vukoti{\'c}} B.,  2013, \mn@doi [\apj]
  {10.1088/0004-637X/777/1/31}, \href
  {http://adsabs.harvard.edu/abs/2013ApJ...777...31A} {777, 31}

\bibitem[\protect\citeauthoryear{{Boji{\v c}i{\'c}}}{{Boji{\v
  c}i{\'c}}}{2010}]{2010PhDT.......399B}
{Boji{\v c}i{\'c}} I.,  2010, PhD thesis, Macquarie University

\bibitem[\protect\citeauthoryear{{Boji{\v c}i{\'c}}, {Filipovi{\'c}}  \&
  {Crawford}}{{Boji{\v c}i{\'c}} et~al.}{2010}]{2010SerAJ.181...63B}
{Boji{\v c}i{\'c}} I.~S.,  {Filipovi{\'c}} M.~D.,   {Crawford} E.~J.,  2010,
  \mn@doi [Serbian Astronomical Journal] {10.2298/SAJ1081063B}, \href
  {http://adsabs.harvard.edu/abs/2010SerAJ.181...63B} {181, 63}

\bibitem[\protect\citeauthoryear{{Bozzetto} et~al.,}{{Bozzetto}
  et~al.}{2017}]{2017ApJS..230....2B}
{Bozzetto} L.~M.,  et~al., 2017, \mn@doi [\apjs] {10.3847/1538-4365/aa653c},
  \href {http://adsabs.harvard.edu/abs/2017ApJS..230....2B} {230, 2}

\bibitem[\protect\citeauthoryear{{Ciardullo}}{{Ciardullo}}{2010}]{2010PASA...27..149C}
{Ciardullo} R.,  2010, \mn@doi [\pasa] {10.1071/AS09022}, \href
  {http://adsabs.harvard.edu/abs/2010PASA...27..149C} {27, 149}

\bibitem[\protect\citeauthoryear{{Ciardullo}, {Jacoby}, {Ford}  \&
  {Neill}}{{Ciardullo} et~al.}{1989}]{1989ApJ...339...53C}
{Ciardullo} R.,  {Jacoby} G.~H.,  {Ford} H.~C.,   {Neill} J.~D.,  1989, \mn@doi
  [\apj] {10.1086/167275}, \href
  {http://adsabs.harvard.edu/abs/1989ApJ...339...53C} {339, 53}

\bibitem[\protect\citeauthoryear{{Cioni} et~al.,}{{Cioni}
  et~al.}{2011}]{2011A&A...527A.116C}
{Cioni} M.-R.~L.,  et~al., 2011, \mn@doi [\aap] {10.1051/0004-6361/201016137},
  \href {http://adsabs.harvard.edu/abs/2011A%26A...527A.116C} {527, A116}

\bibitem[\protect\citeauthoryear{{Clarke}}{{Clarke}}{1976}]{1976MNRAS.174..393C}
{Clarke} J.~N.,  1976, \mn@doi [\mnras] {10.1093/mnras/174.2.393}, \href
  {http://adsabs.harvard.edu/abs/1976MNRAS.174..393C} {174, 393}

\bibitem[\protect\citeauthoryear{{Collier}}{{Collier}}{2016}]{2016PhDT.......266C}
{Collier} J.,  2016, PhD thesis, Western Sydney University (Australia

\bibitem[\protect\citeauthoryear{{Collier} et~al.,}{{Collier}
  et~al.}{2018}]{2018MNRAS.477..578C}
{Collier} J.~D.,  et~al., 2018, \mn@doi [\mnras] {10.1093/mnras/sty564}, \href
  {http://adsabs.harvard.edu/abs/2018MNRAS.477..578C} {477, 578}

\bibitem[\protect\citeauthoryear{{Condon}}{{Condon}}{1997}]{1997PASP..109..166C}
{Condon} J.~J.,  1997, \mn@doi [\pasp] {10.1086/133871}, \href
  {http://adsabs.harvard.edu/abs/1997PASP..109..166C} {109, 166}

\bibitem[\protect\citeauthoryear{{Cornwell}, {Humphreys}, {Lenc}, {Voronkov}
  \& {Whiting}}{{Cornwell} et~al.}{2011}]{Cornwell11}
{Cornwell} T.~J.,  {Humphreys} B.,  {Lenc} E.,  {Voronkov} M.,   {Whiting}
  M.~T.,  2011, Technical Report~028, {Askap-sw-0020: ASKAP science processing}

\bibitem[\protect\citeauthoryear{{Crawford}, {Filipovi{\'c}}, {de Horta},
  {Wong}, {Tothill}, {Draskovic}, {Collier}  \& {Galvin}}{{Crawford}
  et~al.}{2011}]{2011SerAJ.183...95C}
{Crawford} E.~J.,  {Filipovi{\'c}} M.~D.,  {de Horta} A.~Y.,  {Wong} G.~F.,
  {Tothill} N.~F.~H.,  {Draskovic} D.,  {Collier} J.~D.,   {Galvin} T.~J.,
  2011, \mn@doi [Serbian Astronomical Journal] {10.2298/SAJ1183095C}, \href
  {http://adsabs.harvard.edu/abs/2011SerAJ.183...95C} {183, 95}

\bibitem[\protect\citeauthoryear{{Crawford}, {Filipovi{\'c}}, {McEntaffer},
  {Brantseg}, {Heitritter}, {Roper}, {Haberl}  \& {Uro{\v
  s}evi{\'c}}}{{Crawford} et~al.}{2014}]{2014AJ....148...99C}
{Crawford} E.~J.,  {Filipovi{\'c}} M.~D.,  {McEntaffer} R.~L.,  {Brantseg} T.,
  {Heitritter} K.,  {Roper} Q.,  {Haberl} F.,   {Uro{\v s}evi{\'c}} D.,  2014,
  \mn@doi [\aj] {10.1088/0004-6256/148/5/99}, \href
  {http://adsabs.harvard.edu/abs/2014AJ....148...99C} {148, 99}

\bibitem[\protect\citeauthoryear{{Davies}, {Elliott}  \& {Meaburn}}{{Davies}
  et~al.}{1976}]{1976MmRAS..81...89D}
{Davies} R.~D.,  {Elliott} K.~H.,   {Meaburn} J.,  1976, \memras, \href
  {http://adsabs.harvard.edu/abs/1976MmRAS..81...89D} {81, 89}

\bibitem[\protect\citeauthoryear{{DeBoer} et~al.,}{{DeBoer}
  et~al.}{2009}]{2009IEEEP..97.1507D}
{DeBoer} D.~R.,  et~al., 2009, \mn@doi [IEEE Proceedings]
  {10.1109/JPROC.2009.2016516}, \href
  {http://adsabs.harvard.edu/abs/2009IEEEP..97.1507D} {97, 1507}

\bibitem[\protect\citeauthoryear{{Di Teodoro} et~al.,}{{Di Teodoro}
  et~al.}{2019}]{2019MNRAS.483..392D}
{Di Teodoro} E.~M.,  et~al., 2019, \mn@doi [\mnras] {10.1093/mnras/sty3095},
  \href {http://adsabs.harvard.edu/abs/2019MNRAS.483..392D} {483, 392}

\bibitem[\protect\citeauthoryear{{Dickey} et~al.,}{{Dickey}
  et~al.}{2013}]{2013PASA...30....3D}
{Dickey} J.~M.,  et~al., 2013, \mn@doi [\pasa] {10.1017/pasa.2012.003}, \href
  {https://ui.adsabs.harvard.edu/abs/2013PASA...30....3D} {30, e003}

\bibitem[\protect\citeauthoryear{{Emerson}, {McPherson}  \&
  {Sutherland}}{{Emerson} et~al.}{2006}]{2006Msngr.126...41E}
{Emerson} J.,  {McPherson} A.,   {Sutherland} W.,  2006, The Messenger, \href
  {http://adsabs.harvard.edu/abs/2006Msngr.126...41E} {126, 41}

\bibitem[\protect\citeauthoryear{{Filipovi{\'c}}, {Jones}, {White}, {Haynes},
  {Klein}  \& {Wielebinski}}{{Filipovi{\'c}}
  et~al.}{1997}]{1997A&AS..121..321F}
{Filipovi{\'c}} M.~D.,  {Jones} P.~A.,  {White} G.~L.,  {Haynes} R.~F.,
  {Klein} U.,   {Wielebinski} R.,  1997, \mn@doi [\aaps] {10.1051/aas:1997317},
  \href {http://adsabs.harvard.edu/abs/1997A%26AS..121..321F} {121, 321}

\bibitem[\protect\citeauthoryear{Filipovi{\'c}, {Haynes}, {White}  \&
  {Jones}}{Filipovi{\'c} et~al.}{1998}]{1998A&AS..130..421F}
Filipovi{\'c} M.~D.,  {Haynes} R.~F.,  {White} G.~L.,   {Jones} P.~A.,  1998,
  \mn@doi [\aaps] {10.1051/aas:1998417}, \href
  {http://adsabs.harvard.edu/abs/1998A%26AS..130..421F} {130, 421}

\bibitem[\protect\citeauthoryear{{Filipovi{\'c}}, {Bohlsen}, {Reid},
  {Staveley-Smith}, {Jones}, {Nohejl}  \& {Goldstein}}{{Filipovi{\'c}}
  et~al.}{2002}]{2002MNRAS.335.1085F}
{Filipovi{\'c}} M.~D.,  {Bohlsen} T.,  {Reid} W.,  {Staveley-Smith} L.,
  {Jones} P.~A.,  {Nohejl} K.,   {Goldstein} G.,  2002, \mn@doi [\mnras]
  {10.1046/j.1365-8711.2002.05702.x}, \href
  {http://adsabs.harvard.edu/abs/2002MNRAS.335.1085F} {335, 1085}

\bibitem[\protect\citeauthoryear{{Filipovi{\'c}}, {Payne}, {Reid}, {Danforth},
  {Staveley-Smith}, {Jones}  \& {White}}{{Filipovi{\'c}}
  et~al.}{2005}]{2005MNRAS.364..217F}
{Filipovi{\'c}} M.~D.,  {Payne} J.~L.,  {Reid} W.,  {Danforth} C.~W.,
  {Staveley-Smith} L.,  {Jones} P.~A.,   {White} G.~L.,  2005, \mn@doi [\mnras]
  {10.1111/j.1365-2966.2005.09554.x}, \href
  {http://adsabs.harvard.edu/abs/2005MNRAS.364..217F} {364, 217}

\bibitem[\protect\citeauthoryear{{Filipovi{\'c}} et~al.,}{{Filipovi{\'c}}
  et~al.}{2008}]{2008A&A...485...63F}
{Filipovi{\'c}} M.~D.,  et~al., 2008, \mn@doi [\aap]
  {10.1051/0004-6361:200809642}, \href
  {http://adsabs.harvard.edu/abs/2008A%26A...485...63F} {485, 63}

\bibitem[\protect\citeauthoryear{{Filipovi{\'c}} et~al.,}{{Filipovi{\'c}}
  et~al.}{2009}]{2009MNRAS.399..769F}
{Filipovi{\'c}} M.~D.,  et~al., 2009, \mn@doi [\mnras]
  {10.1111/j.1365-2966.2009.15307.x}, \href
  {http://adsabs.harvard.edu/abs/2009MNRAS.399..769F} {399, 769}

\bibitem[\protect\citeauthoryear{{Findlay}}{{Findlay}}{1966}]{1966ARA&A...4...77F}
{Findlay} J.~W.,  1966, \mn@doi [\araa] {10.1146/annurev.aa.04.090166.000453},
  \href {https://ui.adsabs.harvard.edu/abs/1966ARA&A...4...77F} {4, 77}

\bibitem[\protect\citeauthoryear{{For} et~al.,}{{For}
  et~al.}{2018}]{2018MNRAS.480.2743F}
{For} B.-Q.,  et~al., 2018, \mn@doi [\mnras] {10.1093/mnras/sty1960}, \href
  {http://adsabs.harvard.edu/abs/2018MNRAS.480.2743F} {480, 2743}

\bibitem[\protect\citeauthoryear{{Franzen} et~al.,}{{Franzen}
  et~al.}{2015}]{2015MNRAS.453.4020F}
{Franzen} T.~M.~O.,  et~al., 2015, \mn@doi [\mnras] {10.1093/mnras/stv1866},
  \href {http://adsabs.harvard.edu/abs/2015MNRAS.453.4020F} {453, 4020}

\bibitem[\protect\citeauthoryear{{Gaensler}, {Haverkorn}, {Staveley-Smith},
  {Dickey}, {McClure-Griffiths}, {Dickel}  \& {Wolleben}}{{Gaensler}
  et~al.}{2005}]{2005Sci...307.1610G}
{Gaensler} B.~M.,  {Haverkorn} M.,  {Staveley-Smith} L.,  {Dickey} J.~M.,
  {McClure-Griffiths} N.~M.,  {Dickel} J.~R.,   {Wolleben} M.,  2005, \mn@doi
  [Science] {10.1126/science.1108832}, \href
  {http://adsabs.harvard.edu/abs/2005Sci...307.1610G} {307, 1610}

\bibitem[\protect\citeauthoryear{{Gaia Collaboration} et~al.,}{{Gaia
  Collaboration} et~al.}{2018}]{2018A&A...616A...1G}
{Gaia Collaboration} et~al., 2018, \mn@doi [\aap]
  {10.1051/0004-6361/201833051}, \href
  {https://ui.adsabs.harvard.edu/abs/2018A%26A...616A...1G} {616, A1}

\bibitem[\protect\citeauthoryear{{Galvin} et~al.,}{{Galvin}
  et~al.}{2018}]{2018MNRAS.474..779G}
{Galvin} T.~J.,  et~al., 2018, \mn@doi [\mnras] {10.1093/mnras/stx2613}, \href
  {http://adsabs.harvard.edu/abs/2018MNRAS.474..779G} {474, 779}

\bibitem[\protect\citeauthoryear{{Gordon} et~al.,}{{Gordon}
  et~al.}{2011}]{2011AJ....142..102G}
{Gordon} K.~D.,  et~al., 2011, \mn@doi [\aj] {10.1088/0004-6256/142/4/102},
  \href {http://adsabs.harvard.edu/abs/2011AJ....142..102G} {142, 102}

\bibitem[\protect\citeauthoryear{{Gvaramadze}, {Kniazev}  \&
  {Oskinova}}{{Gvaramadze} et~al.}{2019}]{2019MNRAS.485L...6G}
{Gvaramadze} V.~V.,  {Kniazev} A.~Y.,   {Oskinova} L.~M.,  2019, \mn@doi
  [\mnras] {10.1093/mnrasl/slz018}, \href
  {http://adsabs.harvard.edu/abs/2019MNRAS.485L...6G} {485, L6}

\bibitem[\protect\citeauthoryear{{Haberl}, {Filipovi{\'c}}, {Pietsch}  \&
  {Kahabka}}{{Haberl} et~al.}{2000}]{2000A&AS..142...41H}
{Haberl} F.,  {Filipovi{\'c}} M.~D.,  {Pietsch} W.,   {Kahabka} P.,  2000,
  \mn@doi [\aaps] {10.1051/aas:2000136}, \href
  {http://adsabs.harvard.edu/abs/2000A%26AS..142...41H} {142, 41}

\bibitem[\protect\citeauthoryear{{Haberl}, {Sturm}, {Filipovi{\'c}}, {Pietsch}
  \& {Crawford}}{{Haberl} et~al.}{2012a}]{2012A&A...537L...1H}
{Haberl} F.,  {Sturm} R.,  {Filipovi{\'c}} M.~D.,  {Pietsch} W.,   {Crawford}
  E.~J.,  2012a, \mn@doi [\aap] {10.1051/0004-6361/201118369}, \href
  {http://adsabs.harvard.edu/abs/2012A%26A...537L...1H} {537, L1}

\bibitem[\protect\citeauthoryear{{Haberl} et~al.,}{{Haberl}
  et~al.}{2012b}]{2012A&A...545A.128H}
{Haberl} F.,  et~al., 2012b, \mn@doi [\aap] {10.1051/0004-6361/201219758},
  \href {http://adsabs.harvard.edu/abs/2012A%26A...545A.128H} {545, A128}

\bibitem[\protect\citeauthoryear{{Hancock}, {Murphy}, {Gaensler}, {Hopkins}  \&
  {Curran}}{{Hancock} et~al.}{2012}]{2012ascl.soft12009H}
{Hancock} P.~J.,  {Murphy} T.,  {Gaensler} B.~M.,  {Hopkins} A.,   {Curran}
  J.~R.,  2012, {Aegean: Compact source finding in radio images}, Astrophysics
  Source Code Library (\mn@eprint {ascl} {1212.009})

\bibitem[\protect\citeauthoryear{{Hancock}, {Trott}  \&
  {Hurley-Walker}}{{Hancock} et~al.}{2018}]{2018PASA...35...11H}
{Hancock} P.~J.,  {Trott} C.~M.,   {Hurley-Walker} N.,  2018, \mn@doi [\pasa]
  {10.1017/pasa.2018.3}, \href
  {http://adsabs.harvard.edu/abs/2018PASA...35...11H} {35, e011}

\bibitem[\protect\citeauthoryear{{Haynes}, {Klein}, {Wielebinski}  \&
  {Murray}}{{Haynes} et~al.}{1986}]{1986A&A...159...22H}
{Haynes} R.~F.,  {Klein} U.,  {Wielebinski} R.,   {Murray} J.~D.,  1986, \aap,
  \href {http://adsabs.harvard.edu/abs/1986A%26A...159...22H} {159, 22}

\bibitem[\protect\citeauthoryear{{Henize}}{{Henize}}{1956}]{1956ApJS....2..315H}
{Henize} K.~G.,  1956, \mn@doi [\apjs] {10.1086/190025}, \href
  {http://adsabs.harvard.edu/abs/1956ApJS....2..315H} {2, 315}

\bibitem[\protect\citeauthoryear{{Henize} \& {Westerlund}}{{Henize} \&
  {Westerlund}}{1963}]{1963ApJ...137..747H}
{Henize} K.~G.,  {Westerlund} B.~E.,  1963, \mn@doi [\apj] {10.1086/147552},
  \href {http://adsabs.harvard.edu/abs/1963ApJ...137..747H} {137, 747}

\bibitem[\protect\citeauthoryear{{Hern{\'a}ndez-Garc{\'{\i}}a}
  et~al.,}{{Hern{\'a}ndez-Garc{\'{\i}}a} et~al.}{2017}]{2017A&A...603A.131H}
{Hern{\'a}ndez-Garc{\'{\i}}a} L.,  et~al., 2017, \mn@doi [\aap]
  {10.1051/0004-6361/201730530}, \href
  {https://ui.adsabs.harvard.edu/abs/2017A%26A...603A.131H} {603, A131}

\bibitem[\protect\citeauthoryear{{Hilditch}, {Howarth}  \&
  {Harries}}{{Hilditch} et~al.}{2005}]{2005MNRAS.357..304H}
{Hilditch} R.~W.,  {Howarth} I.~D.,   {Harries} T.~J.,  2005, \mn@doi [\mnras]
  {10.1111/j.1365-2966.2005.08653.x}, \href
  {http://adsabs.harvard.edu/abs/2005MNRAS.357..304H} {357, 304}

\bibitem[\protect\citeauthoryear{{Hotan} et~al.,}{{Hotan}
  et~al.}{2014}]{2014PASA...31...41H}
{Hotan} A.~W.,  et~al., 2014, \mn@doi [\pasa] {10.1017/pasa.2014.36}, \href
  {http://adsabs.harvard.edu/abs/2014PASA...31...41H} {31, e041}

\bibitem[\protect\citeauthoryear{{Inoue}, {Koyama}  \& {Tanaka}}{{Inoue}
  et~al.}{1983}]{1983IAUS..101..535I}
{Inoue} H.,  {Koyama} K.,   {Tanaka} Y.,  1983, in {Danziger} J.,  {Gorenstein}
  P.,  eds,  IAU Symposium Vol. 101, Supernova Remnants and their X-ray
  Emission. pp 535--540

\bibitem[\protect\citeauthoryear{{Ivanov} et~al.,}{{Ivanov}
  et~al.}{2016}]{2016A&A...588A..93I}
{Ivanov} V.~D.,  et~al., 2016, \mn@doi [\aap] {10.1051/0004-6361/201527398},
  \href {http://adsabs.harvard.edu/abs/2016A%26A...588A..93I} {588, A93}

\bibitem[\protect\citeauthoryear{{Jacoby} \& {De Marco}}{{Jacoby} \& {De
  Marco}}{2002}]{2002AJ....123..269J}
{Jacoby} G.~H.,  {De Marco} O.,  2002, \mn@doi [\aj] {10.1086/324737}, \href
  {http://adsabs.harvard.edu/abs/2002AJ....123..269J} {123, 269}

\bibitem[\protect\citeauthoryear{{Johnston} et~al.,}{{Johnston}
  et~al.}{2008}]{2008ExA....22..151J}
{Johnston} S.,  et~al., 2008, \mn@doi [Experimental Astronomy]
  {10.1007/s10686-008-9124-7}, \href
  {http://adsabs.harvard.edu/abs/2008ExA....22..151J} {22, 151}

\bibitem[\protect\citeauthoryear{{Kushwaha}, {de Gouveia Dal Pino}, {Gupta}  \&
  {Wiita}}{{Kushwaha} et~al.}{2019}]{2019arXiv190110768K}
{Kushwaha} P.,  {de Gouveia Dal Pino} E.~M.,  {Gupta} A.~C.,   {Wiita} P.~J.,
  2019, arXiv e-prints, \href
  {http://adsabs.harvard.edu/abs/2019arXiv190110768K} {}

\bibitem[\protect\citeauthoryear{{Kwok}}{{Kwok}}{2005}]{2005JKAS...38..271K}
{Kwok} S.,  2005, \mn@doi [Journal of Korean Astronomical Society]
  {10.5303/JKAS.2005.38.2.271}, \href
  {http://adsabs.harvard.edu/abs/2005JKAS...38..271K} {38, 271}

\bibitem[\protect\citeauthoryear{{Kwok}}{{Kwok}}{2015}]{2015HiA....16..623K}
{Kwok} S.,  2015, \mn@doi [Highlights of Astronomy]
  {10.1017/S1743921314012526}, \href
  {http://adsabs.harvard.edu/abs/2015HiA....16..623K} {16, 623}

\bibitem[\protect\citeauthoryear{{Leverenz}, {Filipovi{\'c}}, {Boji{\v
  c}i{\'c}}, {Crawford}, {Collier}, {Grieve}, {Dra{\v s}kovi{\'c}}  \&
  {Reid}}{{Leverenz} et~al.}{2016}]{2016Ap&SS.361..108L}
{Leverenz} H.,  {Filipovi{\'c}} M.~D.,  {Boji{\v c}i{\'c}} I.~S.,  {Crawford}
  E.~J.,  {Collier} J.~D.,  {Grieve} K.,  {Dra{\v s}kovi{\'c}} D.,   {Reid}
  W.~A.,  2016, \mn@doi [\apss] {10.1007/s10509-016-2686-3}, \href
  {http://adsabs.harvard.edu/abs/2016Ap%26SS.361..108L} {361, 108}

\bibitem[\protect\citeauthoryear{{Leverenz}, {Filipovi{\'c}}, {Vukoti{\'c}},
  {Uro{\v s}evi{\'c}}  \& {Grieve}}{{Leverenz}
  et~al.}{2017}]{2017MNRAS.468.1794L}
{Leverenz} H.,  {Filipovi{\'c}} M.~D.,  {Vukoti{\'c}} B.,  {Uro{\v s}evi{\'c}}
  D.,   {Grieve} K.,  2017, \mn@doi [\mnras] {10.1093/mnras/stx555}, \href
  {http://adsabs.harvard.edu/abs/2017MNRAS.468.1794L} {468, 1794}

\bibitem[\protect\citeauthoryear{{Li} et~al.,}{{Li}
  et~al.}{2018}]{2018ApJS..235....1L}
{Li} D.,  et~al., 2018, \mn@doi [\apjs] {10.3847/1538-4365/aaa762}, \href
  {http://adsabs.harvard.edu/abs/2018ApJS..235....1L} {235, 1}

\bibitem[\protect\citeauthoryear{{Maggi} et~al.,}{{Maggi}
  et~al.}{2016}]{2016A&A...585A.162M}
{Maggi} P.,  et~al., 2016, \mn@doi [\aap] {10.1051/0004-6361/201526932}, \href
  {http://adsabs.harvard.edu/abs/2016A%26A...585A.162M} {585, A162}

\bibitem[\protect\citeauthoryear{{Mao}, {Gaensler}, {Stanimirovi{\'c}},
  {Haverkorn}, {McClure-Griffiths}, {Staveley-Smith}  \& {Dickey}}{{Mao}
  et~al.}{2008}]{2008ApJ...688.1029M}
{Mao} S.~A.,  {Gaensler} B.~M.,  {Stanimirovi{\'c}} S.,  {Haverkorn} M.,
  {McClure-Griffiths} N.~M.,  {Staveley-Smith} L.,   {Dickey} J.~M.,  2008,
  \mn@doi [\apj] {10.1086/590546}, \href
  {http://adsabs.harvard.edu/abs/2008ApJ...688.1029M} {688, 1029}

\bibitem[\protect\citeauthoryear{{Mao} et~al.,}{{Mao}
  et~al.}{2012}]{2012ApJ...759...25M}
{Mao} S.~A.,  et~al., 2012, \mn@doi [\apj] {10.1088/0004-637X/759/1/25}, \href
  {http://adsabs.harvard.edu/abs/2012ApJ...759...25M} {759, 25}

\bibitem[\protect\citeauthoryear{{Marigo}, {Girardi}, {Groenewegen}  \&
  {Weiss}}{{Marigo} et~al.}{2001}]{2001AandA...378..958M}
{Marigo} P.,  {Girardi} L.,  {Groenewegen} M.~A.~T.,   {Weiss} A.,  2001,
  \mn@doi [\aap] {10.1051/0004-6361:20011270}, 378, 958

\bibitem[\protect\citeauthoryear{{Mauch}, {Murphy}, {Buttery}, {Curran},
  {Hunstead}, {Piestrzynski}, {Robertson}  \& {Sadler}}{{Mauch}
  et~al.}{2003}]{2003MNRAS.342.1117M}
{Mauch} T.,  {Murphy} T.,  {Buttery} H.~J.,  {Curran} J.,  {Hunstead} R.~W.,
  {Piestrzynski} B.,  {Robertson} J.~G.,   {Sadler} E.~M.,  2003, \mn@doi
  [\mnras] {10.1046/j.1365-8711.2003.06605.x}, \href
  {http://adsabs.harvard.edu/abs/2003MNRAS.342.1117M} {342, 1117}

\bibitem[\protect\citeauthoryear{{McClure-Griffiths}
  et~al.,}{{McClure-Griffiths} et~al.}{2015}]{2015aska.confE.130M}
{McClure-Griffiths} N.~M.,  et~al., 2015, Advancing Astrophysics with the
  Square Kilometre Array (AASKA14), \href
  {http://adsabs.harvard.edu/abs/2015aska.confE.130M} {p.~130}

\bibitem[\protect\citeauthoryear{{McClure-Griffiths}
  et~al.,}{{McClure-Griffiths} et~al.}{2018}]{2018NatAs...2..901M}
{McClure-Griffiths} N.~M.,  et~al., 2018, \mn@doi [Nature Astronomy]
  {10.1038/s41550-018-0608-8}, \href
  {http://adsabs.harvard.edu/abs/2018NatAs...2..901M} {2, 901}

\bibitem[\protect\citeauthoryear{{McConnell}}{{McConnell}}{2017}]{mcconnell17}
{McConnell} D.,  2017, Technical report, {ACES memo 15: Observing with ASKAP:
  Optimisation for survey speed}.
CSIRO Australia Telescope National Facility

\bibitem[\protect\citeauthoryear{{McConnell} et~al.,}{{McConnell}
  et~al.}{2016}]{2016PASA...33...42M}
{McConnell} D.,  et~al., 2016, \mn@doi [\pasa] {10.1017/pasa.2016.37}, \href
  {http://adsabs.harvard.edu/abs/2016PASA...33...42M} {33, e042}

\bibitem[\protect\citeauthoryear{{McGee}, {Newton}  \& {Butler}}{{McGee}
  et~al.}{1976}]{1976AuJPh..29..329M}
{McGee} R.~X.,  {Newton} L.~M.,   {Butler} P.~W.,  1976, \mn@doi [Australian
  Journal of Physics] {10.1071/PH760329}, \href
  {http://adsabs.harvard.edu/abs/1976AuJPh..29..329M} {29, 329}

\bibitem[\protect\citeauthoryear{{McMullin}, {Waters}, {Schiebel}, {Young}  \&
  {Golap}}{{McMullin} et~al.}{2007}]{mcmullin2007casa}
{McMullin} J.~P.,  {Waters} B.,  {Schiebel} D.,  {Young} W.,   {Golap} K.,
  2007, in Astronomical data analysis software and systems XVI. p.~127

\bibitem[\protect\citeauthoryear{{Meixner} et~al.,}{{Meixner}
  et~al.}{2006}]{2006AJ....132.2268M}
{Meixner} M.,  et~al., 2006, \mn@doi [\aj] {10.1086/508185}, \href
  {http://adsabs.harvard.edu/abs/2006AJ....132.2268M} {132, 2268}

\bibitem[\protect\citeauthoryear{{Middelberg} et~al.,}{{Middelberg}
  et~al.}{2008}]{2008AJ....135.1276M}
{Middelberg} E.,  et~al., 2008, \mn@doi [\aj] {10.1088/0004-6256/135/4/1276},
  \href {http://adsabs.harvard.edu/abs/2008AJ....135.1276M} {135, 1276}

\bibitem[\protect\citeauthoryear{{Murphy} et~al.,}{{Murphy}
  et~al.}{2013}]{2013PASA...30....6M}
{Murphy} T.,  et~al., 2013, \mn@doi [\pasa] {10.1017/pasa.2012.006}, \href
  {http://adsabs.harvard.edu/abs/2013PASA...30....6M} {30, e006}

\bibitem[\protect\citeauthoryear{{Norris} et~al.,}{{Norris}
  et~al.}{2006}]{2006AJ....132.2409N}
{Norris} R.~P.,  et~al., 2006, \mn@doi [\aj] {10.1086/508275}, \href
  {http://adsabs.harvard.edu/abs/2006AJ....132.2409N} {132, 2409}

\bibitem[\protect\citeauthoryear{{Norris} et~al.,}{{Norris}
  et~al.}{2011}]{2011PASA...28..215N}
{Norris} R.~P.,  et~al., 2011, \mn@doi [\pasa] {10.1071/AS11021}, \href
  {http://adsabs.harvard.edu/abs/2011PASA...28..215N} {28, 215}

\bibitem[\protect\citeauthoryear{{O'Brien}, {Norris}, {Tothill}  \&
  {Filipovi{\'c}}}{{O'Brien} et~al.}{2018}]{2018MNRAS.481.5247O}
{O'Brien} A.~N.,  {Norris} R.~P.,  {Tothill} N.~F.~H.,   {Filipovi{\'c}} M.~D.,
   2018, \mn@doi [\mnras] {10.1093/mnras/sty2642}, \href
  {http://adsabs.harvard.edu/abs/2018MNRAS.481.5247O} {481, 5247}

\bibitem[\protect\citeauthoryear{{Oliveira} et~al.,}{{Oliveira}
  et~al.}{2013}]{2013MNRAS.428.3001O}
{Oliveira} J.~M.,  et~al., 2013, \mn@doi [\mnras] {10.1093/mnras/sts250}, \href
  {http://adsabs.harvard.edu/abs/2013MNRAS.428.3001O} {428, 3001}

\bibitem[\protect\citeauthoryear{{Olnon}}{{Olnon}}{1975}]{1975AandA....39..217O}
{Olnon} F.~M.,  1975, \aap, 39, 217

\bibitem[\protect\citeauthoryear{{Owen} et~al.,}{{Owen}
  et~al.}{2011}]{2011A&A...530A.132O}
{Owen} R.~A.,  et~al., 2011, \mn@doi [\aap] {10.1051/0004-6361/201116586},
  \href {http://adsabs.harvard.edu/abs/2011A%26A...530A.132O} {530, A132}

\bibitem[\protect\citeauthoryear{{Pavlovi{\'c}}, {Uro{\v s}evi{\'c}},
  {Arbutina}, {Orlando}, {Maxted}  \& {Filipovi{\'c}}}{{Pavlovi{\'c}}
  et~al.}{2018}]{2018ApJ...852...84P}
{Pavlovi{\'c}} M.~Z.,  {Uro{\v s}evi{\'c}} D.,  {Arbutina} B.,  {Orlando} S.,
  {Maxted} N.,   {Filipovi{\'c}} M.~D.,  2018, \mn@doi [\apj]
  {10.3847/1538-4357/aaa1e6}, \href
  {http://adsabs.harvard.edu/abs/2018ApJ...852...84P} {852, 84}

\bibitem[\protect\citeauthoryear{{Payne}, {Filipovi{\'c}}, {Reid}, {Jones},
  {Staveley-Smith}  \& {White}}{{Payne} et~al.}{2004}]{2004MNRAS.355...44P}
{Payne} J.~L.,  {Filipovi{\'c}} M.~D.,  {Reid} W.,  {Jones} P.~A.,
  {Staveley-Smith} L.,   {White} G.~L.,  2004, \mn@doi [\mnras]
  {10.1111/j.1365-2966.2004.08287.x}, \href
  {http://adsabs.harvard.edu/abs/2004MNRAS.355...44P} {355, 44}

\bibitem[\protect\citeauthoryear{{Payne}, {White}, {Filipovi{\'c}}  \&
  {Pannuti}}{{Payne} et~al.}{2007}]{2007MNRAS.376.1793P}
{Payne} J.~L.,  {White} G.~L.,  {Filipovi{\'c}} M.~D.,   {Pannuti} T.~G.,
  2007, \mn@doi [\mnras] {10.1111/j.1365-2966.2007.11561.x}, \href
  {http://adsabs.harvard.edu/abs/2007MNRAS.376.1793P} {376, 1793}

\bibitem[\protect\citeauthoryear{{Payne}, {Filipovi{\'c}}, {Crawford}, {de
  Horta}, {White}  \& {Stootman}}{{Payne} et~al.}{2008}]{2008SerAJ.176...65P}
{Payne} J.~L.,  {Filipovi{\'c}} M.~D.,  {Crawford} E.~J.,  {de Horta} A.~Y.,
  {White} G.~L.,   {Stootman} F.~H.,  2008, \mn@doi [Serbian Astronomical
  Journal] {10.2298/SAJ0876065P}, \href
  {http://adsabs.harvard.edu/abs/2008SerAJ.176...65P} {176, 65}

\bibitem[\protect\citeauthoryear{{Pietrzy{\'n}ski} et~al.,}{{Pietrzy{\'n}ski}
  et~al.}{2019}]{2019Natur.567..200P}
{Pietrzy{\'n}ski} G.,  et~al., 2019, \mn@doi [\nat]
  {10.1038/s41586-019-0999-4}, \href
  {http://adsabs.harvard.edu/abs/2019Natur.567..200P} {567, 200}

\bibitem[\protect\citeauthoryear{{Reid} \& {Parker}}{{Reid} \&
  {Parker}}{2010}]{2010MNRAS.405.1349R}
{Reid} W.~A.,  {Parker} Q.~A.,  2010, \mn@doi [\mnras]
  {10.1111/j.1365-2966.2010.16635.x}, \href
  {http://adsabs.harvard.edu/abs/2010MNRAS.405.1349R} {405, 1349}

\bibitem[\protect\citeauthoryear{{Reid}, {Payne}, {Filipovi{\'c}}, {Danforth},
  {Jones}, {White}  \& {Staveley-Smith}}{{Reid}
  et~al.}{2006}]{2006MNRAS.367.1379R}
{Reid} W.~A.,  {Payne} J.~L.,  {Filipovi{\'c}} M.~D.,  {Danforth} C.~W.,
  {Jones} P.~A.,  {White} G.~L.,   {Staveley-Smith} L.,  2006, \mn@doi [\mnras]
  {10.1111/j.1365-2966.2006.10017.x}, \href
  {http://adsabs.harvard.edu/abs/2006MNRAS.367.1379R} {367, 1379}

\bibitem[\protect\citeauthoryear{{Roper}, {McEntaffer}, {DeRoo},
  {Filipovi{\'c}}, {Wong}  \& {Crawford}}{{Roper}
  et~al.}{2015}]{2015ApJ...803..106R}
{Roper} Q.,  {McEntaffer} R.~L.,  {DeRoo} C.,  {Filipovi{\'c}} M.,  {Wong}
  G.~F.,   {Crawford} E.~J.,  2015, \mn@doi [\apj]
  {10.1088/0004-637X/803/2/106}, \href
  {http://adsabs.harvard.edu/abs/2015ApJ...803..106R} {803, 106}

\bibitem[\protect\citeauthoryear{{Sano} et~al.,}{{Sano}
  et~al.}{2019}]{2019arXiv190404836S}
{Sano} H.,  et~al., 2019, arXiv e-prints, \href
  {http://adsabs.harvard.edu/abs/2019arXiv190404836S} {}

\bibitem[\protect\citeauthoryear{{Sault}, {Teuben}  \& {Wright}}{{Sault}
  et~al.}{1995}]{1995ASPC...77..433S}
{Sault} R.~J.,  {Teuben} P.~J.,   {Wright} M.~C.~H.,  1995, in {Shaw} R.~A.,
  {Payne} H.~E.,   {Hayes} J.~J.~E.,  eds,  Astronomical Society of the Pacific
  Conference Series Vol. 77, Astronomical Data Analysis Software and Systems
  IV. p.~433 (\mn@eprint {} {astro-ph/0612759})

\bibitem[\protect\citeauthoryear{{Sch{\"o}nberner}, {Jacob}, {Steffen}  \&
  {Sandin}}{{Sch{\"o}nberner} et~al.}{2007}]{2007AandA...473..467S}
{Sch{\"o}nberner} D.,  {Jacob} R.,  {Steffen} M.,   {Sandin} C.,  2007, \mn@doi
  [\aap] {10.1051/0004-6361:20077437}, 473, 467

\bibitem[\protect\citeauthoryear{{Sturm} et~al.,}{{Sturm}
  et~al.}{2013}]{2013A&A...558A...3S}
{Sturm} R.,  et~al., 2013, \mn@doi [\aap] {10.1051/0004-6361/201219935}, \href
  {https://ui.adsabs.harvard.edu/abs/2013A&A...558A...3S} {558, A3}

\bibitem[\protect\citeauthoryear{{Takekoshi} et~al.,}{{Takekoshi}
  et~al.}{2013}]{2013ApJ...774L..30T}
{Takekoshi} T.,  et~al., 2013, \mn@doi [\apjl] {10.1088/2041-8205/774/2/L30},
  \href {http://adsabs.harvard.edu/abs/2013ApJ...774L..30T} {774, L30}

\bibitem[\protect\citeauthoryear{{Turtle}, {Ye}, {Amy}  \& {Nicholls}}{{Turtle}
  et~al.}{1998}]{1998PASA...15..280T}
{Turtle} A.~J.,  {Ye} T.,  {Amy} S.~W.,   {Nicholls} J.,  1998, \mn@doi [\pasa]
  {10.1071/AS98280}, \href {http://adsabs.harvard.edu/abs/1998PASA...15..280T}
  {15, 280}

\bibitem[\protect\citeauthoryear{Uro{\v{s}}evi{\'{c}}, Pavlovi{\'{c}}  \&
  Arbutina}{Uro{\v{s}}evi{\'{c}} et~al.}{2018}]{Uro_evi__2018}
Uro{\v{s}}evi{\'{c}} D.,  Pavlovi{\'{c}} M.~Z.,   Arbutina B.,  2018, \mn@doi
  [The Astrophysical Journal] {10.3847/1538-4357/aaac2d}, 855, 59

\bibitem[\protect\citeauthoryear{{Winkler} et~al.,}{{Winkler}
  et~al.}{2005}]{2005AAS...20713203W}
{Winkler} P.~F.,  et~al., 2005, in American Astronomical Society Meeting
  Abstracts. p.~1380

\bibitem[\protect\citeauthoryear{{Wong}, {Filipovi{\'c}}, {Crawford}, {de
  Horta}, {Galvin}, {Draskovic}  \& {Payne}}{{Wong}
  et~al.}{2011a}]{2011SerAJ.182...43W}
{Wong} G.~F.,  {Filipovi{\'c}} M.~D.,  {Crawford} E.~J.,  {de Horta} A.~Y.,
  {Galvin} T.,  {Draskovic} D.,   {Payne} J.~L.,  2011a, \mn@doi [Serbian
  Astronomical Journal] {10.2298/SAJ1182043W}, \href
  {http://adsabs.harvard.edu/abs/2011SerAJ.182...43W} {182, 43}

\bibitem[\protect\citeauthoryear{{Wong} et~al.,}{{Wong}
  et~al.}{2011b}]{2011SerAJ.183..103W}
{Wong} G.~F.,  et~al., 2011b, \mn@doi [Serbian Astronomical Journal]
  {10.2298/SAJ1183103W}, \href
  {http://adsabs.harvard.edu/abs/2011SerAJ.183..103W} {183, 103}

\bibitem[\protect\citeauthoryear{{Wong} et~al.,}{{Wong}
  et~al.}{2012a}]{2012SerAJ.184...93W}
{Wong} G.~F.,  et~al., 2012a, \mn@doi [Serbian Astronomical Journal]
  {10.2298/SAJ1284093W}, \href
  {http://adsabs.harvard.edu/abs/2012SerAJ.184...93W} {184, 93}

\bibitem[\protect\citeauthoryear{{Wong}, {Filipovi{\'c}}, {Crawford},
  {Tothill}, {De Horta}  \& {Galvin}}{{Wong}
  et~al.}{2012b}]{2012SerAJ.185...53W}
{Wong} G.~F.,  {Filipovi{\'c}} M.~D.,  {Crawford} E.~J.,  {Tothill} N.~F.~H.,
  {De Horta} A.~Y.,   {Galvin} T.~J.,  2012b, \mn@doi [Serbian Astronomical
  Journal] {10.2298/SAJ1285053W}, \href
  {http://adsabs.harvard.edu/abs/2012SerAJ.185...53W} {185, 53}

\bibitem[\protect\citeauthoryear{{Wright} \& {Otrupcek}}{{Wright} \&
  {Otrupcek}}{1990}]{1990PKS...C......0W}
{Wright} A.,  {Otrupcek} R.,  1990, in PKS Catalog (1990).

\bibitem[\protect\citeauthoryear{{Ye}, {Turtle}  \& {Kennicutt}}{{Ye}
  et~al.}{1991}]{1991MNRAS.249..722Y}
{Ye} T.,  {Turtle} A.~J.,   {Kennicutt} R.~C. J.,  1991, \mn@doi [\mnras]
  {10.1093/mnras/249.4.722}, \href
  {https://ui.adsabs.harvard.edu/abs/1991MNRAS.249..722Y} {249, 722}

\bibitem[\protect\citeauthoryear{{{\.Z}ywucka}, {Goyal}, {Jamrozy}, {Stawarz},
  {Ostrowski}, {Koz{\l}owski}  \& {Udalski}}{{{\.Z}ywucka}
  et~al.}{2018}]{2018ApJ...867..131Z}
{{\.Z}ywucka} N.,  {Goyal} A.,  {Jamrozy} M.,  {Stawarz} {\L}.,  {Ostrowski}
  M.,  {Koz{\l}owski} S.,   {Udalski} A.,  2018, \mn@doi [\apj]
  {10.3847/1538-4357/aae36d}, \href
  {http://adsabs.harvard.edu/abs/2018ApJ...867..131Z} {867, 131}

\makeatother
\end{thebibliography}


\bsp	
\end{document}